\documentclass[aps,pra,onecolumn,superscriptaddress,preprintnumbers,secnumroman,nofootinbib,12pt]{revtex4-2}
\usepackage[T1]{fontenc}
\usepackage{epsfig}
\usepackage{graphicx}
\usepackage[english]{babel}
\usepackage{hyphenat}
\usepackage{amsmath}
\usepackage{amssymb}
\usepackage{bbold}
\usepackage{mathtools}
\usepackage{mathrsfs}
\usepackage{slashed}
\usepackage{epstopdf}
\usepackage{xcolor}

\usepackage{braket}
\usepackage{booktabs}
\definecolor{lcolor}{rgb}{0.,0.0,0.}
\definecolor{citcolor}{rgb}{0,0.,0.5}
\usepackage[breaklinks,colorlinks,urlcolor=blue,citecolor=blue,linkcolor=blue]{hyperref}
\usepackage{multirow}
\usepackage{ltablex}
\usepackage[normalem]{ulem} %makes underlining possible: \uline \uwave \sout
\usepackage{changepage} % optional (not required)
\usepackage{lipsum}

\usepackage[a4paper,margin=1in]{geometry}
\usepackage{setspace}
\usepackage[e]{esvect}

\usepackage{media9}

\def\cP{{\cal P}}
\def\cK{{\cal K}}
\def\cM{{\cal M}}
\def\cD{{\cal D}}

\def\cO{{\cal O}}

\def\cS{{\cal S}}

\def\bell{{\boldsymbol \ell}}

\newcommand{\secn}[1]{Section~1}
\newcommand{\appn}[1]{Appendix~1}

\long\def\comment#1{ }

\def\cS{{\cal S}}

\def\and{\quad\text{and}\quad}

\def\q{{\boldsymbol q}}
\def\0{{\boldsymbol 0}}
\def\1{{\boldsymbol 1}}
\def\p{{\boldsymbol p}}
\def\l{{\boldsymbol l}}
\def\k{{\boldsymbol k}}
\def\K{{\boldsymbol K}}
\def\x{{\boldsymbol x}}
\def\y{{\boldsymbol y}}
\def\X{{\boldsymbol X}}

\def\r{{\boldsymbol r}}

\def\u{{\boldsymbol u}}
\def\v{{\boldsymbol v}}
\def\w{{\boldsymbol w}}
\def\b{{\boldsymbol b}}
\def\0{{\boldsymbol 0}}

\def\P{{\boldsymbol P}}
\def\X{{\boldsymbol X}}

\def\bDelta{{\boldsymbol \Delta}}

\def\bkappa{{\boldsymbol \kappa}}

\def\max{{\rm max}}

\renewcommand{\l}{\boldsymbol{l}}

\newcommand{\cC}{\mathcal{C}}

\renewcommand\u{\upsilon}

\newcommand\s{\sigma}

\def\u{{\boldsymbol u}}

\renewcommand{\part}{{\rm part}}
\newcommand{\be}{\begin{equation}}
\newcommand{\ee}{\end{equation}}
\newcommand{\bes}{\begin{subequations}}
\newcommand{\ees}{\end{subequations}}
\newcommand{\bea}{\begin{eqnarray}}
\newcommand{\eea}{\end{eqnarray}}

\newcommand{\nn}{\nonumber \\}

\def\bea#1\eea{\begin{align}#1\end{align}}
\newcommand{\bef}{\begin{figure}[h!tb]\centering}
\newcommand{\eef}{\end{figure}}

\allowdisplaybreaks

\newcommand\n{\vec n}

\begin{document}
\newgeometry{
  top=.5in,bottom=.9in,
  left=.5in,right=.5in,
  hcentering,
  includeheadfoot
}

\title{Observing Transverse Momentum Broadening from TEECs 
\\ in Heavy Ion Collisions}

\author{Jo\~{a}o Barata}
\email{jlourenco@bnl.gov}
\affiliation{Physics Division, Brookhaven National Laboratory, Upton, NY 11973, USA}

\author{Hai Tao Li}
\email{haitao.li@sdu.edu.cn}
\affiliation{School of Physics, Shandong University, Jinan, Shandong 250100, China}

\author{Andrey V. Sadofyev}
\email{andrey.sadofyev@ehu.eus}
\affiliation{Department of Physics, University of the Basque Country UPV/EHU, P.O. Box 644, 48080 Bilbao, Spain}
\affiliation{EHU Quantum Center, University of the Basque Country UPV/EHU, P.O. Box 644, 48080 Bilbao, Spain}
\affiliation{IKERBASQUE, Basque Foundation for Science, Plaza Euskadi 5, 48009 Bilbao, Spain}

\author{Jo\~{a}o M. Silva}
\email{jsilva@physics.ucla.edu}
\affiliation{Department of Physics and Astronomy, University of California, Los Angeles, CA 90095, USA}
\affiliation{Mani L. Bhaumik Institute for Theoretical Physics, University of California, Los Angeles, CA 90095, USA}

\begin{abstract}
We study the azimuthal dependence of transverse energy-energy correlators (TEECs) in dijet events in heavy-ion collisions. In the back-to-back region this observable resolves the energy flow around $\phi=\pi$ and is sensitive to medium-induced momentum broadening and the response of the medium. To isolate final-state effects, we consider a quark-antiquark pair produced in a colour-singlet hard process and propagating through a homogeneous background medium. Starting from the vacuum TMD factorization formula, we combine NLL$'$ resummation with medium modifications at first order in the strong coupling.  
 For the parameters studied, (classical) momentum broadening lowers the back-to-back peak and leaves the distribution elsewhere nearly unchanged, while radiative corrections significantly suppress the peak further and enhance larger angular decorrelations. We additionally include the medium response sourced by each jet, as a mean energy flow. This spreads the deposited energy over the whole azimuth, so that the contribution to the TEEC varies slowly across the back-to-back region, is small at the peak and grows toward larger angles.
\end{abstract}

\maketitle

\clearpage
\restoregeometry
\tableofcontents

\clearpage

\section{Introduction}
High energy partons propagating through the bulk matter produced in heavy-ion collisions undergo repeated interactions with this medium, acquiring transverse momentum and radiating additional partons. These processes modify the angular distribution of the hadronic final state energy flow, providing a way to probe the microscopic properties of the underlying matter. 
While substantial progress has been made in understanding medium modifications of jet substructure and energy correlators in collinear configurations in nuclear collisions~\cite{Andres:2024ksi,Yang:2023dwc,Barata:2023bhh,Budhraja:2025ulx,Apolinario:2025vtx,Singh:2026qsa,Yang:2026bsk,Bossi:2026mdz,Barata:2024wsu,Barata:2025uxp}, back-to-back topologies offer another setting in which medium modifications can be theoretically computed and experimentally tested through angular decorrelations, see e.g.~\cite{Mueller:2016xoc,Mueller:2016gko,Chen:2016vem,STAR:2017hhs}. These angular distributions are, however, not exclusively sensitive to medium effects, as they also receive significant contributions from perturbative and non-perturbative vacuum QCD processes, see discussions in~\cite{Mueller:2016xoc,Mueller:2016gko}. Related studies have also explored forward dijet azimuthal correlations as a probe of medium-induced transverse-momentum broadening~\cite{Jia:2019qbl}.  Therefore, developing a clear understanding of back-to-back configurations requires controlled observables that allow the relative size of the different physical effects to be assessed.

Transverse energy-energy correlators (TEECs)~\cite{Ali:1984yp,Gao:2019ojf,Li:2020bub} provide a natural framework for this purpose. These observables measure azimuthal correlations between final-state hadronic transverse energy flows, and their inclusive definition does not require assigning the measured particles to individual jets, see Fig.~\ref{fig:cartoon}. They can be understood as azimuthal projections of two-point functions of the energy flow detector operator~\cite{Hofman:2008ar,Basham:1978bw,Sterman:1975xv},
\begin{equation}\label{eq:energy_flow_operator}
 \mathcal E(\vec n)
 =\lim_{r\to\infty}r^2\int_0^\infty dt\,
 n_iT^{0i}(t,r\vec n)\, ,
\end{equation}
where $T^{\mu\nu}$ is the energy momentum tensor, and $\vec n$ labels the detector position on a large $r$ sphere. Its action on a state $|X\rangle$, $\mathcal E(\vec n)|X\rangle =\sum_{a\in X}E_a\,\delta^{(2)}(\vec n-\vec n_a)|X\rangle$, with the sum over single particle states, allows for clear experimental interpretation of the observable in terms of asymptotic calorimetric detectors, and as a result, the $\mathcal{E}(\vec n)$ operator provides a direct connection between quantum field theory and collider phenomenology. Taking the beam as the polar axis, one defines the total transverse-energy operator as $\widehat E_T=\int d\Omega\,\sin\theta\,\mathcal E(\vec n)$, with $\widehat E_T(X)|X\rangle = \sum_{a\in X}E_a\sin\theta_a|X\rangle$, such that the normalized TEEC distribution is then
\begin{align}\label{eq:teec_operator_definition}
 \frac{1}{\sigma}\frac{d\Sigma}{d\tau}
 &=
 \int d\Omega_1\,d\Omega_2\,
 \sin\theta_1\sin\theta_2\,
\frac{ \left\langle
 \mathcal E(\vec n_1)\mathcal E(\vec n_2)
 \right\rangle_\sigma}{(\sum_iE_{T,i})^2}\, 
 \delta\!\left(\tau-\frac{1+\cos\phi_{12}}{2}\right)\\ \notag
 &=
 \frac{1}{\sigma}\sum_X\int d\sigma_X
 \sum_{a,b\in X}
 \frac{E_{T,a}E_{T,b}}{(\sum_iE_{T,i})^2} \, 
 \delta\!\left(\tau-\frac{1+\cos\phi_{ab}}{2}\right),
\end{align}
where $\phi_{12}\in[0,\pi]$ is the relative azimuthal angle, $E_T=\sum_iE_{T,i}$ is the total transverse energy of the event, and we
denote the expectation value as the cross section averaged over final states, i.e. $ \langle\widehat{\mathcal O}\rangle_\sigma
 \equiv\frac{1}{\sigma}\sum_X\int d\sigma_X\,\mathcal O(X)$. Equation~\eqref{eq:teec_operator_definition} provides three equivalent ways to compute the TEEC distribution, either from a cross section based formulation, from the correlation functions of flow operators, or from a particle based approach; we shall explore these different forms to compute distinct contributions to the observable, see~\cite{Belin:2026wkc,Belitsky:2013xxa} for further discussion on this aspect. Indeed, in the back-to-back limit, the cross section based formulation is particularly
well suited to high energy QCD calculations, as it allows the separation of hard, collinear and soft dynamics within transverse momentum dependent (TMD) factorization and the resummation of large logarithms associated with
small angular decorrelations~\cite{Moult:2018jzp,Gao:2019ojf}. This framework has been applied to EECs at leptonic colliders~\cite{Moult:2018jzp} and TEECs at hadronic colliders~\cite{Gao:2019ojf} and in deep-inelastic scattering~\cite{Li:2020bub,Li:2021txc}, reaching high perturbative
accuracy~\cite{Ebert:2020sfi,Gao:2023ivm}, with extensions to higher-point correlators~\cite{Budhraja:2026pyi}. The azimuthal dependence of back-to-back EECs has also been explored as a probe of TMD partonic structures~\cite{Kang:2024otf}. Related developments have extended the study of back-to-back TEECs to the small-$x$ regime at the EIC and to vector-boson-tagged hadron production in $pp$ and $pA$ collisions~\cite{Kang:2023oqj,Kang:2025vjk}.
Phenomenological studies of transverse energy correlators in QCD  have enabled the experimental determinations of the strong coupling and its scale dependence~\cite{ATLAS:2023tgo}.

In this work, we study the TEEC in the back-to-back limit, for a colour-singlet quark–antiquark pair produced in a hard process and propagating through a homogeneous background medium. This setup simplifies the case of \textit{realistic} hadronic/nuclear scattering processes where nuclear initial state effects have to be considered, allowing us to isolate final state modifications. Under a set of additional assumptions specified below, we retain the vacuum factorized structure and incorporate medium effects through modifications to the functions entering the back-to-back TMD factorization formula. We combine the vacuum distribution at next-to-leading logarithmic prime (NLL$'$) accuracy with medium-induced corrections at first order in the strong coupling in the dense medium limit. Using the definition of the TEEC as a correlator of asymptotic detectors, we further estimate the non-perturbative contributions associated with the medium response to the dijet system, following~\cite{Barata:2024ukm,Barata:2025fzd,Zhao:2025ogc}. For the TEEC, a response spread over a broad range of azimuthal angles is expected to vary slowly across a narrow back-to-back window. Its angular dependence can consequently differ from that of the perturbative contribution, which is concentrated near the back-to-back limit. Including both contributions allows their relative importance to be assessed as a function of the angular decorrelation, providing guidance for experimental studies. Numerical studies of the resulting TEEC distribution show that momentum broadening lowers the back-to-back peak, while radiative contributions produce a stronger suppression and enhance larger angular decorrelations. The broad medium response gives a slowly varying contribution whose relative importance grows away from the peak. These results illustrate the sensitivity of the TEEC to distinct components of the in-medium evolution. They also show that a quantitative interpretation of the peak requires control over medium-induced radiative corrections, which were neglected in previous studies of the back-to-back limit in heavy-ion collisions~\cite{Mueller:2016gko,Mueller:2016xoc}.

The paper is organized as follows. Section~\ref{sec:vac} reviews the vacuum factorization and resummation of the back-to-back TEEC. Section~\ref{sec:medium} introduces the medium-modified jet functions and separates (classical) momentum broadening from the medium-induced radiative correction. The analytic treatment of these contributions is presented in Sec.~\ref{sec:time_regions}. Section~\ref{sec:n_response} describes the contribution from the medium response. Numerical results are presented in Sec.~\ref{sec:numerics}. We summarize our findings and conclude in Sec.~\ref{sec:conclusion}.

\section{Back-to-Back limit of the TEEC in vacuum}\label{sec:vac}

\begin{figure}[h!]
    \centering
    \includegraphics[width=0.8\linewidth]{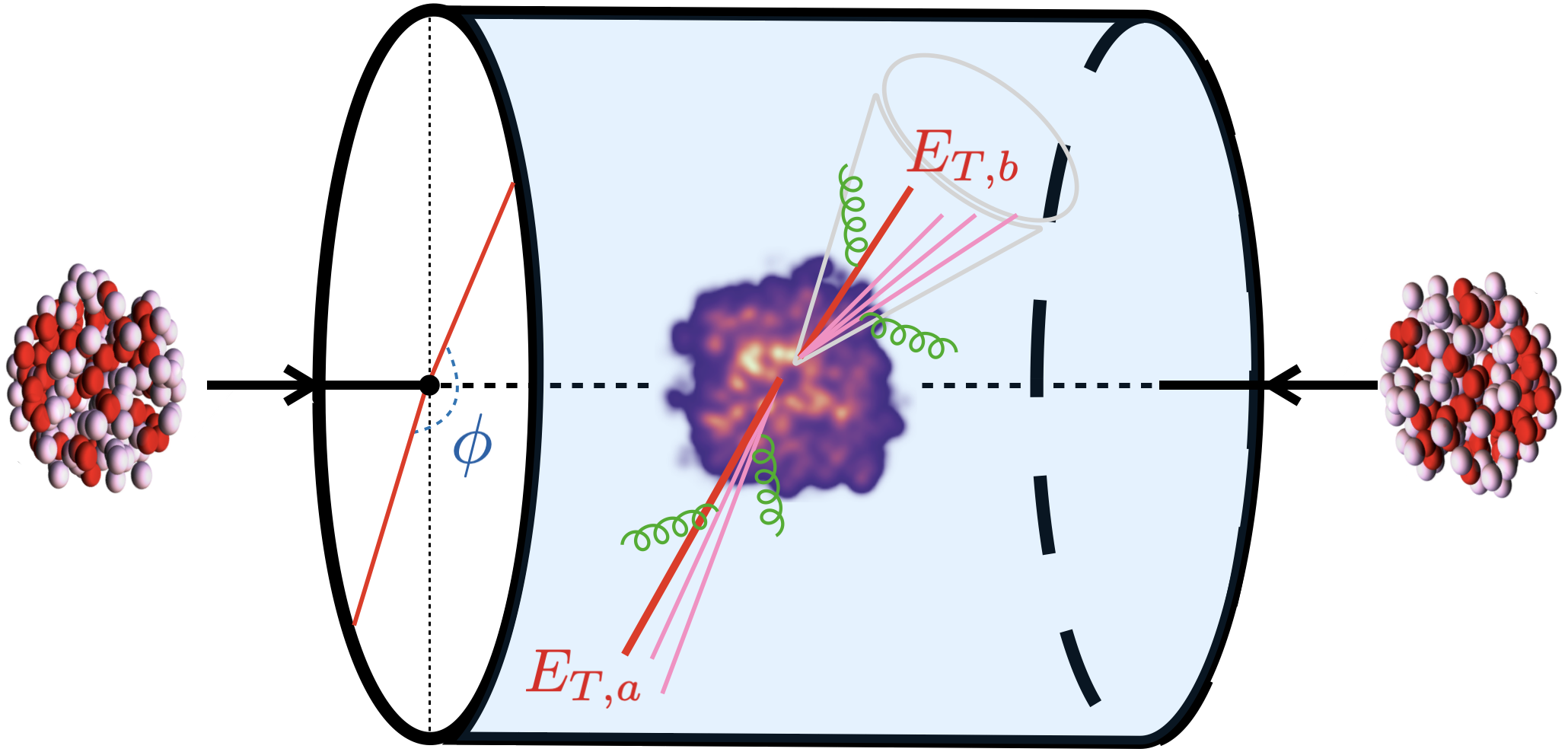}
    \caption{Illustration of the event topology being considered. Two nearly back-to-back jets are produced from a large center of mass hard interaction, and their production and TEEC measurement are represented by Eq.~\eqref{eq:factorized}. Collinear hard radiation is indicated by pink lines, while the hadrons entering the measurement are identified by red lines with their respective transverse energies. Soft radiation is indicated by green gluon lines.}
    \label{fig:cartoon}
\end{figure}

In QCD, the TEEC can be defined in terms of the transverse energy weighted differential inclusive cross section for producing two outgoing hadrons from an initial hadronic collision~\cite{Ali:1984yp,Gao:2019ojf,Gao:2023ivm}:
\begin{align}
    \frac{1}{\sigma}\frac{d \Sigma}{d \tau}
    &= \frac{1}{\sigma}\sum_{a, b} \int d \sigma_{AA \rightarrow a+b+X} \frac{E_{T, a} E_{T, b}}{(\sum_iE_{T,i})^2} \delta\!\left(\tau-\frac{1+\cos\phi_{ab}}{2}\right)\, ,
\end{align}
with $\phi_{ab}$ the relative azimuthal angle between the measured hadrons, and $E_{T,j}$ the transverse energy of hadrons $j$ (see Fig.~\ref{fig:cartoon} and Eq.~\eqref{eq:teec_operator_definition}). In the $\tau\to 0$ back-to-back limit, and assuming a large center of mass energy, $d\sigma_{AA\to a+b+X}$ is described by the TMD factorization formula~\cite{Gao:2019ojf,Gao:2023ivm, Collins:1981uk}, which at leading power gives the TEEC distribution
\begin{align}\label{eq:factorized}
  \frac{d \Sigma^{\rm LP}}{d \tau}& \propto 
     {\rm tr} \left[\mathbf{H}^{f_{1} f_{2} \rightarrow f_{3} f_{4}}\mathbf{S}\right]  \otimes B_{f_{1} / N_{1}}  \otimes B_{f_{2} / N_{2}}  \otimes J_{f_{3}} \otimes J_{f_{4}} \, .
\end{align}
The beam ($B$), hard ($\mathbf{H}$), soft ($\mathbf{S}$), and jet ($J$) functions depend on the factorization scale $\mu$, while the beam, soft, and jet functions also depend on the rapidity scale $\nu$, see~\cite{Gao:2019ojf} for further details. In the context of heavy ion collisions, Eq.~\eqref{eq:factorized} can be employed to describe the production of high energy hadrons/jets which survive the propagation through the plasma at a high intrinsic momentum scale~\cite{Mueller:2016gko} (see e.g.~\cite{Singh:2025scb,Singh:2024pwr,Singh:2024vwb} for recent related works).
Nonetheless, as the presence of the bulk matter introduces new scales, one is required to compute medium dependent corrections to the different functions entering the above factorization formula. Of course such a computation is only sensible assuming that Eq.~\eqref{eq:factorized} still holds in heavy ion collisions, which is a reasonable phenomenological assumption if we consider high $p_T$ dijets. 

The origins of the modifications to the different functions entering the factorization formula are rather distinct. Corrections to the beam and hard functions will enter via cold nuclear matter effects, and shall not be discussed further in this work. In contrast, the soft factor, due to the production of stimulated soft \textit{bremsstrahlung} radiation, and the jet function, where hard/collinear medium-induced branching contributions change the jet fragmentation pattern, are modified due to the late time evolution of the jet cascade. These modifications can be computed within a perturbative QCD picture, and Sec.~\ref{sec:medium} specifies which of them are included in this work. The full calculation of all these corrections, in realistic heavy-ion environments, can only be performed using fully numerical methods (see e.g.~\cite{Leitao:2026fgh, Barata:2026icn} for recent examples). However, qualitative effects of bulk dependent physics can be explored using some simplifying assumptions, which we discuss below, while the structure of the corrections being computed remains generic.

In this work, as we focus on resolving the effects on final state fragmentation, we consider a simplified idealized scenario of an $e^+e^-$ collision occurring in the presence of a bulk background state, representing the quark-gluon plasma produced in heavy-ion collisions.  The vacuum process is $e^+e^-\to\gamma^*\to q\bar q+X$
for massless quarks, with $Q=\sqrt{s}$.
With transverse energies measured relative to the lepton beam axis
$\hat z$, the TEEC is written as
\begin{equation}\label{eq:ee_teec_definition}
 \frac{1}{\sigma}\frac{d\Sigma}{d\tau}
 =\frac{1}{\sigma}\sum_{a,b}\int d\sigma_{\gamma^*\to a+b+X}
 \frac{E_{T,a}E_{T,b}}{(\sum_{i}E_{T,i})^2}
 \delta\!\left(\tau-\frac{1+\cos\phi_{ab}}{2}\right).
\end{equation}
%The pair sum over $a$ and $b$ is unordered 
%The normalization is to the total cross section.  Note that the observable is defined for leading-dijet configurations with finite transverse momentum. 
At Born level, we denote the momenta of the final state quark and antiquark by
$\vec p_0$ and $-\vec p_0$, with $|\vec p_0|=Q/2$, write $c=\cos\theta_0$ for the polar angle of the quark
with respect to the beam, and
\begin{equation}
 p_T=\frac Q2\sin\theta_0,\qquad
 \hat n=\frac{\vec p_0}{|\vec p_0|},\qquad
 \hat y=\frac{\hat z\times\hat n}{|\hat z\times\hat n|}.
\end{equation}
The TEEC is defined with finite jet $p_T$ where $\sin\theta_0$ is of order one in the power expansion.
The quantity $p_T$ labels 
the parton in leading order of the reconstructed jet transverse momentum, since to define the TEEC two leading jets are required in the event.
%For a photon exchange, the normalized angular weight is
%$B(c)=3(1+c^2)/8$, and we factor it out by defining
%\begin{equation}\label{eq:ee_born_normalization}
% \frac{1}{\sigma_{\rm Born}}\frac{d^2\Sigma_{\rm LP}}{dc\,d\tau}
% =B(c)F^{\rm TEEC}(\tau,p_T).
%\end{equation}
%{\color{red}JS: I suggest using this: 
For a photon exchange, $d\sigma_{\rm Born}/dc=\sigma_{\rm Born}B(c)$, with
$B(c)=3(1+c^2)/8$, and we normalize the distribution at fixed $c$ to it,
\[
 \frac{1}{\sigma}\frac{d\Sigma}{d\tau}
 \equiv\Big(\frac{d\sigma_{\rm Born}}{dc}\Big)^{-1}\frac{d^2\Sigma_{\rm LP}}{dc\,d\tau},
\]
that is, the TEEC of the events at fixed $c$, normalized to their leading power cross section $d\sigma_{\rm Born}/dc$.
At $c=0$, hence $p_T=Q/2$, this conditional kernel is finite, while a density in $p_T^2$ instead contains the angular Jacobian. From here on, distributions are at fixed $c$, equivalently fixed $p_T$, and the figures use $c=0$, that is $p_T=Q/2$ with both jets at zero rapidity.

We define $n^\mu=(1,\hat n)$, $\bar n^\mu=(1,-\hat n)$ along the direction of final state jets, so that
$n\cdot\bar n=2$, and a small expansion parameter
$\lambda\sim q_y/Q\sim\sqrt{\tau}$ with $q_y=2p_T\sqrt{\tau}$.
The relevant vacuum modes have the SCET$_{\rm II}$ scaling~\cite{Bauer:2003mga,Chiu:2012ir}
\begin{equation}\label{eq:ee_power_counting}
 (n\cdot p,\bar n\cdot p,\mathbf p_\perp)\sim
 \begin{cases}
 Q(\lambda^2,1,\lambda),&n\text{-collinear},\\
 Q(1,\lambda^2,\lambda),&\bar n\text{-collinear},\\
 Q(\lambda,\lambda,\lambda),&\text{soft}.
 \end{cases}
\end{equation}
The collinear and soft modes therefore have the same characteristic
virtuality $q_y^2$ but different rapidities. Their separation thus requires both
virtuality and rapidity renormalization scales, denoted by $\mu$ and $\nu$. For hadrons in the  two collinear sectors, at Born level, the momentum fractions of final state hadrons with respect to their parent partons are $x=2E_a/Q$ and
$\bar x=2E_b/Q$. To leading power,
$E_{T,a}=xp_T$, $E_{T,b}=\bar x p_T$ and $\sum_iE_{T,i}=2p_T$, so their pair weight is $x\bar x/2$.  Let $\bell$ and $\bar{\bell}$ be the fragmentation momenta relative to the collinear sectors after separating their total soft recoil, and let $k_{s,y}$ be the total soft momentum in the measured direction, so that $k_{a,y}=\ell_y/x$ and $k_{b,y}=\bar\ell_y/\bar x$.
For $\phi_{ab}=\pi-\delta\phi$ the measurement at leading power is
\begin{equation}\label{eq:teec_measurement_lp}
 \tau=\frac{\delta\phi^2}{4}+\mathcal O(\delta\phi^4),\qquad
 \tau\big|_{\rm LP}
 =\frac{(k_{a,y}+k_{b,y}-k_{s,y})^2}{4p_T^2}.
\end{equation}

In the $\tau\to 0$ limit, a TMD factorization analogous to the one alluded for hadronic collisions exists~\cite{Moult:2018jzp}:
\begin{align}\label{eq:vacuum_teec_code}
 %F^{\rm TEEC}(\tau,p_T)
 \frac{1}{\sigma}\frac{d\Sigma}{d\tau}
 &=\frac{p_T}{\sqrt\tau}\int_{-\infty}^{\infty}\frac{db}{2\pi}
 e^{-2ibp_T\sqrt\tau}\, H(Q,\mu)J_q(b,Q;\mu,\nu)J_{\bar q}(b,Q;\mu,\nu)S(b;\mu,\nu),
\end{align} %\flag{JS: I suggest replacing by $\frac{1}{\sigma}\frac{d\Sigma}{d\tau}$ as per Eq.~\eqref{eq:ee_born_normalization}} 
where the functions of $b$ are evaluated along $\hat y$, as in Eq.~\eqref{eq:ee_teec_bspace}.\footnote{We note that the functions entering factorization theorem are nonetheless different from the ones in the hadronic case, see~\cite{Neill:2022lqx} for a summary.} As the soft function is not affected by the measurement, we shall focus on the energy weighted inclusive jet function, which can be written as
\begin{equation}\label{eq:ee_jet_tmdff}
 J_q(\b,Q;\mu,\nu)
 =\sum_h\int_0^1dx\,x\int_{\bell}
 e^{i\b\cdot\bell/x}F_{q\to h}(x,\bell;\mu,\nu)
 =\sum_h\int_0^1dx\,x\,
 \widetilde F_{q\to h}(x,\b/x;\mu,\nu)\, ,
\end{equation}
and by charge conjugation $J_{\bar q}=J_q$. Here we have defined the jet function in terms of the TMD fragmentation function (TMDFF) $F_{i\to h}(x,\bell;\mu,\nu)$, which describes a hadron carrying fraction $x$ and transverse momentum $\bell$ relative to its parent parton. This can be perturbatively matched to the collinear fragmentation function 
\begin{align}\label{eq:tmd_ope}
F_{i\to a}(x,\bell)
=\sum_j\int\frac{d\zeta}{\zeta^3}\,
I_{ij}\!\left(\frac{x}{\zeta},\frac{\bell}{\zeta}\right)
f_{j\to a}(\zeta,\mu_c)+\cO\!\left(\frac{\Lambda_{\rm QCD}^2}{\bell^2}\right)\, ,
\end{align}
such that, using $x=z\zeta$, $\bell=\zeta\k$ and
the momentum sum rule $\sum_h\int_0^1d\zeta\,\zeta f_{j\to h}(\zeta,\mu)=1$, one finds
\begin{equation}\label{eq:ee_jet_matching}
 J_q(\b,Q;\mu,\nu)
 =\sum_j\int_0^1dz\,z\int_{\k}
 e^{i\b\cdot\k/z}I_{qj}(z,\k;\mu,\nu)
 =\sum_j\int_0^1dz\,z\,\widetilde I_{qj}(z,\b/z;\mu,\nu),
\end{equation}
Thus, the jet function is a moment of the matching coefficient due to the momentum sum rule. This reflects the inclusive nature of the energy flow measurement, in contrast with the particle based observables, which would depend on the full distribution. Combining this with Eq.~\eqref{eq:ee_jet_tmdff}, we find the simplified form\footnote{In momentum space we use $\int_{\q_1,\ldots,\q_n}\equiv\prod_{j=1}^n\int d^2\q_j/(2\pi)^2$. Coordinate space integrations are written with the ordinary measure $d^2\b$.}
\begin{align}\label{eq:ee_tmd_convolution}
 \frac{1}{\sigma}\frac{d\Sigma}{d\tau}
 ={}&\frac{H(Q,\mu)}{2}\sum_{h,\bar h}\int_0^1dx\,d\bar x\,x\bar x
 \int_{\bell,\bar{\bell},\k_s}
 F_{q\to h}(x,\bell;\mu,\nu)
 F_{\bar q\to\bar h}(\bar x,\bar{\bell};\mu,\nu)
 S(\k_s;\mu,\nu)\\
 &\times\delta\!\left(\tau-
 \frac{(\ell_y/x+\bar\ell_y/\bar x-k_{s,y})^2}{4p_T^2}\right)\nn
={}&
\frac{H(Q,\mu)}{2}
\sum_{h,\bar h}
\int_0^1 dx
\int_0^1 d\bar x\,x\bar x
\int_{-\infty}^{+\infty}\frac{db}{2\pi} \frac{2p_T}{\sqrt{\tau}}
\cos\left(2bp_T\sqrt{\tau}\right)
\nonumber\\
&\times
\widetilde F_{q\to h}
\left(x,\frac{b}{x}\hat\y;\mu,\nu\right)
\widetilde F_{\bar q\to\bar h}
\left(\bar x,\frac{b}{\bar x}\hat\y;\mu,\nu\right)
\widetilde S\left(-b\hat\y;\mu,\nu\right).
\label{eq:ee_teec_bspace}
\end{align}
%\flag{JS: I suggest replacing by $\frac{1}{\sigma}\frac{d\Sigma}{d\tau}$ as per Eq.~\eqref{eq:ee_born_normalization}}
The dependence on the transverse momenta
\(\bell\), \(\bar{\bell}\), and
\(\k_s\) has been factorized into a product of
Fourier-transformed TMD fragmentation functions and the soft
function, while the hard factor has a trivial convolution structure. Since only the \(\hat y\) components are measured, the remaining
integral is over the impact parameter \(b\) along \(\hat y\), and its
kernel \(\cos(2bp_T\sqrt{\tau})\) implements the constraint
on the measured variable \(\tau\).

The resummation in the TMD limit is achieved by solving the renormalization group (RG) equations of the hard, jet, and soft functions:
\begin{align}\label{eq:ee_mu_rg}
 \frac{d\ln H}{d\ln\mu}
 &=2\Gamma\ln\frac{Q^2}{\mu^2}+2\gamma_H,\nn
 \frac{d\ln J_q}{d\ln\mu}
 &=-\Gamma\ln\frac{Q^2}{\nu^2}+2\gamma_J,\nn
 \frac{d\ln S}{d\ln\mu}
 &=2\Gamma\ln\frac{\mu^2}{\nu^2}-2\gamma_s.
\end{align}
RG consistency requires $\gamma_H+2\gamma_J-\gamma_s=0$. The RG evolution in the rapidity scale reads
\begin{align}\label{eq:ee_nu_rg}
 \mathcal A(b,\mu)
 &=\int_{\mu_b^2}^{\mu^2}\frac{dm^2}{m^2}\Gamma(\alpha_s(m))
   -\gamma_r(\alpha_s(\mu_b)),\nn
 \frac{d\ln J_q}{d\ln\nu}&=\mathcal A(b,\mu)\,, \quad 
 \frac{d\ln S}{d\ln\nu}=-2\mathcal A(b,\mu).
\end{align}
with $\mu_b=b_0/b$, $b_0=2e^{-\gamma_E}$. 
Since the primary goal of this work is to investigate medium-induced effects, we perform the vacuum resummation at next-to-leading logarithmic prime (NLL$^\prime$) accuracy, retaining the cusp anomalous dimension through two loops, the non-cusp anomalous dimensions at one loop, and the one-loop remainders in the functions. We evaluate the perturbative functions at the profiled transverse
separation $\bar b$, defined by
\begin{equation}\label{eq:ee_b_profile}
 b_* = \frac{b}{\sqrt{1+b^2/b_{\max}^2}},\qquad
 \bar b = \sqrt{b_*^2+\frac{b_0^2}{Q^2}},\qquad
 \bar{\mu}_b = \frac{b_0}{\bar b},
\end{equation}
where $b_{\max}=b_0\,\mathrm{GeV}^{-1}$~\cite{Kang:2024dja}.
The $b_*$ prescription freezes the perturbative scale at large $b$,
and the additional modification at small $b$ ensures $\bar{\mu}_b\leq Q$.
The Fourier phase retains the physical impact parameter $b$.
Our central scale choices are
\begin{equation}\label{eq:ee_central_scales}
 \mu_H=\nu_J=Q,\qquad \mu_J=\mu_s\equiv\mu=\bar{\mu}_b, \qquad \nu_s=\mu_b. 
\end{equation}
The invariant mass $Q$ sets the hard scale, and for Born level jets perpendicular to the beam, $Q=2p_T$. We use fixed $n_f=5$ and $\alpha_s(m_Z)=0.118$, with $m_Z=91.1876\,\mathrm{GeV}$. The coupling is held constant on any part of an evolution interval below $1\,\mathrm{GeV}$, while the scale logarithms and evolution intervals are retained.  We estimate perturbative uncertainties by varying $\mu_H$, the common scale $\mu$, and $\nu_s$ independently by factors of two about Eq.~\eqref{eq:ee_central_scales}. Each variation changes a single scale, with the remaining scales at their central values and $\nu_J=Q$ fixed. The band is the envelope of these six variations and the central prediction. 

Non-perturbative effects enter as a multiplicative factor in the impact parameter integrand. We adopt the functional form of Fit~1 in~\cite{Kang:2024dja},
\begin{equation}\label{eq:ee_np_model}
 F_{\rm NP}(b,Q)=
 \exp\!\left[-2a_1b^{a_2}-4D_{\rm NP}(b)\ln\frac{Q}{\bar{\mu}_b}\right],
 \qquad D_{\rm NP}(b)=g_2bb_*,
\end{equation}
with $a_1=0.530\,\mathrm{GeV}^{a_2}$, $a_2=1.152$ and
$g_2=0.076\,\mathrm{GeV}^2$. The first term in the exponent accounts for
the two jet factors, each parametrized by
$\exp[-S_{\rm NP}(b)]$ with $S_{\rm NP}(b)=a_1b^{a_2}$.
The second follows from Collins--Soper evolution between
$\zeta_i=\bar{\mu}_b^2$ and $\zeta_f=Q^2$, in the convention
$K_{\rm NP}=-2D_{\rm NP}$.
The parametrization $D_{\rm NP}\propto bb_*$ is also used in~\cite{Scimemi:2019cmh}.

\section{Final state medium modifications to the TEEC}\label{sec:medium}

In this section, we construct the medium-modified jet functions that enter
the back-to-back TEEC, in the simplified $e^+e^-$ configuration of Sec.~\ref{sec:vac}
with a medium traversed by the outgoing quark and antiquark.
In Sec.~\ref{sec:n_factorization} the
jet function is matched onto collinear fragmentation functions.
Section~\ref{sec:n_broadening} gives the classical broadening term, and
Sec.~\ref{sec:n_radiative} gives the $q\to qg$ cross section for dense matter.
That subsection also splits the $\cO(\alpha_s)$ correction into time
regions, separating a broadened vacuum term from a medium-induced
radiative correction, discusses the large $|\p|$ behaviour of the latter
and introduces the numerical regulator used in
Sec.~\ref{sec:time_regions}.

\subsection{Assumptions and factorized form}
\label{sec:n_factorization}

We make the following assumptions on the medium dependence of the
factorization in Eq.~\eqref{eq:ee_tmd_convolution}, see, e.g.,~\cite{Singh:2025scb,Singh:2024pwr,Singh:2024vwb,Lamas:2026ntq,Ke:2025ibt} for related approaches to in-medium factorization.
\begin{itemize}
\item \emph{Independent propagation.} 
The outgoing quark and antiquark
interact independently with the medium. A treatment of correlated
propagation would require the complete medium-dependent back-to-back
production cross section.\footnote{Relaxing this approximation would require medium-induced correlations between the outgoing quark and antiquark, described by joint Wilson-line correlators of the two legs rather than products of single-leg correlators. Their relation to the double light-cone limit~\cite{Korchemsky:2019nzm} of the correlation functions underlying energy correlators~\cite{Belitsky:2013xxa,Belitsky:2013bja}, and its modification in a thermal state, would be interesting to investigate.} We expect this approximation to hold because the
spatial separation of the two eikonal
trajectories grows linearly with propagation time and the medium resolves
the two colour charges on a time scale short compared with the medium
length $L^+$.
\item \emph{Medium dependence only in the final state TMDFFs.} 
The hard and soft functions are evaluated in vacuum. A medium correction to the soft function would require
the wide-angle radiation off a back-to-back colour dipole, which lies outside
the collinear regime considered below.
In particular, the recoil of the collinear partons against soft
medium-induced gluons is not included. As we discuss in Sec.~\ref{sec:n_radiative}, the attribution of the medium modifications to the jet function has some degree of ambiguity, but the final distributions are independent of this choice at the order considered.
\item \emph{Medium geometry.} The medium is transversely homogeneous, and its
momentum-transfer distribution is isotropic in the plane transverse to the
parent direction. Both parents traverse the same length $L^+$, so that the
quark and antiquark jet functions carry the same medium parameters. In
particular, the medium is invariant under reflection through the plane
spanned by $\hat z$ and $\vec p_0$.
This idealization sets aside effects of more general medium structure and evolution on jet propagation (see, e.g.,~\cite{Sadofyev:2021ohn,Bahder:2024jpa,Barata:2022krd,Barata:2023qds,Kuzmin:2023hko,Barata:2024bqp,Barata:2024xwy,Barata:2025zku,Altenburger:2025iqa,Lindenbauer:2025ctw,Danhoni:2026gve,Avramescu:2024poa}).

\item \emph{Hadronization outside the medium.} In-medium partonic evolution
is separated from hadronization. The hadronization of partons exiting the medium is described by the vacuum collinear fragmentation functions.
\end{itemize}

To keep the transverse recoil of the collinear sector explicit in the TMD
matching, we introduce a parent-resolved TMDFF
$F_{q\to a}(x,\bell,\K;p_0)$, where $p_0$ is the four-momentum of the
parent quark, with three-momentum $\vec p_0$, and
\begin{equation}
\K=\P-\p_0,
\qquad
\bell=\p_a-x\P .
\end{equation}
Here $\P$ is the final total transverse momentum of the collinear sector,
$\K$ its recoil relative to the original parent, and $\bell$ the momentum of
the measured particle relative to the final collinear sector. From here on,
boldface vectors are transverse to the fixed light-cone direction $n$
of Eq.~\eqref{eq:ee_power_counting}. All partons of the sector, including the parent, are
collinear to $n$, $|\p_\perp|\sim\lambda\,p^+$ with $\lambda\ll1$. This is
the power counting of the BDMPS--Z formalism of Sec.~\ref{sec:n_bdmps}, in
which the medium is described along $n$. It fixes $n$ only up to rotations by
angles of order $\lambda$, which leave the jet function unchanged in a
transversely homogeneous medium. Since $\hat y$ is
orthogonal to the beam and to $\hat n$, the $\hat y$ components are those of
Sec.~\ref{sec:vac}. In vacuum, $\P=\p_0$, so the parent direction
and the jet axis coincide and the parent-resolved TMDFF reduces to the usual
one,
\begin{equation}
F_{q\to a}^{\rm vac}(x,\bell,\K;p_0)
=
F_{q\to a}^{\rm vac}(x,\bell)\,(2\pi)^2\delta^{(2)}(\K).
\end{equation}
In the medium, the jet axis is deflected from the parent direction by $\K$.
In terms of these variables the measured displacements are generalized such that Eq.~\eqref{eq:teec_measurement_lp} becomes
\begin{equation}\label{eq:teec_measurement_parent_resolved}
\tau
=
\frac{1}{4p_T^2}
\left(
K_y+\frac{\ell_y}{x}+\bar K_y+\frac{\bar\ell_y}{\bar x}-k_{s,y}
\right)^2 .
\end{equation}
Thus, the parent-resolved TEEC is Eq.~\eqref{eq:ee_tmd_convolution} with the
TMDFFs replaced by the parent-resolved ones, with the additional
integrations over $\K$ and $\bar\K$, and with the measurement of
Eq.~\eqref{eq:teec_measurement_parent_resolved}. The jet function of Eq.~\eqref{eq:ee_jet_tmdff} changes accordingly, with
the additional integration over $\K$ and the phase $e^{i\b\cdot\K}$.
In the perturbative regime, the parent-resolved TMDFF is matched onto
collinear fragmentation functions through Eq.~\eqref{eq:tmd_ope}, with $\K$
as an additional argument of $F_{i\to a}$ and $I_{ij}$. Since hadronization
takes place outside the medium, the $f_{j\to a}$ are the vacuum fragmentation
functions, and all medium effects are contained in $I_{ij}$. Therefore, the
momentum sum rule still removes the fragmentation functions from the energy
weighted jet function,\footnote{Energy transferred to the medium is not
described here. Its contribution, the medium response, is discussed in
Sec.~\ref{sec:n_response}.} and Eq.~\eqref{eq:ee_jet_matching} becomes
\begin{equation}\label{eq:jet_function_def}
J_q(b)
=
\sum_i\int_0^1 dx\,x\int_{\bell,\K}
e^{i\b\cdot(\K+\bell/x)}\,I_{qi}(x,\bell,\K),
\end{equation}
with $\b=b\hat\y$, and analogously for $J_{\bar q}$. With $b$ the impact
parameter along $\hat y$ of Eq.~\eqref{eq:ee_teec_bspace}, the reflection
symmetry of the medium geometry gives $J_q(b)=J_q(-b)$, as in vacuum, so
that Eq.~\eqref{eq:vacuum_teec_code} holds with these jet functions. Their
dependence on the parent energy and the medium parameters is left implicit.
The matching coefficient follows from Eq.~\eqref{eq:tmd_ope} evaluated for
partonic final states, for which both sides can be computed in perturbation
theory, $I_{qi}=I_{qi}^{(0)}+I_{qi}^{(1)}+\cO(\alpha_s^2)$, which we discuss
in Secs.~\ref{sec:n_broadening} and~\ref{sec:n_radiative}.

%===========================================================================
\subsection{Medium-induced broadening of the jet function}
\label{sec:n_broadening}\label{sec:broadening_rad}

This subsection evaluates the zeroth-order matching coefficient
$I^{(0)}_{qi}$ and the jet function of
Eq.~\eqref{eq:jet_function_def} that it gives.
Let $d\widehat\sigma_{q|p_0}$ denote the partonic cross section conditional
on producing a parent quark with momentum $p_0$, with the production process
absorbed into the hard factor.\footnote{In a transversely homogeneous medium,
the cross section depends on the final momentum of the collinear sector only
through $\K=\P-\p_0$, and so does the measurement. Therefore, the Born cross section for
the parent factorizes into the hard function.} Without resolved
radiation, $f^{(0)}_{j\to i}(\zeta)=\delta_{ij}\delta(1-\zeta)$ and
\begin{align}\label{eq:matching_coeff_0}
I_{qi}^{(0)}(x,\bell,\K)
=(2\pi)^2\frac{d\widehat\sigma_{q|p_0}}{d^2\K}\,
\delta_{iq}\,\delta(x-1)\,(2\pi)^2\delta^{(2)}(\bell).
\end{align}
In a transversely homogeneous
medium occupying the light-cone time interval $0<s^+<L^+$, the conditional
cross section without resolved radiation describes pure broadening of the
parent quark,
\begin{align}
(2\pi)^2\frac{d\widehat\sigma_{q|p_0}}{d^2\K}=\cP_F(\K;L^+,0).
\end{align}
The broadening distribution of a parton in colour representation $R$ is the
Fourier transform
\begin{align}\label{eq:PR_simplified}
\cP_R(\q;x_f^+,x_i^+)=\int d^2\u\,e^{i\q\cdot\u}\,D_R[\u;x_f^+,x_i^+]
\end{align}
of the dipole function
\begin{align}
D_R[\x;x_f^+,x_i^+]
=\exp\left\{-\frac{C_R}{2}\int_{x_i^+}^{x_f^+}ds^+\,n(s^+)\,
\sigma\bigl(\x(s^+)\bigr)\right\},
\end{align}
where $n(s^+)$ is the density of scattering centres and $\sigma(\x)$ the
dipole cross section.
A constant argument, as in
Eq.~\eqref{eq:PR_simplified}, denotes $\x(s^+)=\u$. Since
$D_R[\boldsymbol 0]=1$, the distribution is normalized,
$\int_{\q}\cP_R(\q)=1$. Substituting this distribution into Eq.~\eqref{eq:matching_coeff_0}
and using Eq.~\eqref{eq:jet_function_def} gives
\begin{align}\label{eq:DJ0_medium_correction}
J^{(0)}(b)=D_F[\b;L^+,0],
\end{align}
with $J^{(0),{\rm vac}}(b)=1$. From Sec.~\ref{sec:time_regions} onwards we evaluate the dipole function in
the harmonic-oscillator approximation~\cite{Baier:1996sk,Zakharov:1997uu}. It keeps the small-$\u$ behaviour of
the dipole cross section in quadratic form and drops the accompanying
logarithm,
\begin{align}\label{eq:HO_qhat_definition}
&\frac{C_R}{2}\,n(s^+)\,\sigma(\u)\ \longrightarrow\
\frac{C_R}{C_F}\,\frac{\hat q(s^+)}{4}\,\u^2,
\\ \notag
&D_R[\u;x_f^+,x_i^+]
=\exp\left\{-\frac{C_R}{C_F}\frac{\u^2}{4}\int_{x_i^+}^{x_f^+}ds^+\,\hat q(s^+)\right\},
\end{align}
which defines the quark transport coefficient $\hat q\equiv\hat q_F$, with
$\hat q_A=(C_A/C_F)\hat q$. In the medium-induced terms we work at leading
order in $1/N_c$, with $C_F=N_c/2$ and $C_A=N_c$. Then $\hat q_A=2\hat q$, and
a gluon line is represented by a pair of fundamental lines, so that its
dipole is $D_F^2$ (Sec.~\ref{sec:n_bdmps}). Corrections of relative order
$1/N_c^2$ to the medium-induced terms are neglected, while the vacuum
functions $J^{\rm vac}$ and $S_{\rm vac}$ keep their full colour dependence.

%===========================================================================
\subsection{Medium-induced radiative correction to the jet function}
\label{sec:n_bdmps}\label{sec:n_radiative}\label{sec:n_time_regions}

At $\cO(\alpha_s)$, Eq.~\eqref{eq:tmd_ope} gives
$I^{(1)}_{qi}=F^{(1)}_{q\to i}-\sum_jI^{(0)}_{qj}\otimes f^{(1)}_{j\to i}$.
The first term contains the real emission $q\to qg$, conditional on the
parent, and the virtual correction. The second term is the broadening of the
parent, contained in $I^{(0)}$, followed by a vacuum collinear splitting in
$f^{(1)}$, which does not resolve the transverse momentum of the fragmenting
parton. Let $z$ be the energy fraction of the final quark in $q\to qg$, and let $\p=(1-z)\p_1-z\p_2$ be the relative transverse momentum of the daughters, where $\p_1$ and
$\p_2$ are the final quark and gluon transverse momenta. We denote by
$d\widehat\sigma_{qg|p_0}$ the conditional cross section for gluon emission
off the quark, and by $d\widehat\sigma_{qg|p_0}^{\rm vac}$ its vacuum value.
With $f^{(1)}$ defined by a sharp cutoff $\mu_c$ on the relative momentum,
\begin{align}\label{eq:matching_coeff_1}
I_{qi}^{(1)}(x,\bell,\K)
={}&\int_0^1dz\int_{\p}(2\pi)^4\Big[
\frac{d\widehat\sigma_{qg|p_0}}{dz\,d^2\p\,d^2\K}\,\cM_{\rm res.}
-\Theta(\mu_c^2-\p^2)\,
\frac{d\widehat\sigma_{q|p_0}}{d^2\K}\,
\frac{d\widehat\sigma_{qg|p_0}^{\rm vac}}{dz\,d^2\p}\,\cM_{\rm loc.}
\Big]
\nn
&+\big[\text{terms}\propto\delta_{iq}\,\delta(1-x)\big],
\end{align}
where the first term may be evaluated in vacuum or in the medium, and the
last term contains the virtual corrections. The resolved
measurement retains the relative momentum of the two daughters, while the
local one retains only their longitudinal fractions,
\begin{align}\label{eq:measurement_functions}
\cM_{\rm res.}&=(2\pi)^2\left[\delta_{iq}\delta(x-z)\delta^{(2)}(\bell-\p)
+\delta_{ig}\delta(x-(1-z))\delta^{(2)}(\bell+\p)\right],
\nn
\cM_{\rm loc.}&=(2\pi)^2\left[\delta_{iq}\delta(x-z)\delta^{(2)}(\bell)
+\delta_{ig}\delta(x-(1-z))\delta^{(2)}(\bell)\right].
\end{align}
As in vacuum, the soft singularity of the real emission at $z\to1$ is treated
with a plus-prescription,
\begin{align}\label{eq:measurement_functions_plus}
\cM_{\rm res.}^{[+]}&=\cM_{\rm res.}-(2\pi)^2\delta_{iq}\delta(x-1)\delta^{(2)}(\bell-\p),
\nn
\cM_{\rm loc.}^{[+]}&=\cM_{\rm loc.}-(2\pi)^2\delta_{iq}\delta(x-1)\delta^{(2)}(\bell)\, ,
\end{align}
and the plus-subtracted forms are understood from this point
on. In
vacuum, the remaining coefficient of $\delta(1-x)$, with a rapidity regulator
and renormalization, and combined with the soft function, gives the standard
$J^{\rm vac}(b,\mu,\nu)$ and $S_{\rm vac}$~\cite{Moult:2018jzp}, and the
scales $\mu$ and $\nu$ enter the jet function only through this coefficient.
In the medium, it also receives the recoil of the quark against soft
medium-induced gluons, which is rapidity divergent as well, since the
medium-induced spectrum grows like $(1-z)^{-3/2}$ as $z\to1$ in the limit of dense matter; see~\cite{Ke:2023ixa} for related discussion in the dilute limit. This divergence
is expected to cancel only against a medium-modified soft function, which lies outside the
collinear BDMPS--Z formalism used here, and which we leave for future work. Consistently with the vacuum soft
function, we omit this medium recoil term and use the plus-prescription
without it for the medium-induced radiative correction. The
double-logarithmic part of the omitted recoil is the radiative
transverse momentum broadening of~\cite{Liou:2013qya,Iancu:2014kga,Blaizot:2014bha}.

We evaluate the first-order matching coefficient of
Eq.~\eqref{eq:matching_coeff_1} using the in-medium $q\to qg$ cross section.
In the BDMPS--Z
formalism~\cite{Baier:1996kr,Baier:1996sk,Zakharov:1996fv,Zakharov:1997uu},
the fixed-order cross section for $q(p_0)\to q(p_1)g(p_2)$
at finite $z$ reads~\cite{Blaizot:2012fh,Apolinario:2014csa,Isaksen:2023nlr}
\begin{align}\label{eq:full_qg_cross_section}
(2\pi)^4\frac{d\widehat\sigma_{qg|p_0}}{dz\,d^2\p\,d^2\K}
={}&\frac{2C_F\alpha_s}{(2\omega^+)^2}\,P^{\rm vac}_{q\to qg}(z)\;
2\Re\int_{\q_0,\q_1,\bar\q_0,\bar\q_1,\l_1,x^+<\bar x^+}
e^{-\epsilon(\bar x^++x^+)}\,(\bar\bkappa\cdot\bkappa)
\nn
&\times
\cP_F(\q_0-\p_0;x^+,0)\,
\cS^{(3)}_{q\to qg}(\l_1-z\bar\q_0,\bkappa,\q_0-\bar\q_0;\bar x^+,x^+)
\nn
&\times
\cS^{(4)}_{q\to qg}(x_\infty^+,\p_1,\p_2;\bar x^+,
\bar\q_0-\l_1,\l_1,\bar\q_0-\bar\q_1,\bar\q_1),
\end{align}
with $z=p_1^+/p_0^+$, $\omega^+=z(1-z)p_0^+$,
$P^{\rm vac}_{q\to qg}(z)=(1+z^2)/(1-z)$, the total transverse momentum $\P=\p_1+\p_2=\p_0+\K$ and the relative transverse momentum $\p=(1-z)\p_1-z\p_2$. Unbarred internal momenta belong to the amplitude and barred ones to the conjugate amplitude. The splitting vertices occur at $x^+$ in the amplitude
and at $\bar x^+>x^+$ in the conjugate amplitude, and the time shorthand
denotes $\int_0^\infty dx^+\int_{x^+}^\infty d\bar x^+$. The parent and
daughter-quark momenta adjacent to the amplitude vertex are $\q_0$ and
$\q_1$, with relative momentum $\bkappa=\q_1-z\q_0$. On the conjugate side
they are $\bar\q_0$ and $\bar\q_1$, with $\bar\bkappa=\bar\q_1-z\bar\q_0$. The
momentum $\l_1$ is that of the amplitude daughter quark at $\bar x^+$. The
common final time $x_\infty^+$ lies after the medium, where $\p_1$ and $\p_2$
are specified on both sides of the cut. The convergence factor is removed
after the late-time integration by taking $\epsilon\to0^+$. The three factors describe the successive stages of the process. Between $0$
and $x^+$ the parent propagates on both sides of the cut and broadens with
$\cP_F$. Between $x^+$ and $\bar x^+$ the amplitude contains the quark-gluon
pair and the conjugate amplitude the parent, which gives the three-point
function. After $\bar x^+$ the pair propagates on both sides, which gives the
four-point function. As a check of the normalization, in vacuum
$\cP_F\to(2\pi)^2\delta^{(2)}$, the correlators reduce to free propagation,
$\bkappa=\bar\bkappa=\p$, and the time integrals give $(2\omega^+/\p^2)^2$.
Equation~\eqref{eq:full_qg_cross_section} then reduces to
$(2\pi)^2\delta^{(2)}(\K)\,2C_F\alpha_sP^{\rm vac}_{q\to qg}(z)/\p^2$, which
is the leading-order splitting
$(2\pi)^2d\widehat\sigma^{\rm vac}_{qg|p_0}/(dz\,d^2\p)$.

In the large $N_c$ approximation, the momentum space three-point correlator
is
\begin{align}\label{eq:S3_full_Fourier_transform}
&\cS^{(3)}_{q\to qg}(\l_1-z\bar\q_0,\bkappa,\q_0-\bar\q_0;x_f^+,x_i^+)=\int d^2\b_3\,d^2\r_f\,d^2\r_i\,
e^{-i\r_f\cdot(\l_1-z\bar\q_0)}\,e^{i\r_i\cdot\bkappa}\,
e^{i\b_3\cdot(\q_0-\bar\q_0)}
\nn
&\times\int_{\r_i}^{\r_f}\cD\r\,
\exp\left\{i\frac{\omega^+}{2}\int_{x_i^+}^{x_f^+}ds^+\,\dot\r^{\,2}\right\}
D_F[\b_3-z\r;x_f^+,x_i^+]\,D_F[\r;x_f^+,x_i^+].
\end{align}
Here $\r(s^+)$ is the daughter relative coordinate with endpoints $\r_i$ and
$\r_f$, and $\b_3$ is the separation of the daughter transverse centre from
the conjugate-amplitude parent. The dipole with argument $\r$ connects the
amplitude quark and gluon, and the dipole with argument $\b_3-z\r$ connects
the amplitude gluon and the conjugate parent. Transverse translation invariance of the four-point function allows its
common centre coordinate to be integrated out. This gives
$(2\pi)^2\delta^{(2)}(\k_1+\k_2-\bar\k_1-\bar\k_2)$, which is satisfied in
Eq.~\eqref{eq:full_qg_cross_section} because both incoming total momenta are
$\bar\q_0$ and both outgoing total momenta are $\P$. After this factor is
removed, the reduced correlator is
\begin{align}\label{eq:S4_full_Fourier_transform}
&\cS^{(4)}_{q\to qg}(x_\infty^+,\p_1,\p_2;\bar x^+,
\bar\q_0-\l_1,\l_1,\bar\q_0-\bar\q_1,\bar\q_1)
\nn
&=\int d^2\b_4\,d^2\bDelta\,d^2\u\,d^2\w\;
e^{-i(\P-\bar\q_0)\cdot\b_4}\,e^{-i\p\cdot\bDelta}\,
e^{i(\l_1-z\bar\q_0)\cdot\u}\,e^{-i(\bar\q_1-z\bar\q_0)\cdot\w}
\nn
&\times\widetilde{\cS}^{(4)}_{q\to qg}(x_\infty^+,\bDelta;\bar x^+,\u,\w;\b_4,z).
\end{align}
Here $\u$ and $\w$ are the initial daughter relative coordinates in the
amplitude and conjugate amplitude, $\bDelta$ is the difference of the final
ones, and $\b_4$ is the difference of the two transverse centre coordinates.
The coordinate $\b_4$ is conjugate to the total momentum transferred after
$\bar x^+$, and $\bDelta$ to the final relative momentum. The
coordinate-space kernel is
\begin{align}\label{eq:S4_general_definition}
&\widetilde{\cS}^{(4)}_{q\to qg}(x_\infty^+,\bDelta;\bar x^+,\u,\w;\b_4,z)
\nn
& = \int d^2\r
\int_{\v(\bar x^+)=\u}^{\v(x_\infty^+)=\r}\cD\v
\int_{\bar\v(\bar x^+)=\w}^{\bar\v(x_\infty^+)=\r-\bDelta}\cD\bar\v
\exp\Bigg\{\frac{i\omega^+}{2}\int_{\bar x^+}^{x_\infty^+}ds^+
\left(\dot\v^{\,2}-\dot{\bar\v}^{\,2}\right)\Bigg\}
\cC^{(4)}_{q\to qg}[\v,\bar\v,\b_4],
\end{align}
where the $\r$ integration sums over one final relative coordinate at fixed
$\bDelta$. In the same large $N_c$ representation the gluon is a pair of
fundamental Wilson lines, and the six lines of the amplitude and conjugate
amplitude organize into a dipole and a quadrupole,
\begin{align}\label{eq:C4_general_definition}
\cC^{(4)}_{q\to qg}[\v,\bar\v,\b_4]
=D_F\!\left[\b_4-z(\v-\bar\v);x_\infty^+,\bar x^+\right]
\frac{1}{N_c}\left\langle\operatorname{Tr}\!\left[
U_F(\r_1)U_F^\dagger(\r_2)U_F(\bar\r_2)U_F^\dagger(\bar\r_1)
\right]\right\rangle,
\end{align}
with Wilson lines from $\bar x^+$ to $x_\infty^+$ along the paths
\begin{align}\label{eq:C4_coordinate_definitions}
\v=\r_1-\r_2,
\qquad
\bar\v=\bar\r_1-\bar\r_2,
\qquad
\b_4=z(\r_1-\bar\r_1)+(1-z)(\r_2-\bar\r_2).
\end{align}
Since $\b_4-z(\v-\bar\v)=\r_2-\bar\r_2$, the dipole in
Eq.~\eqref{eq:C4_general_definition} connects the gluon in the amplitude and
in the conjugate amplitude. To express the cross section in the variables of the jet function, we
substitute Eqs.~\eqref{eq:S3_full_Fourier_transform} and
\eqref{eq:S4_full_Fourier_transform} into
Eq.~\eqref{eq:full_qg_cross_section} and transform in $\K=\P-\p_0$ with the
phase $e^{i\b\cdot\K}$ of Eq.~\eqref{eq:jet_function_def}. The
momentum integrations set $\b_3=\b_4=\b$, give the parent broadening
$D_F[\b;x^+,0]$, and cancel the $\p_0$ dependence of the phase. Therefore, all medium
factors depend on the impact parameter $\b$. The vertex momenta
become derivatives acting on the kernels, and the remaining integrations are
carried out in Sec.~\ref{sec:time_regions}.

The vertex times $x^+<\bar x^+$ in Eq.~\eqref{eq:full_qg_cross_section} fall
into three regions relative to the medium: in--in with $\bar x^+<L^+$,
in--out with $x^+<L^+<\bar x^+$, and out--out with $L^+<x^+$.
In the out--out region the parent broadens over $[0,L^+]$ and splits during
the subsequent vacuum propagation. Shifting both vertex times by $L^+$ changes
only the convergence factor $e^{-\epsilon(\bar x^++x^+)}$, whose effect
disappears for $\epsilon\to0^+$. Therefore, the out--out contribution is the
broadening of the parent followed by the vacuum splitting,
\begin{align}\label{eq:out_out}
(2\pi)^4\frac{d\widehat\sigma_{\rm out\text{-}out}}{dz\,d^2\p\,d^2\K}
=(2\pi)^4\frac{d\widehat\sigma_{q|p_0}}{d^2\K}\,
\frac{d\widehat\sigma_{qg|p_0}^{\rm vac}}{dz\,d^2\p}
=\cP_F(\K;L^+,0)\,(2\pi)^2\frac{d\widehat\sigma_{qg|p_0}^{\rm vac}}{dz\,d^2\p},
\end{align}
which also multiplies the local subtraction in
Eq.~\eqref{eq:matching_coeff_1}. We
call the sum of the other two regions the medium-induced radiative cross
section,
\begin{align}\label{eq:n_med_cross_section}
d\widehat\sigma_{qg|p_0}^{\rm med}
\equiv
d\widehat\sigma_{\rm in\text{-}in}+d\widehat\sigma_{\rm in\text{-}out},
\end{align}
which vanishes in vacuum. Substituting this decomposition and
Eq.~\eqref{eq:out_out} into Eq.~\eqref{eq:matching_coeff_1}, the out--out
term and the local subtraction share the factor $\cP_F(\K;L^+,0)$ and
leave the vacuum combination
$\cM_{\rm res.}^{[+]}-\Theta(\mu_c^2-\p^2)\cM_{\rm loc.}^{[+]}$.
Together with the virtual term retained above,
this gives
\begin{align}\label{eq:matching_coefficient_medium_split}
I_{qi}^{(1)}(x,\bell,\K)
={}&\cP_F(\K;L^+,0)\,I_{qi}^{(1),{\rm vac}}(x,\bell)
+\int_0^1dz\int_{\p}(2\pi)^4
\frac{d\widehat\sigma_{qg|p_0}^{\rm med}}{dz\,d^2\p\,d^2\K}
\,\cM_{\rm res.}^{[+]} .
\end{align}
Here $I_{qi}^{(1),{\rm vac}}(x,\bell)$ is the usual vacuum matching
coefficient with the factor $(2\pi)^2\delta^{(2)}(\K)$ removed. Its
dependence on $\mu$ and $\nu$ is implicit and is inherited by
$J^{\rm vac}(b,\mu,\nu)$. Taking the
energy-weighted moment of Eq.~\eqref{eq:jet_function_def}, the Fourier
transform of $\cP_F$ gives $D_F$. Including the zeroth-order term of
Eq.~\eqref{eq:DJ0_medium_correction}, we obtain
\begin{align}\label{eq:full_jet_function}
J(b,\mu,\nu)=D_F[\b;L^+,0]\,J^{\rm vac}(b,\mu,\nu)
+J^{(1),\,{\rm med.\,rad.}}(b)+\cO(\alpha_s^2)\,,
\end{align}
with
\begin{align}\label{eq:Jmed_measurement_sum}
J^{(1),\,{\rm med.\,rad.}}(b)
=\int_0^1dz\int_{\K,\p}(2\pi)^4
\frac{d\widehat\sigma_{qg|p_0}^{\rm med}}{dz\,d^2\p\,d^2\K}
\sum_{i=q,g,u}h_i(z)\,e^{i\b\cdot(\K+a_i\p)},
\end{align}
where the energy-weighted, plus-subtracted measurement of
Eq.~\eqref{eq:measurement_functions_plus} gives the three channels
\begin{align}\label{eq:measurement_channel_sum}
(h_q,a_q)=\left(z,\frac1z\right),
\qquad
(h_g,a_g)=\left(1-z,-\frac1{1-z}\right),
\qquad
(h_u,a_u)=(-1,1),
\end{align}
with $u$ denoting the unresolved subtraction channel. At $b=0$ all phases of
the channel sum coincide and $h_q+h_g+h_u=0$, so that
\begin{equation}\label{eq:n_sum_rule}
J^{(1),\,{\rm med.\,rad.}}(0)=0 ,
\end{equation}
which is the energy sum rule for the medium-induced radiative correction.
Integrated over $\p$ and $\K$, the medium-induced cross section is the
harmonic-oscillator BDMPS--Z spectrum~\cite{Baier:1996kr,Zakharov:1996fv},
which grows like $(1-z)^{-3/2}$ as $z\to1$ and like $z^{-1/2}$ as $z\to0$
(App.~\ref{app:endpoint}). At $z\to1$ the energy-weighted measurement summed over channels is suppressed
by one power of $1-z$, since $h_g=1-z$ and
$z\,e^{i\b\cdot\p/z}-e^{i\b\cdot\p}=\cO(1-z)$, so that the $z$ integrand of
the jet function behaves at most as $(1-z)^{-1/2}$. Therefore, the radiative
correction introduces no soft divergence into the jet function. At
$z\to0$ the gluon and unresolved weights tend to $1$ and $-1$, and the
integrable $z^{-1/2}$ behaviour remains.

Equation~\eqref{eq:full_jet_function} is the initial condition of the jet
function at its natural scales, $\mu=\bar{\mu}_b$ of Eq.~\eqref{eq:ee_b_profile} and $\nu=\nu_J$. The hard and
soft functions are those of vacuum, so the full jet function evolves with
the vacuum anomalous dimensions, which is consistent at this order because
the radiative correction has no scale dependence at $\cO(\alpha_s)$. In the
logarithmic counting, $D_F[\b;L^+,0]$ carries no power of $\alpha_s$ and
enters from LL accuracy onwards, while the radiative correction is an
$\cO(\alpha_s)$ term without large logarithms, of the order of the one-loop
constants of the hard, soft and jet functions, and is included with them at
NLL$'$. The product of hard, soft and jet functions at their natural scales
becomes
\begin{equation}\label{eq:medium_insertion_nllp}
D_F[\b;L^+,0]^2\Big[H\,S\,J^{\rm vac}_q\,J^{\rm vac}_{\bar q}\Big]_{\cO(\alpha_s)}
+2\,D_F[\b;L^+,0]\,J^{(1),\,{\rm med.\,rad.}}(b),
\end{equation}
with the vacuum product expanded to first order in $\alpha_s$. Since the broadening factor and the radiative correction carry no scale dependence, assigning them to the jet or to the soft functions leaves the product in Eq.~\eqref{eq:medium_insertion_nllp} unchanged at this order, and both assignments acquire the same vacuum evolution factor. With the broadening factors assigned to the soft function, the broadening term has the form used in earlier studies of dijet decorrelations in heavy ion collisions~\cite{Mueller:2016gko}. Terms of
order $\alpha_s^2$, among them $\big(J^{(1),\,{\rm med.\,rad.}}\big)^2$ and
its products with the one-loop vacuum functions, are of the order of the
two-loop constants that are not included. The radiative correction takes
its coupling at the scale of the one-loop vacuum jet function,
$\alpha_s(\bar{\mu}_b)$, which stays perturbative at large $b$, and follows this
scale in the variations. A medium scale such as $\sqrt{\hat qL}$ would
differ from $\bar{\mu}_b$ by a factor of order one for the parameters considered,
a change of order $\alpha_s^2$. Finally, unlike the one-loop vacuum functions, which are evaluated at $\bar b$, the broadening factor and the radiative correction are evaluated at the physical impact parameter $b$. 

The evaluation of Eq.~\eqref{eq:Jmed_measurement_sum} requires care at
large relative momentum, where the medium-induced cross section falls
slowly. There, the leading $1/\p^2$ tail corresponds to an early splitting
followed by independent broadening of the daughters, with the combined factor
$D_FD_A=D_F^3$ at large $N_c$. Subtracting the out--out contribution gives,
up to terms that oscillate in $\p^2$ or fall faster,
\begin{align}\label{eq:large_p_remainder}
\int\frac{d^2\K}{(2\pi)^2}\,e^{i\b\cdot\K}\,(2\pi)^4
\frac{d\widehat\sigma^{\rm med}_{qg|p_0}}{dz\,d^2\p\,d^2\K}
\;\to\;
\big[D_F^3(\b;L^+,0)-D_F(\b;L^+,0)\big]\,
(2\pi)^2\frac{d\widehat\sigma^{\rm vac}_{qg|p_0}}{dz\,d^2\p}.
\end{align}
The separate time regions have larger tails, which cancel in this sum
(App.~\ref{app:p_integration}). The coefficient in
Eq.~\eqref{eq:large_p_remainder} vanishes at $b=0$. For $b\neq0$ the $\p$
integral converges only through the measurement phases $e^{ia_i\b\cdot\p}$,
which after the angular integration give radial integrands of order
$J_0(|a_i|b|\p|)/|\p|$. Therefore, the integrand is not absolutely
integrable in $\p$ and the vertex times jointly, and the order of
integration matters. The jet function is defined with the time integrals
done first, at fixed $\p$, and the $\p$ integral last. To perform the $\p$
integral analytically before the time integrals
(App.~\ref{app:p_integration}), we include the regulator
$e^{-\eta_{\rm UV}\p^2}$, $\eta_{\rm UV}>0$, the same in every contribution.
For $\eta_{\rm UV}>0$ the integrand is absolutely integrable, so that the two
orders agree, and the regulated result tends to the jet function as
$\eta_{\rm UV}\to0^+$, with the limit taken after the time integrations. The
radiative correction is then
\begin{align}\label{eq:Jmed_fixed_p_strategy}
J^{(1),\,{\rm med.\,rad.}}(b)
&=\lim_{\eta_{\rm UV}\to0^+}\sum_{i=q,g,u}\int_0^1dz\,h_i(z)
\int_{\mathbb R^2}\frac{d^2\p}{(2\pi)^2}\,e^{-\eta_{\rm UV}\p^2}\,
e^{ia_i\b\cdot\p}
\nn
&\times\int\frac{d^2\K}{(2\pi)^2}\,e^{i\b\cdot\K}\,(2\pi)^4
\left[\frac{d\widehat\sigma_{\rm in\text{-}in}}{dz\,d^2\p\,d^2\K}
+\frac{d\widehat\sigma_{\rm in\text{-}out}}{dz\,d^2\p\,d^2\K}\right].
\end{align}
The analytic $\p$ integration and a subtraction that improves the
numerical convergence are given in App.~\ref{app:p_integration}. Section~\ref{sec:time_regions} computes the in--in and in--out
contributions at fixed relative momentum.

%===========================================================================
\section{Analytic evaluation of the medium-induced radiative correction}\label{sec:time_regions}

In this section we analytically evaluate the in--in and in--out contributions to
Eq.~\eqref{eq:Jmed_fixed_p_strategy} at fixed relative momentum $\p$, for a
homogeneous medium in the large $N_c$ harmonic-oscillator approximation of
Eq.~\eqref{eq:HO_qhat_definition}. The remaining $\p$ integration is carried
out in App.~\ref{app:p_integration}. Following Eqs.~\eqref{eq:S3_full_Fourier_transform} and~\eqref{eq:S4_full_Fourier_transform}, the $\K$ transform of the medium-induced radiative cross section at fixed $\p$, i.e., the second line of Eq.~\eqref{eq:Jmed_fixed_p_strategy}, is
\begin{align}\label{eq:fixed_p_master}
	&\int\frac{d^2\K}{(2\pi)^2}\,e^{i\K\cdot\b}
	(2\pi)^4
	\frac{d\widehat\sigma^{\rm med}_{qg|p_0}}
	{dz\,d^2\p\,d^2\K}
	=
	\frac{2C_F\alpha_s}{(2\omega^+)^2}
	P_{q\to qg}^{\rm vac}(z)\,
	2\Re
	\int_{x^+<L^+,\,x^+<\bar x^+}
	e^{-\epsilon(\bar x^++x^+)}
	\nn
	&\quad\times
	\nabla_{\v}	\int d^2\u\,
	D_F[\b;x^+,0]
	\left.
	\widetilde{\cS}^{(3)}_{q\to qg}
	(\u,\bar x^+;\v,x^+;\b,z)
	\right|_{\v=0}
	\nn
	&\quad\cdot
	\left.
		\nabla_{\w}\int d^2\bDelta\,
		e^{-i\bDelta\cdot\p}
		\widetilde{\cS}^{(4)}_{q\to qg}
		(x_\infty^+,\bDelta;\bar x^+,\u,\w;\b,z)
	\right|_{\w=0}.
\end{align}
Here $\widetilde{\cS}^{(3)}_{q\to qg}$ is the coordinate-space representation of the three-point function in Eq.~\eqref{eq:S3_full_Fourier_transform}, with initial relative coordinate $\r_i=\v$ at $x^+$ and final relative coordinate $\r_f=\u$ at $\bar x^+$. The colour structure of the four-point function $\widetilde{\cS}^{(4)}_{q\to qg}$ matters only in the in--in region, $\bar x^+<L^+$, since in the in--out region it is trivially given by its vacuum counterpart. Its colour structure is spanned by the dipole--dipole and quadrupole
correlators
\begin{align}\label{eq:S4_color_basis}
	\mathcal C_1^{(4)}
	&=
	\frac{1}{N_c^2}
	\left\langle
	\operatorname{Tr}\!\left[U_F(\r_1)U_F^\dagger(\r_2)\right]\,
	\operatorname{Tr}\!\left[U_F(\bar\r_2)U_F^\dagger(\bar\r_1)\right]
	\right\rangle,
	\nn
	\mathcal C_2^{(4)}
	&=
	\frac{1}{N_c}
	\left\langle
	\operatorname{Tr}\!\left[
	U_F(\r_1)U_F^\dagger(\r_2)
	U_F(\bar\r_2)U_F^\dagger(\bar\r_1)
	\right]
	\right\rangle.
\end{align}
All Wilson lines run from $\bar x^+$ to the time $t$ along the four paths of
Eq.~\eqref{eq:C4_coordinate_definitions}. The quadrupole in
Eq.~\eqref{eq:C4_general_definition} is $\mathcal C_2^{(4)}$, so that
$\cC^{(4)}_{q\to qg}=D_F[\b-z(\v-\bar\v)]\,\mathcal C_2^{(4)}$. At large
$N_c$ the dipole--dipole component evolves autonomously and sources the
quadrupole, so that the evolution is triangular. In the harmonic-oscillator
approximation it reads~\cite{Isaksen:2020npj,Isaksen:2023nlr}
\begin{align}\label{eq:S4_triangular_evolution}
	\partial_t
	\begin{pmatrix}
	\mathcal C_1^{(4)}\\
	\mathcal C_2^{(4)}
	\end{pmatrix}
	=
	-\frac{\hat q(t)}4
	\begin{pmatrix}
	\mathcal V_1&0\\
	\mathcal T&\mathcal V_2
	\end{pmatrix}
	\begin{pmatrix}
	\mathcal C_1^{(4)}\\
	\mathcal C_2^{(4)}
	\end{pmatrix},
\end{align}
with $\mathcal C_1^{(4)}=\mathcal C_2^{(4)}=1$ at $t=\bar x^+$, where the
Wilson lines have zero extent. In terms of the instantaneous relative
coordinate $\bDelta(t)=\v(t)-\bar\v(t)$, the potentials are
\begin{align}\label{eq:S4_HO_potentials}
	\mathcal V_1&=\v^2+\bar\v^{\,2},
	\nn
	\mathcal V_2
	&=2\b^2+2(1-2z)\b\cdot\bDelta
	+\left[z^2+(1-z)^2\right]\bDelta^2,
	\nn
	\mathcal T(\bDelta,\b)
	&=2z(1-z)\bDelta^2-2\b^2
	-2(1-2z)\b\cdot\bDelta.
\end{align}
The diagonal potential $\mathcal V_2=(\r_1-\bar\r_1)^2+(\r_2-\bar\r_2)^2$
is the sum of the squared sizes of the quark and gluon dipoles that connect
the amplitude and the conjugate amplitude, and the transition potential is
$\mathcal T=\bDelta^2-\mathcal V_2$. As a check, when the amplitude or the
conjugate-amplitude pair is unresolved, $\v=0$ or $\bar\v=0$, one has
$\mathcal T+\mathcal V_2=\mathcal V_1$, and
Eq.~\eqref{eq:S4_triangular_evolution} gives
$\mathcal C_2^{(4)}=\mathcal C_1^{(4)}$. The four-point function then reduces
to two dipoles, which is the structure of the three-point function in
Eq.~\eqref{eq:S3_full_Fourier_transform} with the roles of amplitude and
conjugate amplitude exchanged. Since the evolution is triangular, its
solution separates exactly into diagonal propagation and a single
off-diagonal colour transition,
\begin{align}\label{eq:S4_fac_nonfac_definition}
	\widetilde{\cS}^{(4)}_{q\to qg}
	=
	\widetilde{\cS}^{(4),{\rm fac}}_{q\to qg}
	+
	\widetilde{\cS}^{(4),{\rm nonfac}}_{q\to qg}.
\end{align}
The first term is the homogeneous solution of
Eq.~\eqref{eq:S4_triangular_evolution}. The second contains one insertion of
$\mathcal T$ at an intermediate time, with diagonal propagation on either
side. Since $\mathcal V_2$ is a sum of two dipole potentials, the first term
is a product of a quark and a gluon dipole, and we refer to the two terms as
the factorizable and nonfactorizable parts.

%===========================================================================
\subsection{Factorizable contribution}
\label{sec:inin_fac}

At large $N_c$, the factorizable correlator is
\begin{align}
	\cC^{(4),{\rm fac}}_{q\to qg}
	(\v,\bar\v,\b;L^+,\bar x^+)
	={}&
	D_F\!\left(\b+(1-z)(\v-\bar\v);
	L^+,\bar x^+\right)
	D_F^2\!\left(\b-z(\v-\bar\v);
	L^+,\bar x^+\right).
\end{align}
where $x_\infty^+$ was replaced by $L^+$ since we take $\bar x^+ < L^+$ for the in--in region. It depends on the paths only through $\bDelta$, which the integral over the
common final coordinate keeps constant after $\bar x^+$, equal to $\u-\w$. Therefore, at
fixed $\p$ the four-point function is
\begin{align}\label{eq:S4_fac_fixed_p}
	&\int d^2\bDelta\,e^{-i\bDelta\cdot\p}
	\widetilde{\cS}^{(4),{\rm fac}}_{q\to qg}
	(L^+,\bDelta;\bar x^+,\u,\w;\b,z)
	\nn
	&\quad=
	e^{-i(\u-\w)\cdot\p}
	D_F\!\left(\b+(1-z)(\u-\w);
	L^+,\bar x^+\right)
	D_F^2\!\left(\b-z(\u-\w);
	L^+,\bar x^+\right).
\end{align}
Since Eq.~\eqref{eq:S4_fac_fixed_p} depends on $\u$ and $\w$ only through
$\u-\w$, the derivative $\nabla_\w$ can be traded for $-\nabla_\u$, and
integration by parts in $\u$ gives
\begin{align}\label{eq:fac_integration_by_parts}
	&\nabla_{\v}\int d^2\u\,
	\left.
	\widetilde{\cS}^{(3),{\rm in\text{-}in}}_{q\to qg}
	\right|_{\v=0}
	\cdot
	\left.
	\nabla_{\w}
	\int d^2\bDelta\,e^{-i\bDelta\cdot\p}
	\widetilde{\cS}^{(4),{\rm fac}}_{q\to qg}
	\right|_{\w=0}
	\nn
	&=
	\int d^2\u\,
	e^{-i\u\cdot\p}\,
	D_F\!\left(\b+(1-z)\u;L^+,\bar x^+\right)
	D_F^2\!\left(\b-z\u;L^+,\bar x^+\right)
	\left.
	\left(\nabla_{\u}\cdot\nabla_{\v}\right)
	\widetilde{\cS}^{(3),{\rm in\text{-}in}}_{q\to qg}
	\right|_{\v=0}.
\end{align}
The boundary term vanishes for $\bar x^+<L^+$, where the final state
broadening factors decay at large $|\u|$, and the endpoint $\bar x^+=L^+$
follows by continuity. For a time-independent and homogeneous medium, we denote the two-dimensional harmonic-oscillator propagator with constant frequency $\Omega$ by
\begin{align*}
	\cK_\Omega(\x_f,t_f;\x_i,t_i)
	&=
	\int_{\x(t_i)=\x_i}^{\x(t_f)=\x_f}\!\cD\x\,
	\exp\left\{
	\frac{i\omega^+}{2}\int_{t_i}^{t_f}ds^+
	\left[\dot{\x}^{2}-\Omega^2\x^2\right]
	\right\}
	\nn
	&=
	\frac{\omega^+c_1}{\pi}
	\exp\left\{
	\frac{i\omega^+}{2}c_2\left(\x_f^2+\x_i^2\right)
	+2\omega^+c_1\,\x_f\cdot\x_i
	\right\},
\end{align*}
with
\begin{align*}
	c_1=\frac{\Omega}{2i\sin[\Omega(t_f-t_i)]},
	\qquad
	c_2=\frac{\Omega}{\tan[\Omega(t_f-t_i)]},
\end{align*}
and with $\cK_0$ the free propagator. The in--in three-point function is an oscillator with frequency
$\Omega_3^2=-i(1+z^2)\hat q/(2\omega^+)$, centered at $z\b/(1+z^2)$,
\begin{align}\label{eq:S3_inin_complete}
	\widetilde{\cS}^{(3),{\rm in\text{-}in}}_{q\to qg}
	(\u,\bar x^+;\v,x^+;\b,z)
	={}&
	\exp\left\{
	-\frac{\hat q\b^2}{4(1+z^2)}\tau
	\right\}
	\nn
	&\times
	\cK_{\Omega_3}\!\left(\u-\frac{z}{1+z^2}\b,\bar x^+;\v-\frac{z}{1+z^2}\b,x^+\right),
\end{align}
with $\tau=\bar x^+-x^+$. With the mixed derivative of Eq.~\eqref{eq:S3_inin_complete} inserted, the remaining $\u$ integral in
Eq.~\eqref{eq:fac_integration_by_parts} is Gaussian and we display
its result through the two scalar coefficients
\begin{align}
	A_{\rm G}(\bar x^+,x^+)
	&=
	\frac{\hat q}{4}(L^+-\bar x^+)
	\left((1-z)^2+2z^2\right)
	-\frac{i\omega^+}{2}c_2(\bar x^+-x^+),
	\nn
	B_{\rm G}(\bar x^+,x^+)
	&=
	\frac{\hat q}{2}(L^+-\bar x^+)(1-3z)
	+
	\frac{z\omega^+}{1+z^2}
	\left[
	2c_1(\bar x^+-x^+)
	+ic_2(\bar x^+-x^+)
	\right]\,,
\end{align}
which contain no transverse variables. The transverse dependence is then
explicit in
\begin{align}\label{eq:fixed_p_fac_coordinate_integral}
	&\int d^2\u\,
	e^{-i\u\cdot\p}\,
	D_F\!\left(\b+(1-z)\u;L^+,\bar x^+\right)
	D_F^2\!\left(\b-z\u;L^+,\bar x^+\right)
	\left.
	\left(\nabla_{\u}\cdot\nabla_{\v}\right)
	\widetilde{\cS}^{(3),{\rm in\text{-}in}}_{q\to qg}
	\right|_{\v=0}
	\nn
	&=
	\frac{\omega^+c_1(\bar x^+-x^+)}
	{A_{\rm G}(\bar x^+,x^+)}
	\exp\Bigg\{
	-\frac{\hat q\b^2}{4(1+z^2)}(\bar x^+-x^+)
	-\frac{3\hat q\b^2}{4}(L^+-\bar x^+)
	\nn
	&
	+
	\frac{z^2\omega^+\b^2}{(1+z^2)^2}
	\left[
	2c_1(\bar x^+-x^+)
	+ic_2(\bar x^+-x^+)
	\right] - \frac{
	\p^2
	-2iB_{\rm G}(\bar x^+,x^+)\b\cdot\p
	-B_{\rm G}^2(\bar x^+,x^+)\b^2
	}
	{4A_{\rm G}(\bar x^+,x^+)}
	\Bigg\}
	\nn
	&\times
	\Bigg\{
	4\omega^+c_1
	+
	\frac{z^2(\omega^+)^2\b^2}{(1+z^2)^2}
	\left(4ic_1c_2+4c_1^2-c_2^2\right)
	+
	2i(\omega^+)^2c_1c_2
	\left(
	\frac{1}{A_{\rm G}}
	+
	\frac{
	-\p^2
	+2iB_{\rm G}\b\cdot\p
	+B_{\rm G}^2\b^2
	}
	{4A_{\rm G}^2}
	\right)
	\nn
	&\hspace{0cm}
	+
	\frac{z(\omega^+)^2}{1+z^2}
	\left(4ic_1c_2+4c_1^2-c_2^2\right)
	\frac{
	i\b\cdot\p+B_{\rm G}\b^2
	}
	{2A_{\rm G}}
	\Bigg\}.
\end{align}
In the last two lines $c_1$, $c_2$, $A_{\rm G}$ and $B_{\rm G}$ carry the
arguments displayed above. Before the relative-momentum integration, the
factorizable in--in contribution is
\begin{align}\label{eq:J_inin_fac_fixed_p_complete}
	J^{(1),{\rm med.\,rad.\,fac}}_{{\rm in\text{-}in}}(b)
	& =
	2C_F\alpha_s
	\sum_{i=q,g,u}
	\int_0^1dz\,
	\frac{P_{q\to qg}^{\rm vac}(z)\,h_i(z)}
	{(2\omega^+)^2}
	\int_{\mathbb R^2}
	\frac{d^2\p}{(2\pi)^2}\,
	e^{-\eta_{\rm UV}\p^2}\,
	e^{ia_i\b\cdot\p}
	\nn
	&\times
	2\Re
	\int_0^{L^+}d\tau
	\int_0^{L^+-\tau}dx^+\,
	e^{-\frac{\hat q\b^2}{4}x^+}
	\left[
	\text{RHS of
	Eq.~\eqref{eq:fixed_p_fac_coordinate_integral}}
	\right]_{\bar x^+=x^++\tau}.
\end{align}
The $\p$ integral is performed analytically in App.~\ref{app:p_integration},
which leaves the two time integrations.

%===========================================================================
\subsection{Nonfactorizable contribution}
\label{sec:inin_nonfac}

Nonfactorizable contributions account for medium induced correlations between the daughter partons (see, e.g.,~\cite{Blaizot:2012fh,Isaksen:2023nlr,Andres:2026qrt} for recent related calculations). The off-diagonal solution of Eq.~\eqref{eq:S4_triangular_evolution} has one insertion of $-\hat q\,\mathcal T/4$ at an intermediate time $y^+$, with $\mathcal C_1^{(4)}$ evolving before $y^+$ and $\mathcal C_2^{(4)}$ after it. Between $\bar x^+$ and $y^+$ the amplitude
and conjugate-amplitude relative coordinates are coupled through the
additional dipole $D_F[\b-z(\v-\bar\v)]$ of
Eq.~\eqref{eq:C4_general_definition}. Since the kinetic terms of $\v$ and
$\bar\v$ enter with opposite signs, the coupled system is diagonalized by a
hyperbolic transformation. In the transformed variables, the kernel factorizes
into a harmonic-oscillator propagator and its complex conjugate. The
nonfactorizable four-point function can then be written as
\begin{align}\label{eq:S4_nonfac_generic_delta}
	&\widetilde{\cS}^{(4),{\rm nonfac}}_{q\to qg}
	(L^+,\bDelta;\bar x^+,\u,\w;\b,z)
	\nn
	&=
	-\frac14\,
	\mathcal T(\bDelta,\b)
	\int_{\bar x^+}^{L^+}dy^+\,
	\hat q(y^+)\,
	D_F(\b+(1-z)\bDelta;L^+,y^+)
	D_F^2(\b-z\bDelta;L^+,y^+)
	\nn
	&\times
	\exp\left\{
	-\frac{\b^2}{4(1+2z^2)}
	\int_{\bar x^+}^{y^+}ds^+\,\hat q(s^+)
	\right\}
	\int d^2\x\,
	\cK_\Omega(\X_f,y^+;\X_i,\bar x^+)
	\cK_\Omega^*(\bar\X_f,y^+;\bar\X_i,\bar x^+),
\end{align}
where
\begin{align}
	&\lambda=\sqrt{1+2z^2},
	\qquad
	\beta=\frac{\lambda-1}{\lambda+1},
	\qquad
	\gamma=\frac1{\sqrt{1-\beta^2}},
	\qquad
	\delta=\frac{z}{\lambda^{3/2}},
	\nn
	&\X_i=\gamma(\u-\beta\w)-\delta\b,
	\qquad
	\bar\X_i=\gamma(\w-\beta\u)+\delta\b,
	\nn
	&\X_f=(1-\beta)\gamma\x+\gamma\beta\bDelta-\delta\b,
	\qquad
	\bar\X_f=(1-\beta)\gamma\x-\gamma\bDelta+\delta\b,
	\nn
	&\Omega^2
	=
	-\frac{i\lambda\hat q}{2\omega^+}.
\end{align}
The transformation $(\v,\bar\v)\to(\X,\bar\X)$ leaves the kinetic
combination $\dot\v^{\,2}-\dot{\bar\v}^{\,2}$ invariant, and in the shifted variables the potential
$\v^2+\bar\v^2+(\b-z(\v-\bar\v))^2$ becomes
$\lambda(\X^2+\bar\X^2)+\b^2/(1+2z^2)$. This fixes $\beta$, $\gamma$,
$\delta$ and the frequency $\Omega$, and produces the $\b^2$ term in the
exponent of Eq.~\eqref{eq:S4_nonfac_generic_delta}. To further analytically evaluate this contribution, we define the
intervals $\tau_3=\bar x^+-x^+$ and $\tau_4=y^+-\bar x^+$, such that $L^+-y^+=L^+-x^+-\tau_3-\tau_4$ and the domains read
\begin{align}\label{eq:IV_time_domain}
	0<\tau_3<L^+,
	\qquad
	0<\tau_4<L^+-\tau_3,
	\qquad
	0<x^+<L^+-\tau_3-\tau_4.
\end{align}
For a time-independent, homogeneous medium the oscillator coefficients on the three- and
four-point intervals are
\begin{align}
	&c_{1,3}
	=
	\frac{\Omega_3}{2i\sin(\Omega_3\tau_3)},
	\qquad
	c_{2,3}
	=
	\frac{\Omega_3}{\tan(\Omega_3\tau_3)},
	\nn
	&c_{1,4}
	=
	\frac{\Omega}{2i\sin(\Omega\tau_4)},
	\qquad
	c_{2,4}
	=
	\frac{\Omega}{\tan(\Omega\tau_4)},
	\qquad
	D_4=c_{2,4}-c_{2,4}^*,
	\nn
	&\Omega_3^2
	=
	-\frac{i(1+z^2)\hat q}{2\omega^+},
	\qquad
	\Omega^2
	=
	-\frac{i\lambda\hat q}{2\omega^+}.
\end{align}
At $\w=0$ the four endpoints of the kernels in
Eq.~\eqref{eq:S4_nonfac_generic_delta} are
\begin{align}
	&\X_i=\gamma\u-\delta\b,
	\qquad
	\bar\X_i=-\gamma\beta\u+\delta\b,
	\nn
	&\X_f=(1-\beta)\gamma\x+\gamma\beta\bDelta-\delta\b,
	\qquad
	\bar\X_f=(1-\beta)\gamma\x-\gamma\bDelta+\delta\b.
\end{align}
The Gaussian integral over the common transverse coordinate, including the
derivative $\nabla_\w$ that acts on both kernels, gives
\begin{align}\label{eq:nonfac_x_integrated}
	&\left.
	\nabla_{\w}
	\int d^2\x\,
	\cK_\Omega(\X_f,y^+;\X_i,\bar x^+)
	\cK_\Omega^*(\bar\X_f,y^+;\bar\X_i,\bar x^+)
	\right|_{\w=0}
	\nn
	&\quad=
	-\frac{
	2i(\omega^+)^2c_{1,4}c_{1,4}^*
	}{
	\pi D_4(1-\beta)^2\gamma
	}
	\exp\{\mathcal E_4(\u,\bDelta)\}\,
	\mathcal W(\u,\bDelta),
\end{align}
where the exponent $\mathcal E_4$ is quadratic and the prefactor $\mathcal W$
is linear in $\u$, $\bDelta$ and $\b$. The
remaining $\u$ and $\bDelta$ integrations form a four-dimensional Gaussian.
Combining Eqs.~\eqref{eq:S3_inin_complete},
\eqref{eq:S4_nonfac_generic_delta} and \eqref{eq:nonfac_x_integrated}, the
exponent before the $\u$ and $\bDelta$ integrations is
\begin{align}\label{eq:nonfac_gaussian_form}
	-\frac12
	\begin{pmatrix}\u&\bDelta\end{pmatrix}
	M
	\begin{pmatrix}\u\\\bDelta\end{pmatrix}
	+
	\b\cdot
	\begin{pmatrix}s_u&s_\Delta\end{pmatrix}
	\begin{pmatrix}\u\\\bDelta\end{pmatrix}
	+C_0\b^2-i\bDelta\cdot\p,
\end{align}
where $M$, $(s_u,s_\Delta)$ and $C_0$ are given in
Eq.~\eqref{eq:nonfac_M_s_C}.
We write the inverse matrix as
\begin{align}
	M^{-1}
	=
	\begin{pmatrix}
	C_{uu}&C_{u\Delta}\\
	C_{u\Delta}&C_{\Delta\Delta}
	\end{pmatrix}.
\end{align}
The Gaussian means of $\u$ and $\bDelta$, with the momentum dependence
displayed, are
\begin{align}\label{eq:nonfac_gaussian_means}
	U
	&=
	(C_{uu}s_u+C_{u\Delta}s_\Delta)\b
	-iC_{u\Delta}\p,
	\nn
	V
	&=
	(C_{u\Delta}s_u+C_{\Delta\Delta}s_\Delta)\b
	-iC_{\Delta\Delta}\p.
\end{align}
The polynomial prefactor comes from the transition vertex and the two
kernel derivatives. Its Gaussian average is
\begin{equation}
\mathcal P(\p^2,\b\cdot\p)
=\left\langle\mathcal T(\bDelta,\b)
\left[2\omega^+c_{1,3}\u-\frac{z\omega^+}{1+z^2}
(2c_{1,3}+ic_{2,3})\b\right]\cdot\mathcal W\right\rangle,
\end{equation}
where the average has the means of Eq.~\eqref{eq:nonfac_gaussian_means}
and covariance $M^{-1}$. The coefficients of $\mathcal W$ and the expanded
polynomial are given in App.~\ref{app:nonfac_coefficients}.
Since $U$ and $V$ are given explicitly in Eq.~\eqref{eq:nonfac_gaussian_means},
$\mathcal P$ is a polynomial in $\p^2$ and $\b\cdot\p$ of degree at most four
in the components of $\p$. The $\u$ and $\bDelta$ integrations give
\begin{align}\label{eq:nonfac_fixed_p_complete}
	&\int d^2\u\,d^2\bDelta\,
	e^{-i\bDelta\cdot\p}
	\mathcal T(\bDelta,\b)\,
	\left[
	2\omega^+c_{1,3}\u
	-\frac{z\omega^+}{1+z^2}
	(2c_{1,3}+ic_{2,3})\b
	\right]\cdot\mathcal W
	\nn
	&\qquad\times
	\exp\left\{
	-\frac12
	\begin{pmatrix}\u&\bDelta\end{pmatrix}
	M
	\begin{pmatrix}\u\\\bDelta\end{pmatrix}
	+
	\b\cdot
	\begin{pmatrix}s_u&s_\Delta\end{pmatrix}
	\begin{pmatrix}\u\\\bDelta\end{pmatrix}
	+C_0\b^2
	\right\}
	\nn
	&\quad=
	\frac{(2\pi)^2}{\det M}\,
	\exp\left[
	\Phi\b^2
	-\alpha\p^2
	+i\frac{\kappa_0}{b}\b\cdot\p
	\right]
	\mathcal P(\p^2,\b\cdot\p),
\end{align}
where
\begin{align}\label{eq:nonfac_alpha_kappa_Phi}
	\alpha
	&=
	\frac12C_{\Delta\Delta},
	\nn
	\kappa_0
	&=
	-b\left[
	s_uC_{u\Delta}+s_\Delta C_{\Delta\Delta}
	\right],
	\nn
	\kappa_i
	&=
	b\left[
	a_i-s_uC_{u\Delta}-s_\Delta C_{\Delta\Delta}
	\right],
	\nn
	\Phi
	&=
	C_0
	+\frac12
	\left(
	s_u^2C_{uu}
	+2s_us_\Delta C_{u\Delta}
	+s_\Delta^2C_{\Delta\Delta}
	\right).
\end{align}
Equation~\eqref{eq:nonfac_fixed_p_complete} does not yet include the
measurement factor $e^{ia_i\b\cdot\p}$, which replaces $\kappa_0$ by
$\kappa_i=\kappa_0+a_ib$. The quantities $\alpha$ and $\Phi$ do not depend on
the channel. Collecting the factor $-\hat q/4$ of
Eq.~\eqref{eq:S4_nonfac_generic_delta}, the normalization
$\omega^+c_{1,3}/\pi$ of the three-point function, and the prefactor of
Eq.~\eqref{eq:nonfac_x_integrated}, the nonfactorizable in--in contribution
before the relative-momentum integration is
\begin{align}\label{eq:J_inin_nonfac_fixed_p_complete}
	& J^{(1),{\rm med.\,rad.\,nonfac}}_{{\rm in\text{-}in}}(b)
	=
	2C_F\alpha_s
	\sum_{i=q,g,u}
	\int_0^1dz\,
	\frac{P_{q\to qg}^{\rm vac}(z)\,h_i(z)}
	{(2\omega^+)^2}
	\int_{\mathbb R^2}
	\frac{d^2\p}{(2\pi)^2}\,
	e^{-\eta_{\rm UV}\p^2}\,
	e^{ia_i\b\cdot\p}
	\nn
	&\times
	2\Re
	\int_0^{L^+}d\tau_3
	\int_0^{L^+-\tau_3}d\tau_4
	\int_0^{L^+-\tau_3-\tau_4}dx^+
	\frac{2i\hat q(\omega^+)^3c_{1,3}c_{1,4}c_{1,4}^*}{(2\pi)^2D_4(1-\beta)^2\gamma}
	\left[
	\text{RHS of
	Eq.~\eqref{eq:nonfac_fixed_p_complete}}
	\right].
\end{align}

%===========================================================================
\subsection{In--out interference}
\label{sec:inout}

The in--out region is $0<x^+<L^+<\bar x^+<\infty$. The three-point function
propagates through the medium to $L^+$ and then freely to $\bar x^+$. With
$\r_L$ its relative coordinate at the medium boundary,
\begin{align}
	&\widetilde{\cS}^{(3),{\rm in\text{-}out}}_{q\to qg}
	(\r_f,\r_i,\b;\bar x^+,x^+)= 
	\exp\left\{
	-\frac{L^+-x^+}{4(1+z^2)}
	\hat q\,\b^2
	\right\}\nn
	&\times  \int d^2\r_L\,
	\cK_0\!\left(
	\r_f-\frac{z}{1+z^2}\b-\r_L;
	\bar x^+,L^+
	\right)
	\cK_{\Omega_3}\!\left(
	\r_L,L^+;
	\r_i-\frac{z}{1+z^2}\b,x^+
	\right).
\end{align}
Since the conjugate splitting occurs outside the medium, there is no medium
evolution after $\bar x^+$, and the fixed-$\p$ four-point function is
equal to its vacuum counterpart, $e^{-i(\u-\w)\cdot\p}$.
After integration by parts in $\u$, as in Eq.~\eqref{eq:fac_integration_by_parts},
the free propagation between $L^+$ and $\bar x^+$ is Fourier transformed
directly, and the Gaussian integral over $\r_L$ gives
\begin{align}\label{eq:S3_inout_mixed_fourier}
	&\int d^2\u\,e^{-i\u\cdot\p}
	\left.
	\left(\nabla_{\u}\cdot\nabla_{\v}\right)
	\widetilde{\cS}^{(3),{\rm in\text{-}out}}_{q\to qg}
	(\u,\bar x^+;\v,x^+;\b,z)
	\right|_{\v=0}
	=
	\left[
	\frac{2ic_1(\tau)}{c_2(\tau)}\p^2
	+
	\frac{iz\hat q}{2c_2(\tau)}
	\b\cdot\p
	\right]
	\nn
	&\times
	\frac{2ic_1(\tau)}{c_2(\tau)}
	\exp\Bigg\{
	-\frac{\hat q\b^2}{4(1+z^2)}\tau
	-\frac{i\p^2}{2\omega^+}(\bar x^+-L^+)
	-\frac{i\p^2}{2\omega^+c_2(\tau)}
	\nn
	&
	-\frac{\hat q z^2\b^2}
	{4(1+z^2)c_2(\tau)}
	+\frac{iz\,\b\cdot\p}{1+z^2}
	\left(
	\frac{2ic_1(\tau)}{c_2(\tau)}
	-1
	\right)
	\Bigg\},
\end{align}
with $\tau=L^+-x^+$. The integrand depends on $\bar x^+$ only through the phase
$e^{-i\p^2(\bar x^+-L^+)/(2\omega^+)}$ of the free propagation after the
medium. With the factor $e^{-\epsilon\bar x^+}$ of
Eq.~\eqref{eq:fixed_p_master}, its integral over $L^+<\bar x^+<\infty$ is,
for $\p\neq0$,
\begin{align}
	\lim_{\epsilon\to0^+}
	\int_{L^+}^{\infty}d\bar x^+\,
	e^{-\epsilon(\bar x^+-L^+)}
	e^{-\frac{i\p^2}{2\omega^+}(\bar x^+-L^+)}
	=
	-\frac{2i\omega^+}{\p^2}.
\end{align}
Including the parent broadening
$D_F[\b;x^+,0]=e^{-\hat q\b^2(L^+-\tau)/4}$ of Eq.~\eqref{eq:fixed_p_master},
the fixed-$\p$ in--out contribution is
\begin{align}\label{eq:J_inout_fixed_p_complete}
	& J^{(1),{\rm med.\,rad.}}_{\rm in\text{-}out}(b)=
	2C_F\alpha_s
	\sum_{i=q,g,u}
	\int_0^1dz\,
	\frac{P_{q\to qg}^{\rm vac}(z)\,h_i(z)}
	{(2\omega^+)^2}
	\int_{\mathbb R^2}
	\frac{d^2\p}{(2\pi)^2}\,
	e^{-\eta_{\rm UV}\p^2}\,
	e^{ia_i\b\cdot\p}
	2\Re
	\int_0^{L^+}d\tau
	\nn
	&\times\left(-\frac{2i\omega^+}{\p^2}\right)
	\frac{2ic_1(\tau)}{c_2(\tau)}
	\exp\Bigg\{
	-\frac{\hat q\b^2}{4}(L^+-\tau)
	-\frac{\hat q\b^2}{4(1+z^2)}\tau
	-\frac{i\p^2}{2\omega^+c_2(\tau)}
	-\frac{\hat q z^2\b^2}
	{4(1+z^2)c_2(\tau)}
	\nn
	&
	+\frac{iz\,\b\cdot\p}{1+z^2}
	\left(
	\frac{2ic_1(\tau)}{c_2(\tau)}
	-1
	\right)
	\Bigg\}
	\left[
	\frac{2ic_1(\tau)}{c_2(\tau)}\p^2
	+
	\frac{iz\hat q}{2c_2(\tau)}
	\b\cdot\p
	\right].
\end{align}
Appendix~\ref{app:p_integration} performs the common $\p$ integration for this
expression and for the two in--in contributions.

%===========================================================================
\section{Medium response}
\label{sec:n_response}

We consider the contribution of the medium response to the TEEC, following
recent studies of energy flow and EECs in many body states, see
e.g.~\cite{Barata:2024ukm,Barata:2026pgh,Barata:2025fzd,
Delacretaz:2018cfk,Delacretaz:2020nit,Chicherin:2023gxt,
Cuomo:2025pjp,Grassi:2026tdb}. Throughout this section, expectation
values are taken at fixed dijet geometry, the average over the dijet
geometry being performed afterwards. We decompose the energy flow operator
of Eq.~\eqref{eq:energy_flow_operator} into
jet and medium response (\textit{wake} [$W$]) contributions, $\mathcal E=\mathcal E_J+\mathcal E_W$.
We treat the response
through its classical mean flow
(see, e.g.,~\cite{Casalderrey-Solana:2016jvj,Chen:2017zte,Tachibana:2017syd,Luo:2023nsi,Yang:2022nei,Bossi:2024qho,KunnawalkamElayavalli:2017hxo,Casalderrey-Solana:2026ftw} for phenomenological studies)
and neglect connected correlations
involving it, so that the correlator in Eq.~\eqref{eq:teec_operator_definition} becomes
\begin{align}\label{eq:resp_split}
\langle\mathcal E(\n_1)\mathcal E(\n_2)\rangle
&\simeq\langle\mathcal E_J(\n_1)\mathcal E_J(\n_2)\rangle
+\langle\mathcal E_J(\n_1)\rangle\langle\mathcal E_W(\n_2)\rangle
\nn
&+\langle\mathcal E_W(\n_1)\rangle\langle\mathcal E_J(\n_2)\rangle
+\langle\mathcal E_W(\n_1)\rangle\langle\mathcal E_W(\n_2)\rangle,
\end{align}
where the first term is the perturbative correlator computed in
Secs.~\ref{sec:vac}--\ref{sec:time_regions} and the remaining ones are
products of mean flows. We use the distribution in $\phi$,
$\sigma^{-1}d\Sigma/d\phi$, obtained from
Eq.~\eqref{eq:teec_operator_definition} by replacing
$\delta(\tau-(1+\cos\phi_{12})/2)$ with $\delta(\phi-\phi_{12})$. For the response estimate we
take the two jet axes in the transverse plane, so that $E_T=2p_T$ at leading
power. Since the measurement depends on the azimuthal separation, we first
integrate
over the polar angles,
\begin{equation}\label{eq:resp_profile}
e_A(\varphi)=\int_0^\pi d\theta\,\sin^2\!\theta\,
\big\langle\mathcal E_A(\theta,\varphi)\big\rangle,
\qquad A=J,W,
\end{equation}
the transverse energy per unit beam azimuth. The two separations selected
by the measurement, $\varphi_2=\varphi_1\pm\phi$, give
\begin{equation}\label{eq:resp_total}
\frac{1}{\sigma}\frac{d\Sigma}{d\phi}
=\frac{1}{\sigma}\bigg(\frac{d\Sigma_{\rm pert}}{d\phi}+\frac{d\Sigma_{JW}}{d\phi}+\frac{d\Sigma_{WW}}{d\phi}\bigg),
\end{equation}
with $d\Sigma_{\rm pert}/d\phi$ the first contribution in Eq.~\eqref{eq:resp_split} and
\begin{align}
\frac{1}{\sigma}\frac{d\Sigma_{JW}}{d\phi}
&=\frac{2}{E_T^2}\int_0^{2\pi}d\varphi\,
\big[e_J(\varphi)\,e_W(\varphi+\phi)+e_W(\varphi)\,e_J(\varphi+\phi)\big],
\label{eq:resp_JW}\\\nonumber
\frac{1}{\sigma}\frac{d\Sigma_{WW}}{d\phi}
&=\frac{2}{E_T^2}\int_0^{2\pi}d\varphi\,e_W(\varphi)\,e_W(\varphi+\phi).
\label{eq:resp_WW}
\end{align}
For identical jets and wakes with profiles symmetric about each jet axis,
the total mean flows satisfy $e_A(\varphi+\pi)=e_A(\varphi)$. We assign
the transverse energy $\Delta p_T=\varepsilon p_T$ to each wake and
$(1-\varepsilon)p_T$ to each surviving jet, so that with $E_T=2p_T$
\begin{equation}\label{eq:resp_norm}
\int_0^{2\pi}d\varphi\,e_J(\varphi)=(1-\varepsilon)E_T,\qquad
\int_0^{2\pi}d\varphi\,e_W(\varphi)=\varepsilon E_T .
\end{equation}
Let $\sigma_J$ and $\sigma_W$ be the angular scales on which the jet and
wake profiles vary. For $\sigma_J\ll\sigma_W$ the mixed contribution can
be expanded across each jet,
\begin{equation}\label{eq:resp_expand}
\int_0^{2\pi}d\varphi\,e_J(\varphi)\,e_W(\varphi+\phi)
=(1-\varepsilon)E_T\Big[e_W(\phi)+\mathcal O\big(\sigma_J^2\,e_W''(\phi)\big)\Big],
\end{equation}
where reflection symmetry removes the linear term, so that
\begin{equation}\label{eq:resp_JW_leading}
\frac{1}{\sigma}\frac{d\Sigma_{JW}}{d\phi}
=\frac{4(1-\varepsilon)}{E_T}\,e_W(\phi)\Bigg[1+\mathcal O\Bigg(\frac{\sigma_J^2}{\sigma_W^2}\Bigg)\Bigg]\,,
\end{equation}
i.e., a broad wake probes the integrated jet energy, and the sensitivity to the
angular distribution of the jet is suppressed by the squared ratio of
widths. In this limit, the same property of the profile
controls its variation across the back-to-back window. With
$\phi=\pi-\delta$, both terms are even about $\pi$ and
\begin{equation}\label{eq:resp_window}
\frac{d\Sigma_X}{d\phi}\Bigg|_{\pi-\delta}
=\frac{d\Sigma_X}{d\phi}\Bigg|_{\pi}\left[1+\mathcal O\Bigg(\frac{\delta^2}{\sigma_W^2}\Bigg)\right],
\qquad X=JW,\,WW ,
\end{equation}
so that for $\delta\ll\sigma_W$ the response is approximately constant
across the window.\footnote{In the variable $\tau=(1+\cos\phi)/2$, with
$|d\phi/d\tau|=[\tau(1-\tau)]^{-1/2}$, a contribution with a finite value
at $\phi=\pi$ behaves as $\tau^{-1/2}[\,d\Sigma_X/d\phi|_\pi+\mathcal O(\tau)]$.}
Therefore, its relative importance can grow toward the edge of the window,
where the perturbative distribution is smaller. At small $\varepsilon$
this contribution is linear in $\Delta p_T/p_T$, with its height and
residual angular dependence set by the wake profile. Since the perturbative medium correction
retains the soft partons that may thermalize, their overlap with the
response is not subtracted, and $\Delta p_T$ is an effective response
strength whose relation to the deposited energy is model dependent. A
more complete treatment would exclude thermalized partons from the
perturbative final state and assign their energy to the wake. This requires a consistent separation between the perturbative radiation and
the response, see e.g.~\cite{Barata:2024ukm}, and lies beyond the scope of this work.

For the numerical results of Sec.~\ref{sec:numerics} we specialize
to a Gaussian wake~\cite{Barata:2025fzd} about each jet direction $\pm\hat n$,
with $\hat n$ the Born quark direction of Sec.~\ref{sec:vac},
\begin{equation}\label{eq:resp_gauss}
\big\langle\mathcal E_W(\n_1)\big\rangle\propto
\exp\Big[-\frac{\arccos^2(\n_1\cdot\hat n)}{\theta_W^2}\Big]
+\exp\Big[-\frac{\arccos^2(-\n_1\cdot\hat n)}{\theta_W^2}\Big],
\end{equation}
of width $\theta_W$, perform the polar projection, and fix the
normalization through Eq.~\eqref{eq:resp_norm}. In the limit
$\theta_W\to\infty$, where the expansions above apply, the projected
profile is uniform and the two terms are the constants
$2\varepsilon(1-\varepsilon)/\pi$ and $\varepsilon^2/\pi$. In the opposite
limit, $\theta_W\ll1$, retaining $\sigma_J\ll\theta_W$, the polar
projection is itself Gaussian in beam azimuth about each jet, and the two
terms take the explicit form
\begin{align}\label{eq:resp_gauss_terms}
\frac{1}{\sigma}\frac{d\Sigma_{JW}}{d\phi}&\simeq\frac{2\varepsilon(1-\varepsilon)}{\sqrt\pi\,\theta_W}
\Big[e^{-(\pi-\phi)^2/\theta_W^2}+e^{-\phi^2/\theta_W^2}\Big],
\nonumber\\
\frac{1}{\sigma}\frac{d\Sigma_{WW}}{d\phi}&\simeq\frac{\varepsilon^2}{\sqrt{2\pi}\,\theta_W}
\Big[e^{-(\pi-\phi)^2/2\theta_W^2}+e^{-\phi^2/2\theta_W^2}\Big],
\end{align}
i.e., Gaussians about $\phi=\pi$ and $\phi=0$ of widths $\theta_W$ and
$\sqrt2\,\theta_W$, whose expansion about $\pi$ reproduces
Eq.~\eqref{eq:resp_window} with corrections $-\delta^2/\theta_W^2$ and
$-\delta^2/2\theta_W^2$.

\section{Numerical results}\label{sec:numerics}

\begin{figure}[t]
    \centering
    \includegraphics[width=\linewidth]{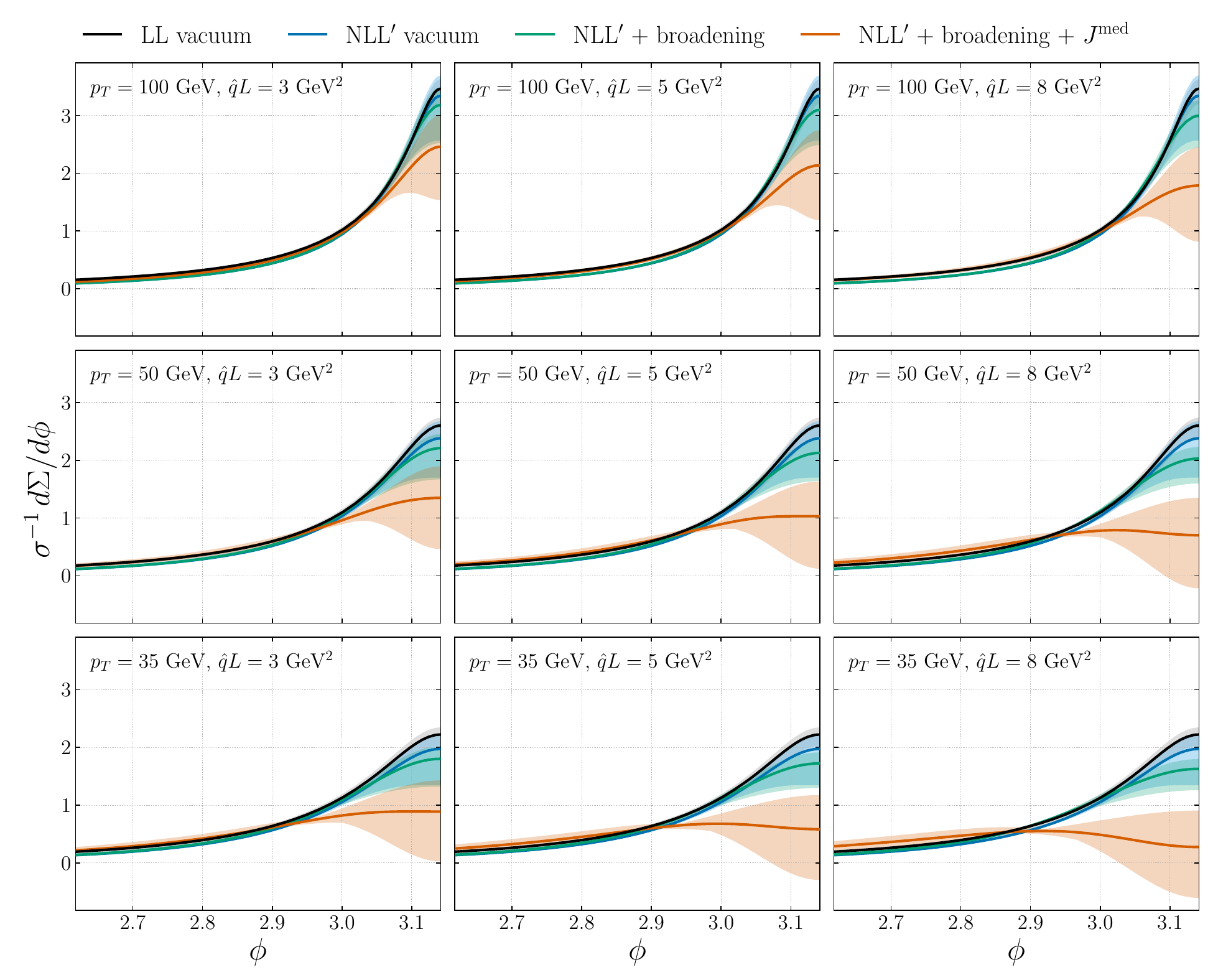}
    \caption{Back-to-back TEEC distribution $\sigma^{-1}d\Sigma/d\phi$
    at fixed $p_T$, for $Q=2p_T$ and $L=5$~fm. Rows: $p_T=100$, $50$ and $35$~GeV. Columns:
    $\hat qL=3$, $5$ and $8$~GeV$^2$. Shown are the vacuum distribution at LL
    and NLL$'$, and the medium-modified distribution at NLL$'$ with broadening
    only and with broadening and medium-induced radiation. The bands denote the scale
    variation discussed in the main text.}
    \label{fig:teec_medium_distributions}
\end{figure}

\begin{figure}[t]
    \centering
    \includegraphics[width=\linewidth]{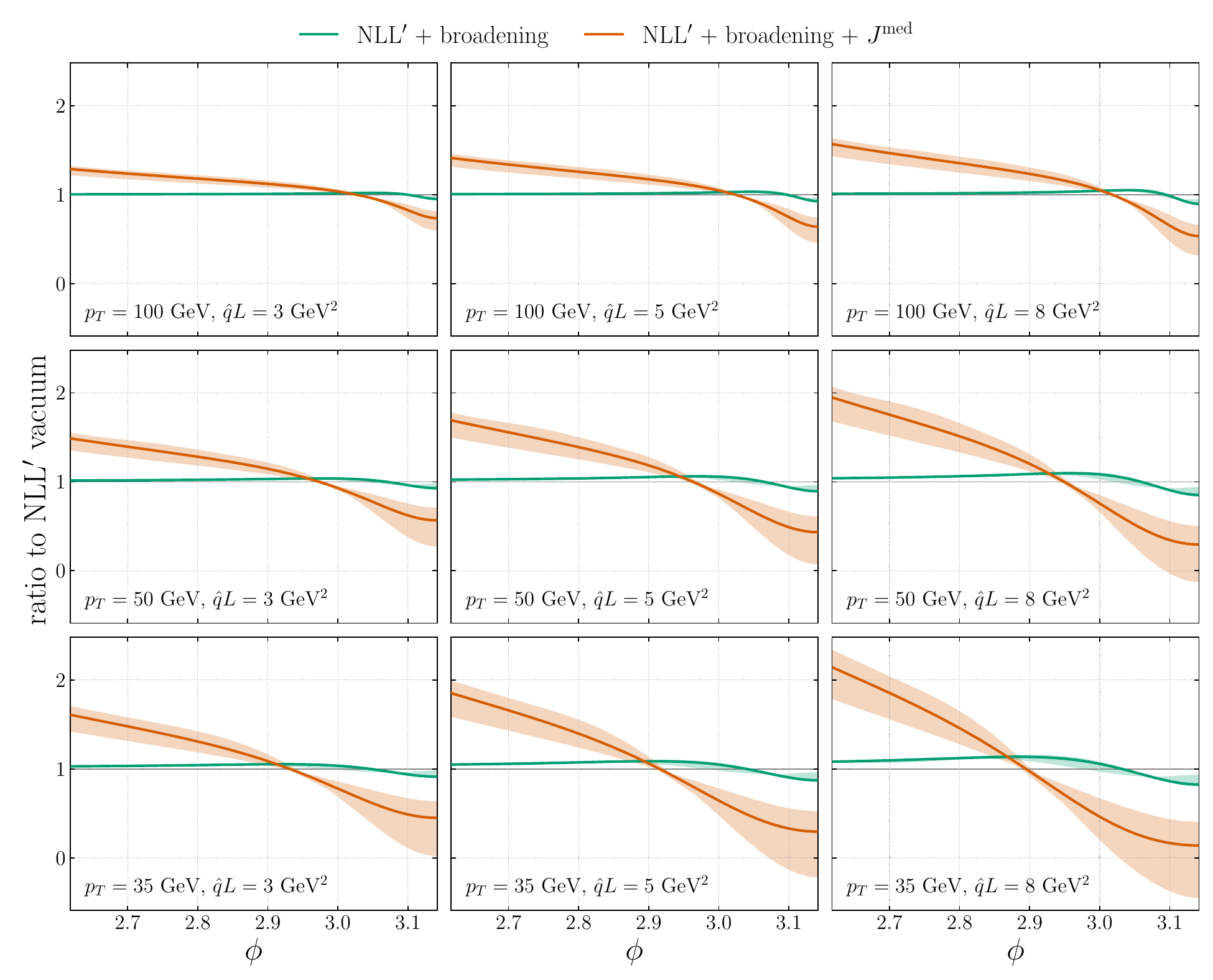}
    \caption{Ratio of the medium-modified distributions of
    Fig.~\ref{fig:teec_medium_distributions} to the vacuum distribution at
    NLL$'$, for the same values of $p_T$ and $\hat qL$. The scale bands
    use the same variations in the numerator and denominator.}
    \label{fig:teec_medium_ratios}
\end{figure}

Figures~\ref{fig:teec_medium_distributions} and~\ref{fig:teec_medium_ratios} show the back-to-back TEEC for transverse jets with parent energy $E_0=p_T$ and $Q=2p_T$, with $p_T=35$, $50$ and $100$~GeV, in a medium of physical length $L=5$~fm and $\hat qL=3$, $5$ and $8$~GeV$^2$.\footnote{The light-cone variables used in the calculation are
$p_0^+=\sqrt2 E_0$ and $L^+=\sqrt2 L$. The quoted $\hat qL$ denotes
$\int_0^{L^+}ds^+\,\hat q(s^+)$, with $\hat q(s^+)$ defined in
Eq.~\eqref{eq:HO_qhat_definition}.} We compare the vacuum distribution at LL and NLL$'$, computed with the numerical prescription of Sec.~\ref{sec:vac}, with the medium-modified distribution at NLL$'$, first with broadening alone and then with the radiative correction added as in Eq.~\eqref{eq:medium_insertion_nllp}. The bands are the scale variations of Sec.~\ref{sec:vac}, with the same variation in the numerator and denominator of the ratios. The propagated numerical uncertainty of the medium-induced radiative correction, below $1\%$ of the vacuum distribution throughout the plotted range, is smaller than these bands and is not shown.

For the parameters studied, broadening lowers the distribution at $\phi=\pi$ by $5$--$17\%$, while the radiative correction is several times larger there, suppressing the peak by $20$--$70\%$ of the vacuum value, and dominates the enhancement toward the edge of the plotted window. Both modifications grow with $\hat qL$ and with decreasing $p_T$, the radiative correction more strongly with $p_T$ and broadening more strongly with $\hat qL$, as expected from the medium scales that control each of them. From $p_T=100$ to $35$~GeV the suppression of the peak grows by a factor of about $2$ for the radiative correction and $1.7$--$1.8$ for broadening, while from $\hat qL=3$ to $8$~GeV$^2$ it grows by $1.5$--$1.7$ and $2.0$--$2.2$, respectively. Broadening suppresses the $b$ integrand beyond $b\sim(\hat qL)^{-1/2}$, which smears the peak by a transverse momentum of order $\sqrt{\hat qL}$, about $2$~GeV for $\hat qL=5$~GeV$^2$. The radiative correction redistributes energy flow toward larger decorrelations. Since $J^{(1),\,{\rm med.\,rad.}}(b)$ is negative over the relevant range of $b$, it lowers the distribution at $\phi=\pi$, and its contribution changes sign at $\pi-\phi\approx0.13$--$0.27$, beyond which it enhances the distribution.
The soft BDMPS--Z spectrum~\cite{Baier:1996kr} scales as $d\widehat\sigma^{\rm med}_{qg|p_0}/dz\sim\alpha_s/(1-z)\sqrt{\hat qL^2/(z(1-z)p_T)}$ for $(1-z)p_T\ll\hat qL^2$, as follows from Eq.~\eqref{eq:n_endpoint_scaling} in App.~\ref{app:endpoint}. This suggests a larger redistribution as $\hat qL$ increases or $p_T$ decreases at fixed $L$, and, since a gluon that broadens over a distance of order $L$ gives $\pi-\phi\sim\sqrt{\hat qL}/[(1-z)p_T]$, that softer emissions can populate larger decorrelations.
The radiative correction is proportional to the coupling evaluated at $\bar{\mu}_b$, whereas broadening at fixed $\hat qL$ contains no explicit factor of $\alpha_s$. A different coupling prescription, such as a fixed value or a medium scale (Sec.~\ref{sec:n_radiative}), changes the size of the correction roughly by the ratio of the couplings in the region of $b$ that dominates the peak, where $\alpha_s(\bar{\mu}_b)$ lies between $0.2$ and $0.3$, and, since the coupling enters at each $b$, it also affects the shape in $\phi$. Near $\phi=\pi$ the scale bands are driven mainly by $\mu\to\mu/2$, which raises the coupling at large $b$. At lower $p_T$ and larger $\hat qL$, where the correction becomes comparable to the vacuum distribution and some scale variations give negative values near the peak, sub-eikonal corrections to the medium-induced radiative correction can become important.

\begin{figure}[t]
    \centering
    \includegraphics[width=\linewidth]{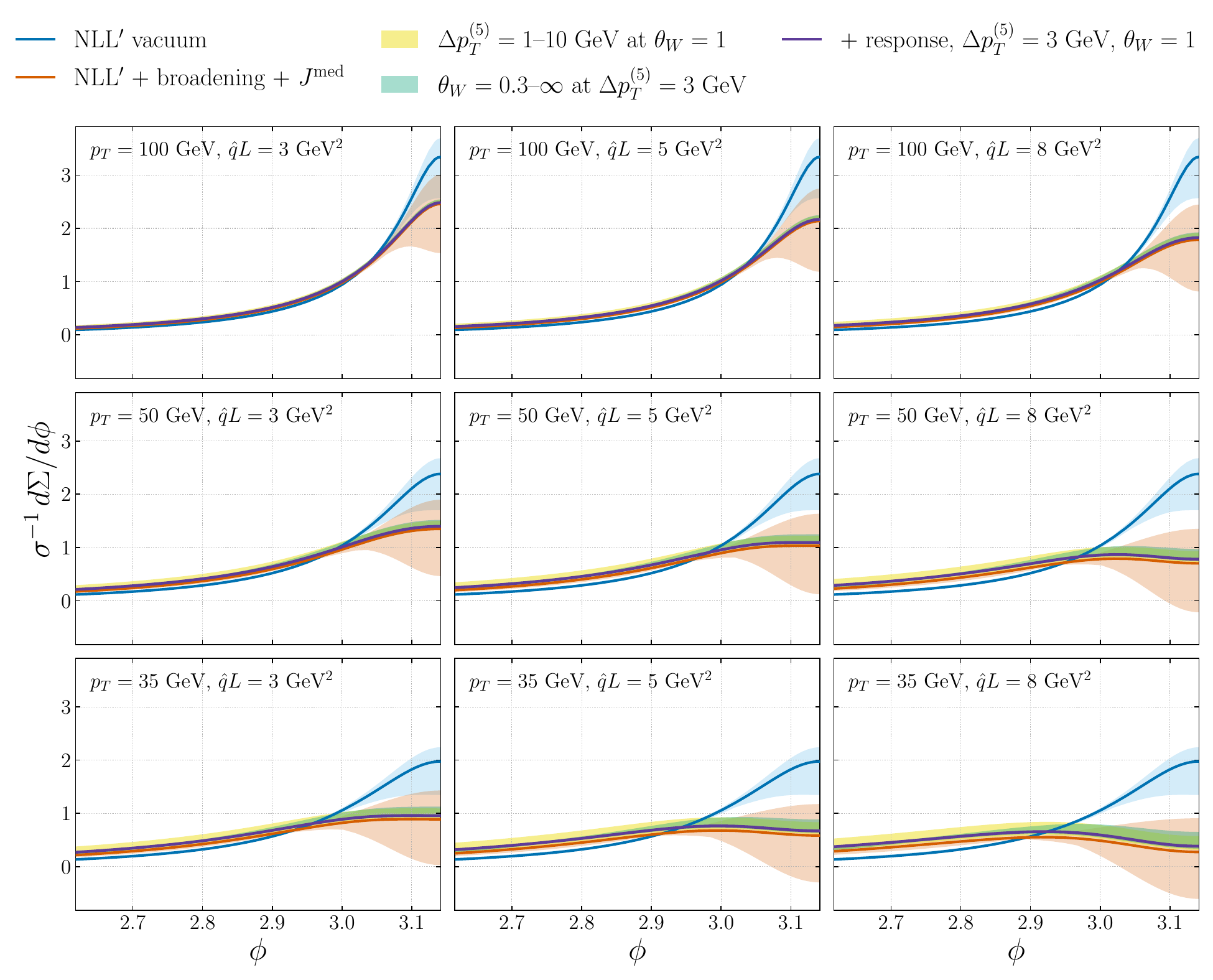}
    \caption{Back-to-back TEEC distribution as in
    Fig.~\ref{fig:teec_medium_distributions}, with the medium response of
    Eq.~\eqref{eq:resp_total} added to the NLL$'$ distribution with
    broadening and medium-induced radiation, at fixed $L=5$~fm, for
    $\Delta p_T=\Delta p_T^{(5)}\sqrt{\hat qL/5~{\rm GeV}^2}$ with
    $\Delta p_T^{(5)}=3$~GeV and $\theta_W=1$. Bands: scale variation for the
    vacuum and medium curves, $\Delta p_T^{(5)}=1$--$10$~GeV at $\theta_W=1$,
    and $\theta_W=0.3$--$\infty$ at $\Delta p_T^{(5)}=3$~GeV.
    Response bands vary these parameters at central perturbative scales.}
    \label{fig:teec_response_distributions}
\end{figure}

\begin{figure}[t]
    \centering
    \includegraphics[width=\linewidth]{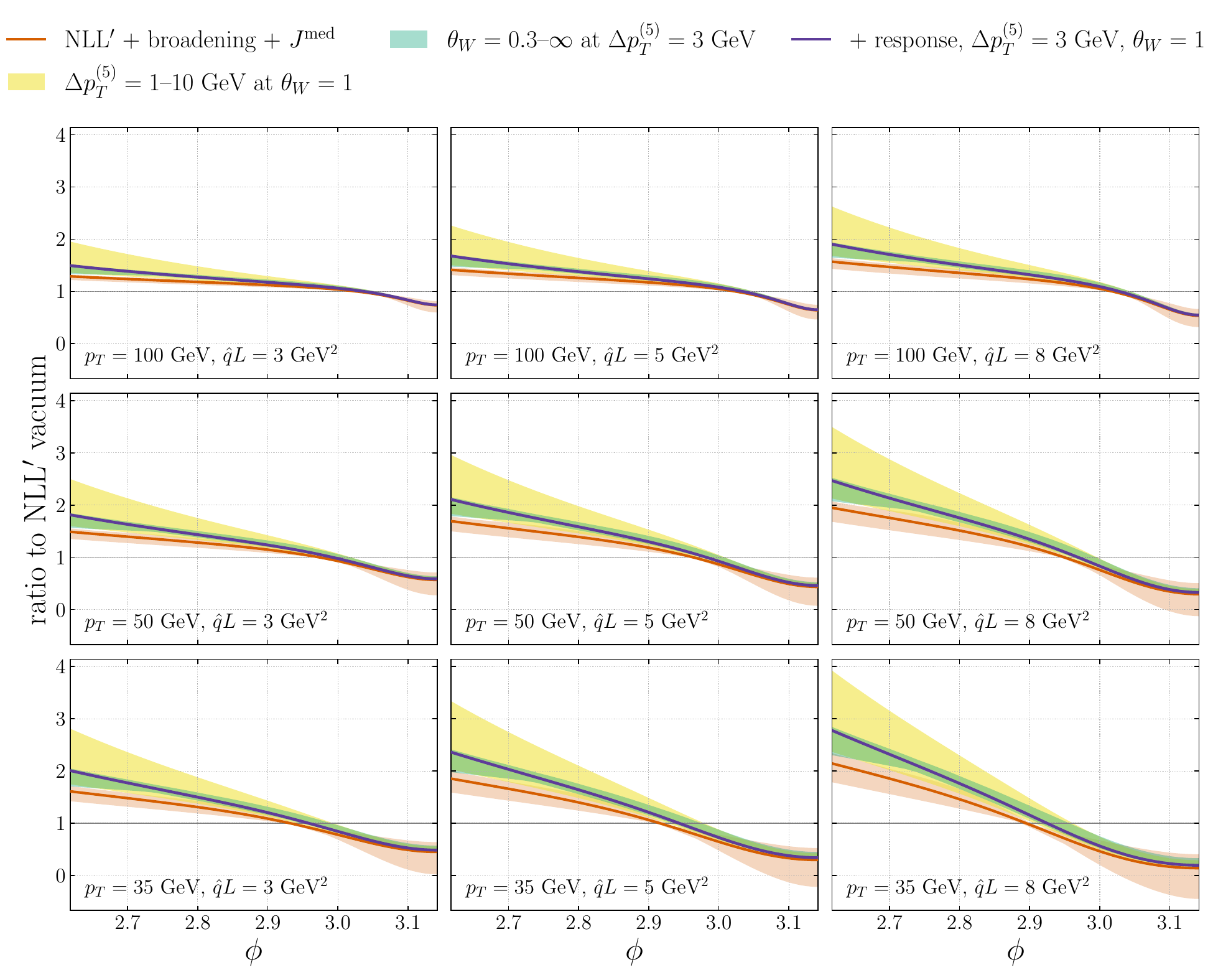}
    \caption{Ratio of the distributions of
    Fig.~\ref{fig:teec_response_distributions} to the vacuum distribution at
    NLL$'$, for the same values of $p_T$ and $\hat qL$. Scale bands use
    the same variations in the numerator and denominator. Response bands
    vary the model parameters at central perturbative scales.}
    \label{fig:teec_response_ratios}
\end{figure}

Figures~\ref{fig:teec_response_distributions}
and~\ref{fig:teec_response_ratios} add the medium response of
Sec.~\ref{sec:n_response}, Eq.~\eqref{eq:resp_total}, to the distribution
with broadening and medium-induced radiation. In the harmonic-oscillator
spectrum, the energy carried by soft gluons below a fixed thermalization
scale scales as $\sqrt{\hat q L^2}$, independently of $p_T$. Motivated by this
behavior, we adopt the response prescription
$\Delta p_T=\Delta p_T^{(5)}\sqrt{\hat qL/5~\text{GeV}^2}$, with
$\Delta p_T^{(5)}=3$~GeV and $\theta_W=1$ for the central curve, and vary
$\Delta p_T^{(5)}$ between $1$~GeV and $10$~GeV,
and $\theta_W$ between $0.3$, a representative jet angular scale,
and $\infty$, with the full polar projection of Eq.~\eqref{eq:resp_gauss}.
The $1$~GeV benchmark is guided by the wake energy inside the jet cone
estimated in~\cite{Barata:2025fzd}, while $10$~GeV is an illustrative
upper benchmark.
The central value of $\varepsilon=\Delta p_T/p_T$ ranges from $0.023$ to
$0.11$, and over the $\Delta p_T^{(5)}$ band from $0.008$ to $0.36$. At small $\varepsilon$, the response scales as $1/p_T$ and as $\sqrt{\hat qL}$ at fixed $L$ under this prescription. The response bands, which vary $\Delta p_T^{(5)}$ or $\theta_W$ separately at central perturbative scales, describe the parameter dependence of the response model and are not combined with the scale variation bands. The energy variation changes the response by roughly a factor of one third to three relative to the central choice, and for the largest energy fractions the quadratic terms in $\varepsilon$ also matter.
 A broad wake gives a
slowly varying addition across the window, so its relative importance
grows as the perturbative distribution falls. Within the chosen parameter ranges, the wake energy has the larger effect
away from the peak, while narrower wakes concentrate more energy near
$\phi=\pi$. At $p_T=100$~GeV and $\hat qL=5$~GeV$^2$, the central
response raises the ratio to vacuum at $\phi=2.64$ from about $1.4$ to $1.6$.
The response becomes more significant at lower $p_T$ and larger $\hat qL$, although its quantitative interpretation is limited by the unsubtracted overlap with perturbative radiation (Sec.~\ref{sec:n_response}). A broad response may
become dominant at larger decorrelations beyond the back-to-back region.

\section{Conclusions}\label{sec:conclusion}

Back-to-back topologies in heavy ion collisions offer a setting,
complementary to collinear configurations, in which medium modifications of
the final state energy flow can be computed and tested through angular
decorrelations. In this work we studied the TEEC in the back-to-back limit in
a simplified scenario of an $e^+e^-$ collision occurring in the presence of a
homogeneous background medium, in which a $q\bar q$ pair produced in
a colour-singlet hard process propagates through the medium. This setup discards the nuclear initial state effects of realistic heavy ion collisions, thus isolating final state modifications. We assumed that the vacuum
factorized structure holds in the medium, with medium effects entering only
through the jet functions of the back-to-back TMD factorization formula. We
combined the vacuum distribution at NLL$'$ with momentum broadening and the
medium-induced radiative correction at first order in the strong coupling,
computed for partons propagating through a
dense medium, in the large $N_c$ harmonic-oscillator approximation. We
further estimated the contribution of the medium response through its mean
flow contribution to the energy flow operator.

For jets with $p_T=35$--$100$~GeV in a medium with $\hat qL=3$--$8$~GeV$^2$,
broadening lowers the back-to-back peak by $5$--$17\%$. The radiative
correction suppresses the peak by $20$--$70\%$ of the vacuum value and
enhances the distribution at larger angular decorrelations. Both
modifications grow with $\hat qL$ and with decreasing $p_T$. The radiative
correction depends more strongly on $p_T$, as expected from the scaling of
the soft BDMPS--Z spectrum with $\sqrt{\hat qL^2/p_T}$, whereas the
broadening factor is mostly sensitive to $\hat qL$. Therefore, a quantitative
description of the peak requires the medium-induced radiative correction in
addition to broadening. Its size is proportional to the coupling evaluated at
the scale $\bar{\mu}_b$, so that it depends directly on the coupling prescription.
For a wake of transverse energy $\Delta p_T$ and angular width $\theta_W$
about each jet, the medium response adds a nearly constant contribution when
the wake is broader than the back-to-back window, linear in $\Delta p_T/p_T$
when this ratio is small. In the numerical results the response populates larger angular
decorrelations, where its relative importance grows as the perturbative
distribution falls, and its size is set mainly by the wake energy, while
narrower wakes concentrate more of it near $\phi=\pi$.

Together, these results provide a starting point for the study of
back-to-back configurations in heavy ion collisions with energy correlators,
and show that momentum broadening and the medium
response leave distinct imprints on the azimuthal decorrelation. Several
extensions are left for future work. A direct application to phenomenology
requires initial state effects and a realistic expanding medium in which the
two jets traverse different path lengths. Beyond momentum broadening, the
assignment of medium modifications to the jet and soft functions requires a
consistent treatment of the medium-dependent soft and collinear sectors, and
relaxing the independent propagation of the quark and antiquark requires accounting for
medium-dependent correlations between them. At low $p_T$ and large $\hat qL$,
where the radiative correction becomes comparable to the vacuum distribution,
sub-eikonal corrections may become relevant. Finally, a more sophisticated
modeling of the background would allow for a better separation between the
perturbative effects and the medium response.

\section*{Acknowledgments}
JB was supported by the U.S. Department of Energy under Contract No. DE-SC0012704. 
HTL is supported by the National Natural Science Foundation of China under Grants No.~12275156 and No.~12321005, and the Shandong Provincial Department of Science and Technology under Project No.~TQ012025001.
The work of AVS is supported by the Spanish Ministry of Science, Innovation and Universities through grant PID2025-174203NB-I00 (MICIU/AEI/10.13039/501100011033) and by the Basque Government through grant IT1977-26. AVS also acknowledges support from Ikerbasque, Basque Foundation for Science. JMS is supported by the National Science Foundation under grant No. PHY-2515057.

\appendix

\section{Coefficients of the non-factorizable contribution}
\label{app:nonfac_coefficients}

This appendix collects the coefficients entering the Gaussian integrals of
Sec.~\ref{sec:inin_nonfac}. We use a $2\times2$ notation on the pair $(\u,\bDelta)$, in which
every entry multiplies the identity in transverse space, and introduce the
coefficient space vectors
\begin{align}
	&\rho_f=(0,\gamma\beta),
	\qquad
	\rho_i=(\gamma,0),
	\qquad
	\bar\rho_f=(0,-\gamma),
	\qquad
	\bar\rho_i=(-\gamma\beta,0),
	\nn
	&\sigma_f=\sigma_i=-\delta,
	\qquad
	\bar\sigma_f=\bar\sigma_i=\delta.
\end{align}
For example,
$\gamma\beta\bDelta-\delta\b=\rho_f\cdot(\u,\bDelta)+\sigma_f\b$. With
\begin{align}
	\chi
	={}&
	c_{2,4}\rho_f-c_{2,4}^*\bar\rho_f
	-2i(c_{1,4}\rho_i+c_{1,4}^*\bar\rho_i),
	\nn
	\sigma_\chi
	={}&
	c_{2,4}\sigma_f-c_{2,4}^*\bar\sigma_f
	-2i(c_{1,4}\sigma_i+c_{1,4}^*\bar\sigma_i),
\end{align}
the quadratic, linear and constant parts of $\mathcal E_4$ are
\begin{align}
	K_4
	={}&
	\frac{i\omega^+}{2}c_{2,4}
	(\rho_f\rho_f^{T}+\rho_i\rho_i^{T})
	+
	\omega^+c_{1,4}
	(\rho_f\rho_i^{T}+\rho_i\rho_f^{T})
	\nn
	&-
	\frac{i\omega^+}{2}c_{2,4}^*
	(\bar\rho_f\bar\rho_f^{T}+\bar\rho_i\bar\rho_i^{T})
	+
	\omega^+c_{1,4}^*
	(\bar\rho_f\bar\rho_i^{T}+\bar\rho_i\bar\rho_f^{T})
	-
	\frac{i\omega^+}{2D_4}\chi\chi^{T},
	\label{eq:nonfac_K4}\\
	\ell_4
	={}&
	i\omega^+c_{2,4}(\sigma_f\rho_f+\sigma_i\rho_i)
	+
	2\omega^+c_{1,4}(\sigma_f\rho_i+\sigma_i\rho_f)
	\nn
	&-
	i\omega^+c_{2,4}^*
	(\bar\sigma_f\bar\rho_f+\bar\sigma_i\bar\rho_i)
	+
	2\omega^+c_{1,4}^*
	(\bar\sigma_f\bar\rho_i+\bar\sigma_i\bar\rho_f)
	-
	\frac{i\omega^+}{D_4}\sigma_\chi\chi,
	\label{eq:nonfac_l4}\\
	k_4
	={}&
	\frac{i\omega^+}{2}c_{2,4}
	(\sigma_f^2+\sigma_i^2)
	+
	2\omega^+c_{1,4}\sigma_f\sigma_i
	\nn
	&-
	\frac{i\omega^+}{2}c_{2,4}^*
	(\bar\sigma_f^2+\bar\sigma_i^2)
	+
	2\omega^+c_{1,4}^*\bar\sigma_f\bar\sigma_i
	-
	\frac{i\omega^+}{2D_4}\sigma_\chi^2.
	\label{eq:nonfac_k4}
\end{align}

The coefficients of the combined exponent in
Eq.~\eqref{eq:nonfac_gaussian_form} are
\begin{align}\label{eq:nonfac_M_s_C}
	M
	={}&
	-2\left[
	K_4+
	\begin{pmatrix}
	\dfrac{i\omega^+}{2}c_{2,3}&0\\[2mm]
	0&
	-\dfrac{\hat q}{4}
	(L^+-x^+-\tau_3-\tau_4)
	\left((1-z)^2+2z^2\right)
	\end{pmatrix}
	\right],
	\nn
	\begin{pmatrix}s_u\\s_\Delta\end{pmatrix}
	={}&
	\ell_4+
	\begin{pmatrix}
	-\dfrac{z\omega^+}{1+z^2}
	(2c_{1,3}+ic_{2,3})
	\\[2mm]
	-\dfrac{\hat q}{2}
	(L^+-x^+-\tau_3-\tau_4)(1-3z)
	\end{pmatrix},
	\nn
	C_0
	={}&
	k_4
	-\frac{\hat q x^+}{4}
	-\frac{\hat q\tau_3}{4(1+z^2)}
	-\frac{\hat q\tau_4}{4\lambda^2}
	-\frac{3\hat q}{4}
	(L^+-x^+-\tau_3-\tau_4)
	\nn
	&+
	\frac{z^2\omega^+}{(1+z^2)^2}
	(2c_{1,3}+ic_{2,3}).
\end{align}

The polynomial prefactor arises from
$\mathcal T(\bDelta,\b)\,\nabla_{\v}\widetilde{\cS}^{(3)}\cdot
\nabla_{\w}\widetilde{\cS}^{(4)}$. We write the two linear forms as
\begin{align}
	&2\omega^+c_{1,3}\u
	-\frac{z\omega^+}{1+z^2}
	(2c_{1,3}+ic_{2,3})\b
	=
	L_u\u+L_b\b,
	\nn
	&\mathcal W=W_u\u+W_\Delta\bDelta+W_b\b,
\end{align}
with
\begin{align}
	L_u=2\omega^+c_{1,3},
	\qquad
	L_b=
	-\frac{z\omega^+}{1+z^2}
	(2c_{1,3}+ic_{2,3}),
\end{align}
and
\begin{align}
	\begin{pmatrix}W_u&W_\Delta\end{pmatrix}
	={}&
	2(\beta c_{1,4}\rho_f-c_{1,4}^*\bar\rho_f)
	-2\frac{\beta c_{1,4}-c_{1,4}^*}{D_4}\,\chi
	+
	i(\beta c_{2,4}\rho_i+c_{2,4}^*\bar\rho_i),
	\nn
	W_b
	={}&
	2(\beta c_{1,4}\sigma_f-c_{1,4}^*\bar\sigma_f)
	-2\frac{\beta c_{1,4}-c_{1,4}^*}{D_4}\,\sigma_\chi
	+
	i(\beta c_{2,4}\sigma_i+c_{2,4}^*\bar\sigma_i).
\end{align}
Their product is
\begin{align}
	&(L_u\u+L_b\b)\cdot
	(W_u\u+W_\Delta\bDelta+W_b\b)
	\nn
	&\qquad=
	q_1\u^2+q_2\u\cdot\bDelta+q_3\b\cdot\u
	+q_4\b\cdot\bDelta+q_5\b^2,
\end{align}
with
\begin{align}
	q_1=L_uW_u,\quad
	q_2=L_uW_\Delta,\quad
	q_3=L_uW_b+L_bW_u,\quad
	q_4=L_bW_\Delta,\quad
	q_5=L_bW_b.
\end{align}
Wick's theorem then fixes all required moments of $\u$ and $\bDelta$.
Denoting Gaussian expectation values by angle brackets, the connected second
moments are $\langle u_ju_k\rangle-U_jU_k=C_{uu}\delta_{jk}$,
$\langle u_j\Delta_k\rangle-U_jV_k=C_{u\Delta}\delta_{jk}$ and
$\langle\Delta_j\Delta_k\rangle-V_jV_k=C_{\Delta\Delta}\delta_{jk}$, so that,
for example,
\begin{align}
	\langle\bDelta^2\u^2\rangle
	=
	(V^2+2C_{\Delta\Delta})(U^2+2C_{uu})
	+4C_{u\Delta}U\cdot V+4C_{u\Delta}^2 .
\end{align}
The complete momentum polynomial is
\begin{align}\label{eq:nonfac_momentum_polynomial}
	\mathcal P(\p^2,\b\cdot\p)
	={}&
	2z(1-z)
	\Big[
	q_1\langle\bDelta^2\u^2\rangle
	+q_2\langle\bDelta^2\,\u\cdot\bDelta\rangle
	+q_3\langle\bDelta^2\,\b\cdot\u\rangle
	\nn
	&\hspace{1.9cm}
	+q_4\langle\bDelta^2\,\b\cdot\bDelta\rangle
	+q_5\b^2\langle\bDelta^2\rangle
	\Big]
	\nn
	&-
	2(1-2z)
	\Big[
	q_1\langle(\b\cdot\bDelta)\u^2\rangle
	+q_2\langle(\b\cdot\bDelta)(\u\cdot\bDelta)\rangle
	\nn
	&\hspace{1.9cm}
	+q_3\langle(\b\cdot\bDelta)(\b\cdot\u)\rangle
	+q_4\langle(\b\cdot\bDelta)^2\rangle
	+q_5\b^2\langle\b\cdot\bDelta\rangle
	\Big]
	\nn
	&-2\b^2
	\Big[
	q_1\langle\u^2\rangle
	+q_2\langle\u\cdot\bDelta\rangle
	+q_3\langle\b\cdot\u\rangle
	+q_4\langle\b\cdot\bDelta\rangle
	+q_5\b^2
	\Big].
\end{align}

%===========================================================================
\section{Relative momentum integration and short time subtraction}
\label{app:p_integration}

\paragraph{Gaussian moments.}
Each fixed-$\p$ result of Sec.~\ref{sec:time_regions} is a polynomial
$\mathcal P(\p^2,\b\cdot\p)$ multiplying
$e^{-\alpha\p^2+i(\kappa/b)\b\cdot\p}$, with complex $\alpha$ and $\kappa$
that depend on the time variables, on $z$ and on the measurement channel.
With the regulator of Eq.~\eqref{eq:Jmed_fixed_p_strategy},
\begin{align}\label{eq:eta_generating_integral}
	\mathcal I_{\eta_{\rm UV}}(\alpha,\kappa)
	&\equiv
	\int_{\mathbb R^2}\frac{d^2\p}{(2\pi)^2}
	\exp\left[
	-(\alpha+\eta_{\rm UV})\p^2
	+i\frac{\kappa}{b}\b\cdot\p
	\right]
	\nn
	&=
	\frac{1}{4\pi(\alpha+\eta_{\rm UV})}
	\exp\left[-\frac{\kappa^2}{4(\alpha+\eta_{\rm UV})}\right],
\end{align}
and the polynomial moments follow by differentiation,
\begin{align}\label{eq:eta_polynomial_integral}
	\int_{\mathbb R^2}\frac{d^2\p}{(2\pi)^2}\,
	e^{-(\alpha+\eta_{\rm UV})\p^2+i(\kappa/b)\b\cdot\p}\,
	\mathcal P(\p^2,\b\cdot\p)
	=
	\mathcal P\!\left(-\partial_\alpha,-ib\,\partial_\kappa\right)
	\mathcal I_{\eta_{\rm UV}}(\alpha,\kappa).
\end{align}
The right hand sides are the analytic continuation from the domain in which
the integrals converge absolutely. The in--out contribution also contains
the inverse moment
\begin{align}\label{eq:eta_inverse_moment}
	\mathcal H_{\eta_{\rm UV}}(\alpha,\kappa)
	&\equiv
	\int_{\mathbb R^2}\frac{d^2\p}{(2\pi)^2}
	\frac{\b\cdot\p}{\p^2}\,
	e^{-(\alpha+\eta_{\rm UV})\p^2+i(\kappa/b)\b\cdot\p}
	=
	\frac{ib}{2\pi\kappa}
	\left[
	1-e^{-\kappa^2/[4(\alpha+\eta_{\rm UV})]}
	\right].
\end{align}

\paragraph{The three contributions.}
In the factorizable in--in contribution,
Eq.~\eqref{eq:J_inin_fac_fixed_p_complete}, the arguments are
\begin{align}\label{eq:eta_fac_arguments}
	\alpha_{\rm fac}=\frac{1}{4A_{\rm G}(x^++\tau,x^+)},
	\qquad
	\kappa_{{\rm fac},i}
	=b\left[
	a_i+\frac{B_{\rm G}(x^++\tau,x^+)}{2A_{\rm G}(x^++\tau,x^+)}
	\right].
\end{align}
In the non-factorizable contribution they are $\alpha$ and $\kappa_i$ of
Eq.~\eqref{eq:nonfac_alpha_kappa_Phi}, with the polynomial $\mathcal P$ of
Eq.~\eqref{eq:nonfac_fixed_p_complete} and the prefactor in
Eq.~\eqref{eq:J_inin_nonfac_fixed_p_complete}. In both cases
Eq.~\eqref{eq:eta_polynomial_integral} gives the $\p$ integral in closed form.
In the in--out contribution, Eq.~\eqref{eq:J_inout_fixed_p_complete}, we set
\begin{align}\label{eq:eta_inout_arguments}
	\alpha_{\rm out}=\frac{i}{2\omega^+c_2(\tau)},
	\qquad
	\kappa_{{\rm out},i}
	=b\left[
	a_i+\frac{z}{1+z^2}\bigl(r(\tau)-1\bigr)
	\right],
	\qquad
	r(\tau)=\frac{2ic_1(\tau)}{c_2(\tau)} ,
\end{align}
and the $\p$ integral of the integrand, without the common normalization, is
\begin{align}\label{eq:eta_inout_p_integrated}
	&\exp\left[
	-\frac{\hat q\b^2}{4}(L^+-\tau)
	-\frac{\hat q\b^2}{4(1+z^2)}\tau
	-\frac{\hat qz^2\b^2}{4(1+z^2)c_2(\tau)}
	\right]
	\nn
	&\quad\times
	\left[
	-2i\omega^+r^2(\tau)\,
	\mathcal I_{\eta_{\rm UV}}(\alpha_{\rm out},\kappa_{{\rm out},i})
	+\frac{z\hat q\,\omega^+}{c_2(\tau)}\,r(\tau)\,
	\mathcal H_{\eta_{\rm UV}}(\alpha_{\rm out},\kappa_{{\rm out},i})
	\right].
\end{align}

After these steps only the time integrations, the $z$ integral and the
channel sum remain.

\paragraph{Short-time subtraction.}
The regulator matters only where the Gaussian coefficient of $\p^2$ is
small, that is for $\tau\to0$ in the factorizable and in--out contributions
and for $\tau_3,\tau_4\to0$ in the nonfactorizable one. These limits carry
the large $|\p|$ tails of the separate contributions. At fixed $\p$ each
contribution is a time integral of the vacuum phase
$e^{-i\p^2(\bar x^+-x^+)/(2\omega^+)}$ times a weight $w(x^+,\bar x^+)$, which
varies on the medium time scales and on the resolution time
$(\omega^{+2}/\hat q\p^2)^{1/3}$. When the formation time $2\omega^+/\p^2$ is
short compared with these times, integration by parts in $\bar x^+-x^+$ gives
non-oscillating tails of order $(2\omega^+/\p^2)^2$, from the boundaries of
the time domains and from the in--in bulk. At large $|\p|$ the pair is
resolved immediately after the splitting, so that the equal-time weight is
$w(\bar x^+,\bar x^+)=D_F(\b;\bar x^+,0)\,D_F^3(\b;L^+,\bar x^+)$. The in--in
region then gives a tail proportional to $w(0,0)+w(L^+,L^+)$, whose bulk part
$w(L^+,L^+)-w(0,0)$ comes from the nonfactorizable contribution, and the
in--out region gives $-2w(L^+,L^+)$ with the same coefficient. The separate
tails are of order one, while their sum, $D_F^3(\b;L^+,0)-D_F(\b;L^+,0)$, is
the tail of Eq.~\eqref{eq:large_p_remainder} and vanishes at $b=0$.
Therefore, we subtract the short-time limits of the factorizable and in--out
integrands, in which the pair is compact and the three-point correlator
reduces to the parent dipole,
\begin{align}\label{eq:colour_transparency}
\lim_{|\r|\to0}D_F[\b-z\r]\,D_F[\r]=D_F[\b].
\end{align}
With $\alpha=i\tau/(2\omega^+)$, the limit of the factorizable integrand, where $\tau=\bar x^+-x^+$, is
\begin{align}\label{eq:short_time_kernels}
D_F(\b;\bar x^+,0)\,D_F^3(\b;L^+,\bar x^+)\,
\frac{\mathcal I_{\eta_{\rm UV}}(\alpha,a_ib)}{\alpha+\eta_{\rm UV}}
\left[1-\frac{a_i^2b^2}{4(\alpha+\eta_{\rm UV})}\right],
\end{align}
and that of the in--out integrand, where $\tau=L^+-x^+$, is
\begin{align}\label{eq:short_time_kernel_inout}
-2i\omega^+D_F(\b;L^+,0)\,\mathcal I_{\eta_{\rm UV}}(\alpha,a_ib).
\end{align}
They are subtracted at the same values of $z$, of the time variables, of the
channel and of $\eta_{\rm UV}$, and their integral is added back. At fixed
$\bar x^+$ the factorizable kernel is $2i\omega^+\partial_\tau$ of the
equal-time weight times $\mathcal I_{\eta_{\rm UV}}(\alpha,a_ib)$, so that the
added term is elementary. At $\eta_{\rm UV}=0$, with the common
normalization and real-part projection of Sec.~\ref{sec:time_regions}, it
reads
\begin{align}\label{eq:short_time_addback}
\frac{2\omega^{+2}}{\pi}\int_0^{L^+}\frac{d\bar x^+}{\bar x^+}
\Big[D_F(\b;\bar x^+,0)\,D_F^3(\b;L^+,\bar x^+)-D_F(\b;L^+,0)\Big]
\cos\frac{a_i^2b^2\omega^+}{2\bar x^+},
\end{align}
which converges at $\bar x^+\to0$ through the oscillation and vanishes at
$b=0$. The nonfactorizable integrand is not subtracted. It takes the
compact-pair form only for $\tau_4\lesssim(\omega^+\tau_3/\hat q)^{1/3}$, and
its tail, the bulk term above, is of the size of the tail of the sum. After
the subtraction the factorizable integrand has no residual logarithm of
$\eta_{\rm UV}$, and the subtracted and unsubtracted evaluations of the jet
function agree within the uncertainty of the regulator removal.

%===========================================================================
\section{Endpoint treatment and numerical implementation}\label{app:endpoint}

\paragraph{Quadrature.}
The time quadrature resolves the regulator time scale,
$t_{\rm UV}\sim2\omega^+\eta_{\rm UV}$. Ordered times are mapped to
triangular domains, and long oscillatory intervals are divided into phase
blocks. Exact and subtraction kernels are evaluated on the same nodes and
subtracted before integration. Each contribution is converged separately,
with a tolerance set by the final jet function, and so is their sum.

\paragraph{Longitudinal endpoints.}
As $\omega^+=z(1-z)p_0^+\to0$, the oscillator frequency of the three-point
function, $|\Omega_3|=[(1+z^2)\hat q/(2\omega^+)]^{1/2}$, grows and the
formation time $1/|\Omega_3|$ becomes short compared with $L^+$. Integrated
over $\p$ and $\K$, the medium-induced cross section is the
harmonic-oscillator BDMPS--Z spectrum~\cite{Baier:1996kr,Zakharov:1996fv},
\begin{equation}\label{eq:n_bdmps_spectrum}
\frac{d\widehat\sigma^{\rm med}_{qg|p_0}}{dz}
=\frac{\alpha_sC_F}{\pi}\,P^{\rm vac}_{q\to qg}(z)\,
\Re\ln\cos\!\left(\Omega_3L^+\right),
\end{equation}
which for $|\Omega_3|L^+\gg1$ becomes
\begin{equation}\label{eq:n_endpoint_scaling}
\frac{d\widehat\sigma^{\rm med}_{qg|p_0}}{dz}
\simeq
\frac{\alpha_sC_F}{\pi}\,L^+\,P^{\rm vac}_{q\to qg}(z)
\sqrt{\frac{(1+z^2)\,\hat q}{4z(1-z)\,p_0^+}} ,
\end{equation}
up to relative corrections of order $(|\Omega_3|L^+)^{-1}$. The $z$
integral is evaluated on three bulk panels, $[0.02,0.2]$ in the variable $\sqrt z$,
$[0.2,0.8]$ in $z$, and $[0.8,0.98]$ in $\sqrt{1-z}$, and on the strips
$10^{-3}\le z\le2\times10^{-2}$ and $10^{-3}\le1-z\le2\times10^{-2}$.
Below $10^{-3}$ each contribution is completed from its asymptotic form.
In units of the $z$ integrand of
Eq.~\eqref{eq:Jmed_fixed_p_strategy}, the factorizable contribution minus
its short-time kernel behaves as
\begin{align}\label{eq:n_endpoint_fac}
&C_{\rm right}\,(1-z)^{-1/2}+d_{\rm fac}\ln\frac{1}{1-z},
\nn
&C_{\rm right}=-\frac{\sqrt2\,C_F\alpha_s}{\pi}\,D_F(\b;L^+,0)\,\ell_\eta
\sqrt{\frac{\hat q}{p_0^+}},
\qquad
\ell_\eta=\frac{\ln(1+2\hat qL^+\eta_{\rm UV})}{2\hat q\,\eta_{\rm UV}},
\end{align}
with
$d_{\rm fac}=-\frac{2C_F\alpha_s}{\pi}
\Big[\frac{D_F(\b;L^+,0)}{2(1+2\hat qL^+\eta_{\rm UV})}-D_F^3(\b;L^+,0)\Big]$,
and the in--out contribution minus its kernel as
$-\frac{2C_F\alpha_s}{\pi}D_F(\b;L^+,0)\ln(|\Omega_3|L^+)$. For
$\eta_{\rm UV}\to0$, $\ell_\eta\to L^+$ and $C_{\rm right}$ follows from
Eq.~\eqref{eq:n_endpoint_scaling} with the weight $h_q+h_u=-(1-z)$, the
gluon channel being suppressed at $b\neq0$ by its phase $\b\cdot\p/(1-z)$.
The nonfactorizable contribution behaves as $A+B\ln[1/(1-z)]$, with
$A$ and $B$ fitted.
At $z\to0$ the factorizable contribution behaves as $C_{\rm left}\,z^{-1/2}$,
with $C_{\rm left}$ obtained by the argument that gives $C_{\rm right}$ and checked
numerically to better than $0.5\%$, the nonfactorizable one as
$A+B\ln(1/z)$ with fitted coefficients, and the in--out contribution
vanishes. The terms not fixed analytically are fitted on the strip nodes in
bases containing $\ln[1/(1-z)]$, $1$, $\sqrt{1-z}$ and $1-z$,
and likewise in $z$, and the spread over the bases is the endpoint
uncertainty.

The endpoint completion is performed at fixed $\eta_{\rm UV}$, before the
regulator is removed, following the order of integrations specified in
Sec.~\ref{sec:n_time_regions}.

\paragraph{Removal of the regulator.}
For each $b$ the jet function is computed, after the endpoint completion,
at $\eta_{\rm UV}=1/\Lambda^2$ with $\Lambda b=10$, $12$ and $14$, and with
$\Lambda=10$, $12$ and $14$~GeV for $b>1\,{\rm GeV}^{-1}$. These
values satisfy $\eta_{\rm UV}\ll b^2/4$, so that the measurement phases cut
off the large $|\p|$ tail of Eq.~\eqref{eq:large_p_remainder} before the
regulator acts and the remaining dependence on $\eta_{\rm UV}$ is
polynomial. The $\eta_{\rm UV}\to0^+$ value is the envelope of a quadratic
fit through the three points and a linear fit through the two smallest,
each enlarged by the endpoint uncertainties propagated with the fit
weights. A fourth point, $\Lambda b=17$ or $\Lambda=17$~GeV, is held out, and the regulator
uncertainty is the larger of the difference between the two fits and the
shift of the extrapolation when this point is included. The extrapolation
is checked for each time region as well as for the sum, so that an
unconverged cancellation between regions cannot hide behind a small total
uncertainty.

\paragraph{Impact parameter grid.}
The only value imposed on the curve is $J^{(1),{\rm med.\,rad.}}(0)=0$,
Eq.~\eqref{eq:n_sum_rule}. The nodes are uniform in $\ln b$ and are refined
until the interpolation error is below $0.1\%$ of the largest
$|J^{(1),{\rm med.\,rad.}}|$. At small $b$ the gluon channel loses the energy
of soft gluons, $1-z\lesssim b\sqrt{\hat qL^+}/2$, whose measurement phase
$\b\cdot\p/(1-z)$ decoheres. With the spectrum of
Eq.~\eqref{eq:n_endpoint_scaling} this gives
$J^{(1),{\rm med.\,rad.}}(b)\simeq-C\sqrt b$, which the nodes confirm at the
smallest $b$. Accordingly, we interpolate $J^{(1),{\rm med.\,rad.}}/\sqrt b$,
and below the smallest node, $b\approx10^{-2}\,{\rm GeV}^{-1}$, we use
$-C\sqrt b+Db^{3/2}$ matched to the smallest nodes. At large $b$ every
contribution decays through its broadening factors. Beyond the last node,
$b=3.16\,{\rm GeV}^{-1}$, where $|J^{(1),{\rm med.\,rad.}}|<3\times10^{-5}$,
the correction is set to zero, and the slowest Gaussian decay of the
broadening factors, $e^{-\hat qL^+b^2/12}$, makes the omitted tail
negligible. The quoted uncertainty combines quadrature, endpoint completion,
regulator removal, interpolation and the small-$b$ form as correlated
envelopes, so that the tabulated correction can be folded with another
impact-parameter kernel.

\bibliographystyle{bibstyle}
\bibliography{Lib}

@article{Mueller:2016gko,
    author = "Mueller, A. H. and Wu, Bin and Xiao, Bo-Wen and Yuan, Feng",
    title = "{Probing Transverse Momentum Broadening in Heavy Ion Collisions}",
    eprint = "1604.04250",
    archivePrefix = "arXiv",
    primaryClass = "hep-ph",
    doi = "10.1016/j.physletb.2016.10.037",
    journal = "Phys. Lett. B",
    volume = "763",
    pages = "208--212",
    year = "2016"
}

@article{Ali:1984yp,
    author = "Ali, Ahmed and Pietarinen, E. and Stirling, W. James",
    title = "{Transverse Energy-energy Correlations: A Test of Perturbative {QCD} for the Proton - Anti-proton Collider}",
    reportNumber = "CERN-TH-3800",
    doi = "10.1016/0370-2693(84)90283-1",
    journal = "Phys. Lett. B",
    volume = "141",
    pages = "447--454",
    year = "1984"
}

@article{Gao:2019ojf,
    author = "Gao, AnJie and Li, Hai Tao and Moult, Ian and Zhu, Hua Xing",
    title = "{Precision QCD Event Shapes at Hadron Colliders: The Transverse Energy-Energy Correlator in the Back-to-Back Limit}",
    eprint = "1901.04497",
    archivePrefix = "arXiv",
    primaryClass = "hep-ph",
    doi = "10.1103/PhysRevLett.123.062001",
    journal = "Phys. Rev. Lett.",
    volume = "123",
    number = "6",
    pages = "062001",
    year = "2019"
}

@article{Basham:1978bw,
    author = "Basham, C. Louis and Brown, Lowell S. and Ellis, Stephen D. and Love, Sherwin T.",
    title = "{Energy Correlations in electron - Positron Annihilation: Testing QCD}",
    reportNumber = "RLO-1388-759",
    doi = "10.1103/PhysRevLett.41.1585",
    journal = "Phys. Rev. Lett.",
    volume = "41",
    pages = "1585",
    year = "1978"
}

@article{Delacretaz:2018cfk,
    author = "Delacr{\'e}taz, Luca V. and Hartman, Thomas and Hartnoll, Sean A. and Lewkowycz, Aitor",
    title = "{Thermalization, Viscosity and the Averaged Null Energy Condition}",
    eprint = "1805.04194",
    archivePrefix = "arXiv",
    primaryClass = "hep-th",
    doi = "10.1007/JHEP10(2018)028",
    journal = "JHEP",
    volume = "10",
    pages = "028",
    year = "2018"
}

@article{Delacretaz:2020nit,
    author = "Delacretaz, Luca V.",
    title = "{Heavy Operators and Hydrodynamic Tails}",
    eprint = "2006.01139",
    archivePrefix = "arXiv",
    primaryClass = "hep-th",
    reportNumber = "EFI 20-10",
    doi = "10.21468/SciPostPhys.9.3.034",
    journal = "SciPost Phys.",
    volume = "9",
    number = "3",
    pages = "034",
    year = "2020"
}

@article{Chicherin:2023gxt,
    author = "Chicherin, Dmitry and Korchemsky, Gregory P. and Sokatchev, Emery and Zhiboedov, Alexander",
    title = "{Energy correlations in heavy states}",
    eprint = "2306.14330",
    archivePrefix = "arXiv",
    primaryClass = "hep-th",
    reportNumber = "CERN-TH-2023-109, IPhT-T23/051, LAPTH-034/23",
    doi = "10.1007/JHEP11(2023)134",
    journal = "JHEP",
    volume = "11",
    pages = "134",
    year = "2023"
}

@article{Cuomo:2025pjp,
    author = "Cuomo, Gabriel and Firat, Eren and Nardi, Filippo and Ricci, Lorenzo",
    title = "{Conformal collider physics at large charge}",
    eprint = "2503.21867",
    archivePrefix = "arXiv",
    primaryClass = "hep-th",
    doi = "10.1007/JHEP02(2026)223",
    journal = "JHEP",
    volume = "02",
    pages = "223",
    year = "2026"
}

@article{Grassi:2026tdb,
    author = "Grassi, Alba and Iossa, Cristoforo and Tizzano, Luigi and Zhiboedov, Alexander",
    title = "{Large Charge Jets in $\mathcal{N}=4$ SYM Energy Correlators}",
    eprint = "2610.10707",
    archivePrefix = "arXiv",
    primaryClass = "hep-th",
    reportNumber = "CERN-TH-2026-245",
    month = "10",
    year = "2026"
}

@article{Li:2021txc,
    author = "Li, Hai Tao and Makris, Yiannis and Vitev, Ivan",
    title = "{Energy-energy correlators in Deep Inelastic Scattering}",
    eprint = "2102.05669",
    archivePrefix = "arXiv",
    primaryClass = "hep-ph",
    doi = "10.1103/PhysRevD.103.094005",
    journal = "Phys. Rev. D",
    volume = "103",
    number = "9",
    pages = "094005",
    year = "2021"
}

@article{Moult:2018jzp,
    author = "Moult, Ian and Zhu, Hua Xing",
    title = "{Simplicity from Recoil: The Three-Loop Soft Function and Factorization for the Energy-Energy Correlation}",
    eprint = "1801.02627",
    archivePrefix = "arXiv",
    primaryClass = "hep-ph",
    doi = "10.1007/JHEP08(2018)160",
    journal = "JHEP",
    volume = "08",
    pages = "160",
    year = "2018"
}

@article{Gao:2023ivm,
    author = "Gao, Anjie and Li, Hai Tao and Moult, Ian and Zhu, Hua Xing",
    title = "{The transverse energy-energy correlator at next-to-next-to-next-to-leading logarithm}",
    eprint = "2312.16408",
    archivePrefix = "arXiv",
    primaryClass = "hep-ph",
    reportNumber = "MIT-CTP 5662",
    doi = "10.1007/JHEP09(2024)072",
    journal = "JHEP",
    volume = "09",
    pages = "072",
    year = "2024"
}

@article{Singh:2025scb,
    author = "Singh, Balbeer",
    title = "{Toward factorization of jet observable in dense media: An EFT approach}",
    eprint = "2505.18070",
    archivePrefix = "arXiv",
    primaryClass = "hep-ph",
    doi = "10.1142/S0217751X25300121",
    journal = "Int. J. Mod. Phys. A",
    volume = "40",
    number = "28",
    pages = "2530012",
    year = "2025"
}

@article{Singh:2024pwr,
    author = "Singh, Balbeer and Vaidya, Varun",
    title = "{Towards factorization with emergent scales for jets in dense media}",
    eprint = "2412.18967",
    archivePrefix = "arXiv",
    primaryClass = "hep-ph",
    month = "12",
    year = "2024"
}

@article{Singh:2024vwb,
    author = "Singh, Balbeer and Vaidya, Varun",
    title = "{Factorization for energy-energy correlator in heavy ion collision}",
    eprint = "2408.02753",
    archivePrefix = "arXiv",
    primaryClass = "hep-ph",
    doi = "10.1007/JHEP06(2025)071",
    journal = "JHEP",
    volume = "06",
    pages = "071",
    year = "2025"
}

@article{Baier:1996kr,
    author = "Baier, R. and Dokshitzer, Yuri L. and Mueller, Alfred H. and Peigne, S. and Schiff, D.",
    title = "{Radiative energy loss of high-energy quarks and gluons in a finite volume quark - gluon plasma}",
    eprint = "hep-ph/9607355",
    archivePrefix = "arXiv",
    reportNumber = "BI-TP-96-21, CUTP-759, LPTHE-ORSAY-96-34",
    doi = "10.1016/S0550-3213(96)00553-6",
    journal = "Nucl. Phys. B",
    volume = "483",
    pages = "291--320",
    year = "1997"
}

@article{Zakharov:1996fv,
    author = "Zakharov, B. G.",
    title = "{Fully quantum treatment of the Landau--Pomeranchuk--Migdal effect in QED and QCD}",
    eprint = "hep-ph/9607440",
    archivePrefix = "arXiv",
    doi = "10.1134/1.567126",
    journal = "JETP Lett.",
    volume = "63",
    pages = "952--957",
    year = "1996"
}

@article{Apolinario:2014csa,
    author = "Apolinario, Liliana and Armesto, Nestor and Milhano, Jose Guilherme and Salgado, Carlos A.",
    title = "{Medium-induced gluon radiation and colour decoherence beyond the soft approximation}",
    eprint = "1407.0599",
    archivePrefix = "arXiv",
    primaryClass = "hep-ph",
    doi = "10.1007/JHEP02(2015)119",
    journal = "JHEP",
    volume = "02",
    pages = "119",
    year = "2015"
}

@article{Liou:2013qya,
    author = "Liou, Tseh and Mueller, A. H. and Wu, Bin",
    title = "{Radiative $p_\bot$-broadening of high-energy quarks and gluons in QCD matter}",
    eprint = "1304.7677",
    archivePrefix = "arXiv",
    primaryClass = "hep-ph",
    doi = "10.1016/j.nuclphysa.2013.08.005",
    journal = "Nucl. Phys. A",
    volume = "916",
    pages = "102--125",
    year = "2013"
}

@article{Barata:2024ukm,
    author = "Barata, Jo{\~a}o and Kuzmin, Matvey V. and Milhano, Jos{\'e} Guilherme and Sadofyev, Andrey V.",
    title = "{Giving wake to energy-energy correlators: Hydrodynamic response on the celestial sphere}",
    eprint = "2412.03616",
    archivePrefix = "arXiv",
    primaryClass = "hep-ph",
    reportNumber = "CERN-TH-2024-180",
    doi = "10.1103/q8pr-djlw",
    journal = "Phys. Rev. D",
    volume = "112",
    number = "1",
    pages = "016005",
    year = "2025"
}

@article{Barata:2026pgh,
    author = "Barata, Jo{\~a}o and Kuzmin, Matvey V. and Moult, Ian and Sadofyev, Andrey V. and Silva, Jo{\~a}o M.",
    title = "{Hydrodynamics and Energy Correlators}",
    eprint = "2604.21971",
    archivePrefix = "arXiv",
    primaryClass = "hep-ph",
    reportNumber = "CERN-TH-2026-094",
    month = "4",
    year = "2026"
}

@article{Barata:2025fzd,
    author = "Barata, Jo{\~a}o and Moult, Ian and Sadofyev, V., Andrey and Silva, Jo{\~a}o M.",
    title = "{Dissecting Jet Modification in the Quark-Gluon Plasma with Multipoint Energy Correlators}",
    eprint = "2503.13603",
    archivePrefix = "arXiv",
    primaryClass = "hep-ph",
    reportNumber = "CERN-TH-2025-029",
    doi = "10.1103/9jzc-7jcv",
    journal = "Phys. Rev. Lett.",
    volume = "137",
    number = "5",
    pages = "052302",
    year = "2026"
}

@article{Baier:1996sk,
    author = "Baier, R. and Dokshitzer, Yuri L. and Mueller, Alfred H. and Peigne, S. and Schiff, D.",
    title = "{Radiative energy loss and p(T) broadening of high-energy partons in nuclei}",
    eprint = "hep-ph/9608322",
    archivePrefix = "arXiv",
    reportNumber = "CU-TP-760, BI-TP-96-26, LPTHE-ORSAY-96-61",
    doi = "10.1016/S0550-3213(96)00581-0",
    journal = "Nucl. Phys. B",
    volume = "484",
    pages = "265--282",
    year = "1997"
}

@article{Zakharov:1997uu,
    author = "Zakharov, B. G.",
    title = "{Radiative energy loss of high-energy quarks in finite size nuclear matter and quark - gluon plasma}",
    eprint = "hep-ph/9704255",
    archivePrefix = "arXiv",
    reportNumber = "LPTHE-ORSAY-97-09",
    doi = "10.1134/1.567389",
    journal = "JETP Lett.",
    volume = "65",
    pages = "615--620",
    year = "1997"
}

@article{Blaizot:2012fh,
    author = "Blaizot, Jean-Paul and Dominguez, Fabio and Iancu, Edmond and Mehtar-Tani, Yacine",
    title = "{Medium-induced gluon branching}",
    eprint = "1209.4585",
    archivePrefix = "arXiv",
    primaryClass = "hep-ph",
    doi = "10.1007/JHEP01(2013)143",
    journal = "JHEP",
    volume = "01",
    pages = "143",
    year = "2013"
}

@article{Isaksen:2023nlr,
    author = "Isaksen, Johannes Hamre and Tywoniuk, Konrad",
    title = "{Precise description of medium-induced emissions}",
    eprint = "2303.12119",
    archivePrefix = "arXiv",
    primaryClass = "hep-ph",
    doi = "10.1007/JHEP09(2023)049",
    journal = "JHEP",
    volume = "09",
    pages = "049",
    year = "2023"
}

@article{Casalderrey-Solana:2016jvj,
    author = "Casalderrey-Solana, Jorge and Gulhan, Doga and Milhano, Guilherme and Pablos, Daniel and Rajagopal, Krishna",
    title = "{Angular Structure of Jet Quenching Within a Hybrid Strong/Weak Coupling Model}",
    eprint = "1609.05842",
    archivePrefix = "arXiv",
    primaryClass = "hep-ph",
    reportNumber = "ICCUB-16-033, CERN-TH-2016-201, MIT-CTP-4830",
    doi = "10.1007/JHEP03(2017)135",
    journal = "JHEP",
    volume = "03",
    pages = "135",
    year = "2017"
}

@article{Chen:2017zte,
    author = "Chen, Wei and Cao, Shanshan and Luo, Tan and Pang, Long-Gang and Wang, Xin-Nian",
    title = "{Effects of jet-induced medium excitation in $\gamma$-hadron correlation in A+A collisions}",
    eprint = "1704.03648",
    archivePrefix = "arXiv",
    primaryClass = "nucl-th",
    doi = "10.1016/j.physletb.2017.12.015",
    journal = "Phys. Lett. B",
    volume = "777",
    pages = "86--90",
    year = "2018"
}

@article{KunnawalkamElayavalli:2017hxo,
    author = "Kunnawalkam Elayavalli, Raghav and Zapp, Korinna Christine",
    title = "{Medium response in JEWEL and its impact on jet shape observables in heavy ion collisions}",
    eprint = "1707.01539",
    archivePrefix = "arXiv",
    primaryClass = "hep-ph",
    reportNumber = "CERN-TH-2017-146, MCNET-17-11",
    doi = "10.1007/JHEP07(2017)141",
    journal = "JHEP",
    volume = "07",
    pages = "141",
    year = "2017"
}

@article{Iancu:2014kga,
    author = "Iancu, Edmond",
    title = "{The non-linear evolution of jet quenching}",
    eprint = "1403.1996",
    archivePrefix = "arXiv",
    primaryClass = "hep-ph",
    doi = "10.1007/JHEP10(2014)095",
    journal = "JHEP",
    volume = "10",
    pages = "095",
    year = "2014"
}

@article{Blaizot:2014bha,
    author = "Blaizot, Jean-Paul and Mehtar-Tani, Yacine",
    title = "{Renormalization of the jet-quenching parameter}",
    eprint = "1403.2323",
    archivePrefix = "arXiv",
    primaryClass = "hep-ph",
    doi = "10.1016/j.nuclphysa.2014.05.018",
    journal = "Nucl. Phys. A",
    volume = "929",
    pages = "202--229",
    year = "2014"
}

@article{Hofman:2008ar,
    author = "Hofman, Diego M. and Maldacena, Juan",
    title = "{Conformal collider physics: Energy and charge correlations}",
    eprint = "0803.1467",
    archivePrefix = "arXiv",
    primaryClass = "hep-th",
    doi = "10.1088/1126-6708/2008/05/012",
    journal = "JHEP",
    volume = "05",
    pages = "012",
    year = "2008"
}

@article{Belitsky:2013xxa,
    author = "Belitsky, A. V. and Hohenegger, S. and Korchemsky, G. P. and Sokatchev, E. and Zhiboedov, A.",
    title = "{From correlation functions to event shapes}",
    eprint = "1309.0769",
    archivePrefix = "arXiv",
    primaryClass = "hep-th",
    reportNumber = "CERN-PH-TH-2013-211, IPHT-T13-210, LAPTH-047-13",
    doi = "10.1016/j.nuclphysb.2014.04.020",
    journal = "Nucl. Phys. B",
    volume = "884",
    pages = "305--343",
    year = "2014"
}

@article{Belitsky:2013bja,
    author = "Belitsky, A. V. and Hohenegger, S. and Korchemsky, G. P. and Sokatchev, E. and Zhiboedov, A.",
    title = "{Event shapes in $\mathcal{N} = 4$ super-Yang-Mills theory}",
    eprint = "1309.1424",
    archivePrefix = "arXiv",
    primaryClass = "hep-th",
    reportNumber = "CERN-PH-TH-2013-212",
    doi = "10.1016/j.nuclphysb.2014.04.019",
    journal = "Nucl. Phys. B",
    volume = "884",
    pages = "206--256",
    year = "2014"
}

@article{Kang:2024dja,
    author = "Kang, Zhong-Bo and Penttala, Jani and Zhang, Congyue",
    title = "{Determination of the strong coupling constant and the Collins-Soper kernel from the energy-energy correlator in $e^+e^-$ collisions}",
    eprint = "2410.21435",
    archivePrefix = "arXiv",
    primaryClass = "hep-ph",
    month = "10",
    year = "2024"
}

@article{Li:2020bub,
    author = "Li, Hai Tao and Vitev, Ivan and Zhu, Yu Jiao",
    title = "{Transverse-Energy-Energy Correlations in Deep Inelastic Scattering}",
    eprint = "2006.02437",
    archivePrefix = "arXiv",
    primaryClass = "hep-ph",
    doi = "10.1007/JHEP11(2020)051",
    journal = "JHEP",
    volume = "11",
    pages = "051",
    year = "2020"
}

@article{Kang:2023oqj,
    author = "Kang, Zhong-Bo and Penttala, Jani and Zhao, Fanyi and Zhou, Yiyu",
    title = "{Transverse energy-energy correlators in the color-glass condensate at the electron-ion collider}",
    eprint = "2311.17142",
    archivePrefix = "arXiv",
    primaryClass = "hep-ph",
    reportNumber = "MIT-CTP/5649",
    doi = "10.1103/PhysRevD.109.094012",
    journal = "Phys. Rev. D",
    volume = "109",
    number = "9",
    pages = "094012",
    year = "2024"
}

@article{Kang:2024otf,
    author = "Kang, Zhong-Bo and Lee, Sookhyun and Penttala, Jani and Zhao, Fanyi and Zhou, Yiyu",
    title = "{Transverse energy-energy correlator for vector boson-tagged hadron production in pp and pA collisions}",
    eprint = "2410.02747",
    archivePrefix = "arXiv",
    primaryClass = "hep-ph",
    reportNumber = "MIT-CTP/5668",
    doi = "10.1103/1djw-8zrh",
    journal = "Phys. Rev. D",
    volume = "112",
    number = "1",
    pages = "014012",
    year = "2025"
}

@article{Kang:2025vjk,
    author = "Kang, Zhong-Bo and Kao, Robert and Li, Meijian and Penttala, Jani",
    title = "{Transverse energy-energy correlators at small x for photon-hadron production}",
    eprint = "2504.00069",
    archivePrefix = "arXiv",
    primaryClass = "hep-ph",
    doi = "10.1103/vllq-fhr7",
    journal = "Phys. Rev. D",
    volume = "112",
    number = "7",
    pages = "076006",
    year = "2025"
}

@article{Jia:2019qbl,
    author = "Jia, Jiangyong and Wei, Shu-Yi and Xiao, Bo-Wen and Yuan, Feng",
    title = "{Medium-Induced Transverse Momentum Broadening via Forward Dijet Correlations}",
    eprint = "1910.05290",
    archivePrefix = "arXiv",
    primaryClass = "hep-ph",
    doi = "10.1103/PhysRevD.101.094008",
    journal = "Phys. Rev. D",
    volume = "101",
    number = "9",
    pages = "094008",
    year = "2020"
}

@article{Andres:2024ksi,
    author = "Andres, Carlota and Dominguez, Fabio and Holguin, Jack and Marquet, Cyrille and Moult, Ian",
    title = "{Towards an interpretation of the first measurements of energy correlators in the quark-gluon plasma}",
    eprint = "2407.07936",
    archivePrefix = "arXiv",
    primaryClass = "hep-ph",
    doi = "10.1007/JHEP03(2025)166",
    journal = "JHEP",
    volume = "03",
    pages = "166",
    year = "2025"
}

@article{Yang:2023dwc,
    author = "Yang, Zhong and He, Yayun and Moult, Ian and Wang, Xin-Nian",
    title = "{Probing the Short-Distance Structure of the Quark-Gluon Plasma with Energy Correlators}",
    eprint = "2310.01500",
    archivePrefix = "arXiv",
    primaryClass = "hep-ph",
    doi = "10.1103/PhysRevLett.132.011901",
    journal = "Phys. Rev. Lett.",
    volume = "132",
    number = "1",
    pages = "011901",
    year = "2024"
}

@article{Barata:2023bhh,
    author = "Barata, Jo{\~a}o and Caucal, Paul and Soto-Ontoso, Alba and Szafron, Robert",
    title = "{Advancing the understanding of energy-energy correlators in heavy-ion collisions}",
    eprint = "2312.12527",
    archivePrefix = "arXiv",
    primaryClass = "hep-ph",
    reportNumber = "CERN-TH-2023-243",
    doi = "10.1007/JHEP11(2024)060",
    journal = "JHEP",
    volume = "11",
    pages = "060",
    year = "2024"
}

@article{Budhraja:2025ulx,
    author = "Budhraja, Ankita and Singh, Balbeer",
    title = "{Exploiting {\ensuremath{\nu}}-dependence of projected energy correlators in HICs}",
    eprint = "2503.20019",
    archivePrefix = "arXiv",
    primaryClass = "hep-ph",
    doi = "10.1016/j.physletb.2025.140079",
    journal = "Phys. Lett. B",
    volume = "872",
    pages = "140079",
    year = "2026"
}

@article{Apolinario:2025vtx,
    author = "Apolin{\'a}rio, Liliana and Kunnawalkam Elayavalli, Raghav and Madureira, Nuno Olavo and Sheng, Jun-Xing and Wang, Xin-Nian and Yang, Zhong",
    title = "{Flavor dependence of energy-energy correlators}",
    eprint = "2502.11406",
    archivePrefix = "arXiv",
    primaryClass = "hep-ph",
    doi = "10.1103/4sfq-315y",
    journal = "Phys. Rev. D",
    volume = "112",
    number = "5",
    pages = "054018",
    year = "2025"
}

@article{Chen:2016vem,
    author = "Chen, Lin and Qin, Guang-You and Wei, Shu-Yi and Xiao, Bo-Wen and Zhang, Han-Zhong",
    title = "{Probing Transverse Momentum Broadening via Dihadron and Hadron-jet Angular Correlations in Relativistic Heavy-ion Collisions}",
    eprint = "1607.01932",
    archivePrefix = "arXiv",
    primaryClass = "hep-ph",
    doi = "10.1016/j.physletb.2017.09.031",
    journal = "Phys. Lett. B",
    volume = "773",
    pages = "672--676",
    year = "2017"
}

@article{STAR:2017hhs,
    author = "Adamczyk, L. and others",
    collaboration = "STAR",
    title = "{Measurements of jet quenching with semi-inclusive hadron+jet distributions in Au+Au collisions at $\sqrt{s_{NN}}$ = 200 GeV}",
    eprint = "1702.01108",
    archivePrefix = "arXiv",
    primaryClass = "nucl-ex",
    doi = "10.1103/PhysRevC.96.024905",
    journal = "Phys. Rev. C",
    volume = "96",
    number = "2",
    pages = "024905",
    year = "2017"
}

@article{Mueller:2016xoc,
    author = "Mueller, A. H. and Wu, Bin and Xiao, Bo-Wen and Yuan, Feng",
    title = "{Medium Induced Transverse Momentum Broadening in Hard Processes}",
    eprint = "1608.07339",
    archivePrefix = "arXiv",
    primaryClass = "hep-ph",
    doi = "10.1103/PhysRevD.95.034007",
    journal = "Phys. Rev. D",
    volume = "95",
    number = "3",
    pages = "034007",
    year = "2017"
}

@article{Singh:2026qsa,
    author = "Singh, Balbeer",
    title = "{Probing jet medium interaction with generalized projected energy correlators}",
    eprint = "2609.01416",
    archivePrefix = "arXiv",
    primaryClass = "hep-ph",
    month = "9",
    year = "2026"
}

@article{Yang:2026bsk,
    author = "Yang, Zhong and Kunnawalkam Elayavalli, Raghav and Wang, Xin-Nian",
    title = "{Energy-energy correlators inside single inclusive jets in heavy-ion collisions with the CoLBT-hydro model}",
    eprint = "2605.28788",
    archivePrefix = "arXiv",
    primaryClass = "hep-ph",
    doi = "10.1103/zpvs-4x68",
    journal = "Phys. Rev. C",
    volume = "114",
    number = "3",
    pages = "034914",
    year = "2026"
}

@article{Belin:2026wkc,
    author = "Belin, Alexandre and Borak, Nathan and Henriksson, Johan and Piron, Romain and Zhiboedov, Alexander",
    title = "{Positivity in energy correlators and the event distribution formula}",
    eprint = "2608.26246",
    archivePrefix = "arXiv",
    primaryClass = "hep-th",
    reportNumber = "CERN-TH-2026-198",
    month = "8",
    year = "2026"
}

@article{ATLAS:2023tgo,
    author = "Aad, Georges and others",
    collaboration = "ATLAS",
    title = "{Determination of the strong coupling constant from transverse energy$-$energy correlations in multijet events at $\sqrt{s} = 13$ TeV with the ATLAS detector}",
    eprint = "2301.09351",
    archivePrefix = "arXiv",
    primaryClass = "hep-ex",
    reportNumber = "CERN-EP-2022-282",
    doi = "10.1007/JHEP07(2023)085",
    journal = "JHEP",
    volume = "07",
    pages = "085",
    year = "2023"
}

@article{Ebert:2020sfi,
    author = "Ebert, Markus A. and Mistlberger, Bernhard and Vita, Gherardo",
    title = "{The Energy-Energy Correlation in the back-to-back limit at N$^{3}$LO and N$^{3}$LL'}",
    eprint = "2012.07859",
    archivePrefix = "arXiv",
    primaryClass = "hep-ph",
    reportNumber = "MIT-CTP/5263, SLAC-PUB-17579, MPP-2020-225",
    doi = "10.1007/JHEP08(2021)022",
    journal = "JHEP",
    volume = "08",
    pages = "022",
    year = "2021"
}

@article{Zhao:2025ogc,
    author = "Zhao, Wenbin and Koch, Volker and Yuan, Feng",
    title = "{Particle Correlations in Jets}",
    eprint = "2507.18790",
    archivePrefix = "arXiv",
    primaryClass = "hep-ph",
    doi = "10.1103/yscq-npz8",
    journal = "Phys. Rev. Lett.",
    volume = "137",
    number = "16",
    pages = "162301",
    year = "2026"
}

@article{Leitao:2026fgh,
    author = "Leit{\~a}o, Marco and Milhano, Jos{\'e} Guilherme and Soto-Ontoso, Alba",
    title = "{In-medium QCD splittings beyond the soft, large-$N_c$ and harmonic-oscillator approximations all at once}",
    eprint = "2606.26039",
    archivePrefix = "arXiv",
    primaryClass = "hep-ph",
    month = "6",
    year = "2026"
}

@inproceedings{Neill:2022lqx,
    author = "Neill, Duff and Vita, Gherardo and Vitev, Ivan and Zhu, Hua Xing",
    title = "{Energy-Energy Correlators for Precision QCD}",
    booktitle = "{Snowmass 2021}",
    eprint = "2203.07113",
    archivePrefix = "arXiv",
    primaryClass = "hep-ph",
    month = "3",
    year = "2022"
}

@article{Collins:1981uk,
    author = "Collins, John C. and Soper, Davison E.",
    title = "{Back-To-Back Jets in QCD}",
    doi = "10.1016/0550-3213(81)90339-4",
    journal = "Nucl. Phys. B",
    volume = "193",
    pages = "381",
    year = "1981",
    note = "[Erratum: Nucl. Phys. B 213, 545 (1983)]"
}

@article{Sterman:1975xv,
	author = {Sterman, George F.},
	month = {12},
	reportnumber = {ILL-TH-75-32},
	title = {{Jet Structure in e+ e- Annihilation with Massless Hadrons}},
	year = {1975}}

@article{Barata:2026icn,
    author = "Barata, Jo{\~a}o and Li, Meijian and Qian, Wenyang and Salgado, Carlos A. and Silva, Jo{\~a}o M.",
    title = "{Quantum simulating multi-particle processes in high energy nuclear physics: dijet production and color (de)coherence}",
    eprint = "2604.11616",
    archivePrefix = "arXiv",
    primaryClass = "hep-ph",
    reportNumber = "CERN-TH-2026-082",
    doi = "10.1007/JHEP07(2026)063",
    journal = "JHEP",
    volume = "07",
    pages = "063",
    year = "2026"
}

@article{Korchemsky:2019nzm,
    author = "Korchemsky, G. P.",
    title = "{Energy correlations in the end-point region}",
    eprint = "1905.01444",
    archivePrefix = "arXiv",
    primaryClass = "hep-th",
    reportNumber = "IPhT-T19/041",
    doi = "10.1007/JHEP01(2020)008",
    journal = "JHEP",
    volume = "01",
    pages = "008",
    year = "2020"
}

@article{Scimemi:2019cmh,
    author = "Scimemi, Ignazio and Vladimirov, Alexey",
    title = "{Non-perturbative structure of semi-inclusive deep-inelastic and Drell-Yan scattering at small transverse momentum}",
    eprint = "1912.06532",
    archivePrefix = "arXiv",
    primaryClass = "hep-ph",
    doi = "10.1007/JHEP06(2020)137",
    journal = "JHEP",
    volume = "06",
    pages = "137",
    year = "2020"
}

@article{Ke:2023ixa,
    author = "Ke, Weiyao and Vitev, Ivan",
    title = "{Understanding parton evolution in matter from renormalization group analysis}",
    eprint = "2301.11940",
    archivePrefix = "arXiv",
    primaryClass = "hep-ph",
    reportNumber = "LA-UR-23-20730",
    doi = "10.1016/j.physletb.2024.138751",
    journal = "Phys. Lett. B",
    volume = "854",
    pages = "138751",
    year = "2024"
}

@article{Bossi:2026mdz,
    author = "Bossi, Hannah and Larose, Maxence and Mehtar-Tani, Yacine",
    title = "{The Colors of Jet Quenching}",
    eprint = "2609.05609",
    archivePrefix = "arXiv",
    primaryClass = "hep-ph",
    month = "9",
    year = "2026"
}

@article{Bauer:2003mga,
    author = "Bauer, Christian W. and Pirjol, Dan and Stewart, Iain W.",
    title = "{On Power suppressed operators and gauge invariance in SCET}",
    eprint = "hep-ph/0303156",
    archivePrefix = "arXiv",
    reportNumber = "MIT-CTP-3349, UCSD-PTH-03-01",
    doi = "10.1103/PhysRevD.68.034021",
    journal = "Phys. Rev. D",
    volume = "68",
    pages = "034021",
    year = "2003"
}

@article{Chiu:2012ir,
    author = "Chiu, Jui-Yu and Jain, Ambar and Neill, Duff and Rothstein, Ira Z.",
    title = "{A Formalism for the Systematic Treatment of Rapidity Logarithms in Quantum Field Theory}",
    eprint = "1202.0814",
    archivePrefix = "arXiv",
    primaryClass = "hep-ph",
    reportNumber = "INT-PUB-11-057, CALT-68-2864",
    doi = "10.1007/JHEP05(2012)084",
    journal = "JHEP",
    volume = "05",
    pages = "084",
    year = "2012"
}

@article{Andres:2026qrt,
    author = "Andres, Carlota and Dominguez, Fabio",
    title = "{Full energy fraction and angular dependence of medium-induced splittings in the large-$N_c$ limit}",
    eprint = "2603.27846",
    archivePrefix = "arXiv",
    primaryClass = "hep-ph",
    month = "3",
    year = "2026"
}

@article{Lamas:2026ntq,
    author = "Lamas, Carlos and Salgado, Carlos A. and Wu, Bin",
    title = "{Factorization and virtuality evolution of jet functions in heavy-ion collisions}",
    eprint = "2609.35653",
    archivePrefix = "arXiv",
    primaryClass = "hep-ph",
    month = "9",
    year = "2026"
}

@article{Casalderrey-Solana:2026ftw,
    author = "Casalderrey-Solana, Jorge and Milhano, Jos{\'e} Guilherme and Pablos, Daniel and Rajagopal, Krishna and Yao, Xiaojun",
    title = "{Computationally Efficient Description of Medium Response to Jets in Heavy Ion Collisions}",
    eprint = "2609.10679",
    archivePrefix = "arXiv",
    primaryClass = "hep-ph",
    month = "9",
    year = "2026"
}

@article{Isaksen:2020npj,
    author = "Isaksen, Johannes Hamre and Tywoniuk, Konrad",
    title = "{Wilson line correlators beyond the large-N$_{c}$}",
    eprint = "2107.02542",
    archivePrefix = "arXiv",
    primaryClass = "hep-ph",
    doi = "10.1007/JHEP11(2021)125",
    journal = "JHEP",
    volume = "11",
    pages = "125",
    year = "2021"
}

@article{Sadofyev:2021ohn,
    author = "Sadofyev, Andrey V. and Sievert, Matthew D. and Vitev, Ivan",
    title = "{Ab~initio coupling of jets to collective flow in the opacity expansion approach}",
    eprint = "2104.09513",
    archivePrefix = "arXiv",
    primaryClass = "hep-ph",
    reportNumber = "LA-UR-21-21420",
    doi = "10.1103/PhysRevD.104.094044",
    journal = "Phys. Rev. D",
    volume = "104",
    number = "9",
    pages = "094044",
    year = "2021"
}

@article{Barata:2022krd,
    author = "Barata, Jo{\~a}o and Sadofyev, Andrey V. and Salgado, Carlos A.",
    title = "{Jet broadening in dense inhomogeneous matter}",
    eprint = "2202.08847",
    archivePrefix = "arXiv",
    primaryClass = "hep-ph",
    doi = "10.1103/PhysRevD.105.114010",
    journal = "Phys. Rev. D",
    volume = "105",
    number = "11",
    pages = "114010",
    year = "2022"
}

@article{Barata:2023qds,
    author = "Barata, Jo{\~a}o and Mayo L{\'o}pez, Xo{\'a}n and Sadofyev, Andrey V. and Salgado, Carlos A.",
    title = "{Medium induced gluon spectrum in dense inhomogeneous matter}",
    eprint = "2304.03712",
    archivePrefix = "arXiv",
    primaryClass = "hep-ph",
    doi = "10.1103/PhysRevD.108.034018",
    journal = "Phys. Rev. D",
    volume = "108",
    number = "3",
    pages = "034018",
    year = "2023"
}

@article{Kuzmin:2023hko,
    author = "Kuzmin, Matvey V. and Mayo L{\'o}pez, Xo{\'a}n and Reiten, Jared and Sadofyev, Andrey V.",
    title = "{Jet quenching in anisotropic flowing matter}",
    eprint = "2309.00683",
    archivePrefix = "arXiv",
    primaryClass = "hep-ph",
    doi = "10.1103/PhysRevD.109.014036",
    journal = "Phys. Rev. D",
    volume = "109",
    number = "1",
    pages = "014036",
    year = "2024"
}

@article{Altenburger:2025iqa,
    author = "Altenburger, Alois and Boguslavski, Kirill and Lindenbauer, Florian",
    title = "{Jet broadening and radiation in the early anisotropic plasma in heavy-ion collisions}",
    eprint = "2509.03868",
    archivePrefix = "arXiv",
    primaryClass = "hep-ph",
    reportNumber = "MIT-CTP/5917",
    month = "9",
    year = "2025"
}

@article{Lindenbauer:2025ctw,
    author = "Lindenbauer, Florian",
    title = "{Gluon splitting rates in an anisotropic plasma in the AMY formalism}",
    eprint = "2509.09897",
    archivePrefix = "arXiv",
    primaryClass = "hep-ph",
    reportNumber = "MIT-CTP/5920",
    month = "9",
    year = "2025"
}

@article{Bossi:2024qho,
    author = "Bossi, Hannah and Kudinoor, Arjun Srinivasan and Moult, Ian and Pablos, Daniel and Rai, Ananya and Rajagopal, Krishna",
    title = "{Imaging the wakes of jets with energy-energy-energy correlators}",
    eprint = "2407.13818",
    archivePrefix = "arXiv",
    primaryClass = "hep-ph",
    reportNumber = "MIT-CTP-5739",
    doi = "10.1007/JHEP12(2024)073",
    journal = "JHEP",
    volume = "12",
    pages = "073",
    year = "2024"
}

@article{Luo:2023nsi,
    author = "Luo, Tan and He, Yayun and Cao, Shanshan and Wang, Xin-Nian",
    title = "{Linear Boltzmann transport for jet propagation in the quark-gluon plasma: Inelastic processes and jet modification}",
    eprint = "2306.13742",
    archivePrefix = "arXiv",
    primaryClass = "nucl-th",
    doi = "10.1103/PhysRevC.109.034919",
    journal = "Phys. Rev. C",
    volume = "109",
    number = "3",
    pages = "034919",
    year = "2024"
}

@article{Yang:2022nei,
    author = "Yang, Zhong and Luo, Tan and Chen, Wei and Pang, Long-Gang and Wang, Xin-Nian",
    title = "{3D Structure of Jet-Induced Diffusion Wake in an Expanding Quark-Gluon Plasma}",
    eprint = "2203.03683",
    archivePrefix = "arXiv",
    primaryClass = "hep-ph",
    doi = "10.1103/PhysRevLett.130.052301",
    journal = "Phys. Rev. Lett.",
    volume = "130",
    number = "5",
    pages = "052301",
    year = "2023"
}

@article{Tachibana:2017syd,
    author = "Tachibana, Yasuki and Chang, Ning-Bo and Qin, Guang-You",
    title = "{Full jet in quark-gluon plasma with hydrodynamic medium response}",
    eprint = "1701.07951",
    archivePrefix = "arXiv",
    primaryClass = "nucl-th",
    doi = "10.1103/PhysRevC.95.044909",
    journal = "Phys. Rev. C",
    volume = "95",
    number = "4",
    pages = "044909",
    year = "2017"
}

@article{Bahder:2024jpa,
    author = "Bahder, Joseph and Rahman, Hasan and Sievert, Matthew D. and Vitev, Ivan",
    title = "{Signatures of jet drift in quark-gluon plasma hard-probe observables}",
    eprint = "2412.05474",
    archivePrefix = "arXiv",
    primaryClass = "nucl-th",
    doi = "10.1103/8yzj-2l5c",
    journal = "Phys. Rev. Res.",
    volume = "8",
    number = "1",
    pages = "L012016",
    year = "2026"
}

@article{Barata:2024bqp,
    author = "Barata, Jo{\~a}o and Salgado, Carlos A. and Silva, Jo{\~a}o M.",
    title = "{Gluon to $ \textrm{q}\overline{\textrm{q}} $ antenna in anisotropic QCD matter: spin-polarized and azimuthal jet observables}",
    eprint = "2407.04774",
    archivePrefix = "arXiv",
    primaryClass = "hep-ph",
    doi = "10.1007/JHEP12(2024)023",
    journal = "JHEP",
    volume = "12",
    pages = "023",
    year = "2024"
}

@article{Barata:2025zku,
    author = "Barata, Jo{\~a}o and Milhano, Jos{\'e} Guilherme and Sadofyev, Andrey V. and Silva, Jo{\~a}o M.",
    title = "{Early-Time Dynamics of Heavy-Ion Collisions through Energy Correlators: celestial blocks and the spacetime structure of out-of-equilibrium QCD matter}",
    eprint = "2512.17009",
    archivePrefix = "arXiv",
    primaryClass = "hep-ph",
    reportNumber = "CERN-TH-2025-257",
    month = "12",
    year = "2025"
}

@article{Danhoni:2026gve,
    author = "Danhoni, Isabella and Mullins, Nicki and Noronha, Jorge",
    title = "{Covariant diffusion tensor for jet momentum broadening out of equilibrium}",
    eprint = "2603.00844",
    archivePrefix = "arXiv",
    primaryClass = "hep-ph",
    month = "3",
    year = "2026"
}

@article{Avramescu:2024poa,
    author = {Avramescu, Dana and Greco, Vincenzo and Lappi, Tuomas and M{\"a}ntysaari, Heikki and M{\"u}ller, David},
    title = "{Impact of glasma on heavy flavor azimuthal correlations and spectra}",
    eprint = "2409.10564",
    archivePrefix = "arXiv",
    primaryClass = "hep-ph",
    doi = "10.1103/PhysRevD.111.074036",
    journal = "Phys. Rev. D",
    volume = "111",
    number = "7",
    pages = "074036",
    year = "2025"
}

@article{Barata:2024wsu,
    author = "Barata, Joao and Kang, Zhong-Bo and Mayo L\'opez, Xo\'an and Penttala, Jani",
    title = "{Energy-Energy Correlator for jet production in $pp$ and $pA$ collisions}",
    eprint = "2411.11782",
    archivePrefix = "arXiv",
    primaryClass = "hep-ph",
    reportNumber = "CERN-TH-2024-201",
    month = "11",
    year = "2024"
}

@article{Barata:2025uxp,
    author = "Barata, Jo{\~a}o and Brewer, Jasmine and Lee, Kyle and Silva, Jo{\~a}o M.",
    title = "{Heavy Quark Pair Energy Correlators: From Profiling Partonic Splittings to Probing Heavy-Flavor Fragmentation}",
    eprint = "2508.19404",
    archivePrefix = "arXiv",
    primaryClass = "hep-ph",
    reportNumber = "CERN-TH-2025-172, MIT-CTP 5888",
    month = "8",
    year = "2025"
}

@article{Ke:2025ibt,
    author = "Ke, Weiyao and Mecaj, Bianka and Vitev, Ivan",
    title = "{Renormalization Group Evolution for In-medium Energy Correlators}",
    eprint = "2512.11952",
    archivePrefix = "arXiv",
    primaryClass = "hep-ph",
    reportNumber = "LA-UR-25-24936",
    month = "12",
    year = "2025"
}

@article{Budhraja:2026pyi,
    author = "Budhraja, Ankita and Pels, Isabelle and Waalewijn, Wouter J.",
    title = "{Higher-point energy correlators: factorization in the back-to-back limit {\&} non-perturbative effects}",
    eprint = "2603.16996",
    archivePrefix = "arXiv",
    primaryClass = "hep-ph",
    doi = "10.1007/JHEP08(2026)154",
    journal = "JHEP",
    volume = "08",
    pages = "154",
    year = "2026"
}

@article{Barata:2024xwy,
    author = "Barata, Jo{\~a}o and Hauksson, Sigtryggur and Mayo L{\'o}pez, Xo{\'a}n and Sadofyev, Andrey V.",
    title = "{Jet quenching in the glasma phase: Medium-induced radiation}",
    eprint = "2406.07615",
    archivePrefix = "arXiv",
    primaryClass = "hep-ph",
    doi = "10.1103/PhysRevD.110.094055",
    journal = "Phys. Rev. D",
    volume = "110",
    number = "9",
    pages = "094055",
    year = "2024"
}

\end{document}